\documentclass[twocolumn]{article}
\pdfoutput=1
\usepackage{cuted}

\usepackage{arxiv}
\usepackage{stfloats}
\usepackage{xcolor}
\usepackage{graphicx}
\usepackage{longtable}
\usepackage{flafter}
\usepackage{amsmath}
\usepackage{amssymb}
\usepackage{pdfpages}
\usepackage{subcaption}
\usepackage{array}
\usepackage[nottoc,notlof,notlot]{tocbibind}
\usepackage{xurl}
\usepackage{booktabs}

\usepackage[ruled,linesnumbered]{algorithm2e}
\usepackage{tabularx}
\usepackage{subcaption}

 \usepackage[numbers]{natbib}
\usepackage[displaymath,switch]{lineno}

\newcommand{\RN}[1]{\uppercase\expandafter{\romannumeral#1}}
\usepackage{setspace}
\usepackage[bottom]{footmisc} % Bilder nicht mehr unter der Fußnote, wenn mit [b] spezifiziert
\usepackage{hyperref}
\usepackage{babel}
\usepackage{cleveref}
\usepackage{tikz}
\usepackage{ifthen}
\usetikzlibrary{calc}
\usepackage{intcalc}
\usepackage{xcolor}
\usepackage{bigints} % to use bigger integral symbols
\usepackage{neuralnetwork} % to draw a neural network
\usepackage{amsmath,mathtools}
\usetikzlibrary{positioning}% To get more advances positioning options
\usetikzlibrary{arrows}% To get more arrow heads
\usepackage{blkarray, bigstrut} % for lgtdata

\usepackage[parfill]{parskip}

\usepackage{pgfplots}
  \pgfplotsset{compat=newest}
  \usetikzlibrary{plotmarks}
  \usetikzlibrary{arrows.meta}
  \usepgfplotslibrary{patchplots}
  \usepackage{grffile}
  \usepackage{amsmath}

  \pgfplotsset{plot coordinates/math parser=false}
  \newlength\figureheight
  \newlength\figurewidth

\usetikzlibrary{decorations.markings}
\usetikzlibrary{decorations.pathmorphing}
\usetikzlibrary{arrows.meta}
\usetikzlibrary{shadows,arrows,positioning,shapes.geometric}
\pgfdeclarelayer{background}
\pgfdeclarelayer{foreground}
\pgfsetlayers{background,main,foreground}
\pgfplotsset{every axis/.append style={
                    label style={font=\scriptsize},
                    tick label style={font=\scriptsize}, 
                    legend style={font=\scriptsize}
                    }}

\graphicspath{{Bilder/}{appendix/bilder/}{appendix/bilder/waveNumberEvolution/}{body/04_benchmarkWithoutWind/bilder/}{body/06_benchmarkWithWind/bilder/}{body/07_flexingMuscles/bilder/}}%{playground/bilder/}}

\newcommand{\mtensor}[1]{\mathbf{#1}}
\AlgoDontDisplayBlockMarkers\AlgoDontDisplayGroupMarkers%   % removes vertical lines along blocks and prevents writing blocks in one single line
\DeclareMathOperator*{\argmin}{arg\,min}

\newlength{\picWidth}
\newcommand{\philipp}[1]{{#1}}

\newcommand{\g}[0]{\mathord{g}}
\newcommand{\G}[0]{\mathbf{G}}
\newcommand{\ampMod}[0]{\tilde{\mathord{a}}}
\newcommand{\phaseShift}[0]{\varphi}
\newcommand{\x}[0]{\mathbf{x}}
\newcommand{\indexMic}[0]{\mathord{n}}
\newcommand{\indexSpeak}[0]{\mathord{m}}
\newcommand{\numberSpeak}[0]{\mathord{M}}
\newcommand{\numberMic}[0]{\mathord{N}}
\newcommand{\indexMicSpeak}[0]{\indexMic\indexSpeak}
\newcommand{\waveNumber}[0]{\mathord{k}}
\newcommand{\waveNumberMod}[0]{\tilde{\mathord{k}}}
\newcommand{\phaseMod}[0]{\tilde{\mathord{\varphi}}}
\newcommand{\Ma}[0]{\textit{Ma}}
\newcommand{\pScalar}[0]{\mathord{p}}
\newcommand{\p}[0]{\mathbf{p}}
\newcommand{\f}[0]{\mathord{f}}
\newcommand{\s}[0]{\mathord{s}}
\newcommand{\speedSound}[0]{\mathord{c}}
\newcommand{\greenFunction}[0]{\mathcal{G}}
\newcommand{\wScalar}[0]{\mathord{w}}
\newcommand{\w}[0]{\mathbf{w}}
\newcommand{\NN}[0]{\text{NN}}
\newcommand{\regParam}[0]{\mathord{\lambda}}
\newcommand{\J}[0]{\mathord{J}}

\newcommand{\iterationStep}[0]{\mathord{l}}
\renewcommand{\S}{\mathord{S}}
\newcommand{\AC}[0]{\text{AC}}
\newcommand{\RE}[0]{\text{RE}}
\newcommand{\REBoundary}[0]{\text{RE}_\partial}
\newcommand{\ampPressure}[0]{\hat{\mathord{p}}}
\newcommand{\phasePressure}[0]{\mathord{\phi}}
\newcommand{\phaseSpeaker}[0]{\mathord{\phi}}
\newcommand{\backgroundFlow}[0]{\mathord{u_0}}

\newcommand{\unit}[1]{\,\text{#1}}

\title{Supervised Learning for Multi Zone Sound Field Reproduction under Harsh Environmental Conditions}

\date{\today}	% Here you can change the date presented in the paper title
 \author{ Henry Sallandt
            \thanks{corresponding author} \\
       Institute of Fluid Mechanics and Engineering Acoustics\\
        Technical University Berlin\\
        M\"uller-Breslau-Str. 15, 10623 Berlin, Germany\\
	\texttt{jannick.h.sallandt@tu-berlin.de} \\
	\And
	Philipp Krah\\
	Institute of Mathematics\\
	Technical University Berlin\\
	Straße des 17. Juni 136, 10623 Berlin, Germany\\
    \texttt{krah@math.tu-berlin.de} \\
	\And
	Mathias Lemke \\
	 Institute of Fluid Mechanics and Engineering Acoustics\\
        Technical University Berlin\\
        M\"uller-Breslau-Str. 15, 10623 Berlin, Germany\\
	\texttt{mathias.lemke@tnt.tu-berlin.de} \\
}

\begin{document}
\begin{strip}
\maketitle

\begin{abstract}
This manuscript presents an approach for multi zone sound field reproduction using supervised learning. Traditional multi zone sound field reproduction methods assume constant speed of sound, neglecting nonlinear effects like wind and temperature stratification. We show how to overcome these restrictions using supervised learning of transfer functions. The quality of the solution is measured by the acoustic contrast and the reproduction error. Our results show that for the chosen setup, even with relatively small wind speeds, the acoustic contrast and reproduction error can be improved by up to $16\unit{dB}$, when wind is considered in the trained model.
\end{abstract}
\vspace{1cm}
\vspace{1cm}
\end{strip}
\newpage

%\pagenumbering{arabic}
%\setstretch{1.2}

\section{Introduction}
Supervised learning is a field in machine learning where possibly complicated dependencies between chosen inputs and outputs are not analytically solved but learned by e.g.~neural networks that try to approximate the relevant correlation. Neural networks are used in various acoustic applications such as speech recognition \cite{y_long_large-scale_2019}, characterisation of reverbing rooms \cite{ciaburro_acoustic_2021}, echo cancellation \cite{lei_deep_2019} and various others \cite{bianco_machine_2019}.

Sound field reproduction is aimed to create different sound experiences for listeners at different locations. There are several ways to perform sound field reproduction such as wave field synthesis \cite{boone_spatial_1995} or lower/higher order ambisonics \cite{gerzon_design_1975,zotter_ambisonics_2019}. Usually nonlinear effects such as wind and temperature statification are neglected, which can lead to errors whose relative sizes are dependent on the application. In addition to these studies there are also pressure matching methodologies \philipp{\cite{choi_generation_2002,shin_controlled_2014,du_multizone_2021,heuchel2018sound}} which are based on the acoustic transfer function and traditionally use the same neglection of wind and temperature stratification. 

Another method for sound field reproduction is the adjoint approach \cite{lemke_adjoint-based_2017,stein_adjoint-based_2019}. Although it copes with nonlinear effects like wind and temperature, it comes with high computational costs because it is based on solving the full set of compressible Euler equations. Furthermore, the optimisation can be trapped in local minima and setting it up for systems with complicated boundary conditions and geometries can be difficult.
%, which renders the method impractical for real world application.

Here, we propose an alternative strategy based on the multi zone sound field reproduction algorithm of \cite{du_multizone_2021}, which makes use of the acoustic transfer function, but incorporates nonlinear effects.
The acoustic transfer function describes how sound propagates between a speaker and a microphone. When generalising the concept to nonlinear acoustics, the transfer function can still be useful although hard to be analytically determined. Here, we make use of supervised learning, to capture nonlinear effects like wind and temperature stratification. Instead of training the neural network directly on the acoustic transfer function, we use the knowledge of sound propagation. By combining the physical information, the structure of the transfer function, with neural networks we simplify the functions that have to be approximated by the networks and makes our approach interpretable.

%but  The trained neural networks can be quickly evaluated for changing nonlinearities and the governing system of equations can also be quickly solved.

The manuscript is organised as follows: \Cref{sec:methodology} introduces the methodology. \Cref{sec:acoustic_transfer_function} defines the acoustic transfer function and the generalisations made for supervised learning. Furthermore we summarise the methodology of \cite{du_multizone_2021} in \cref{subsec:multizone_sound_field_repro} and give insights into our optimisation and training procedure (see \Cref{subsec:opt,subsec:train}). Finally, in \Cref{sec:Results} the methodology is applied to different analytical and numerical test cases.
In \Cref{sec:case_without_wind} we reproduce the results from \cite{du_multizone_2021} and validate the proposed methodology. Thereafter, the setup is generalised to applications where wind is considered in \Cref{sec:case_with_wind}. Here, a uniform wind flow is introduced to have an analytical solution for validation and reproducible results. The effects of the training data's sample density on the results are examined. Lastly, the value of the methodology is shown in \Cref{sec:flexingMuscles} in a setup where no analytical solution is known with a logarithmic wind profile and a temperature gradient. We finish with a discussion in \Cref{sec:discussion} where we recap the results and give a short outlook.
\section{Methodology}
\label{sec:methodology}

In this section we introduce the acoustic transfer function, its generalised form and how it can be learned from data. Furthermore, we show how the learned transfer function can be used inside the multizone sound field reproduction method of \cite{du_multizone_2021} to include nonlinear effects.

\subsection{Recreation of the Acoustic Transfer Function}
\label{sec:acoustic_transfer_function}

The acoustic transfer function $\g_{\indexMicSpeak} \in \mathbb{C}$ relates a measured \philipp{sound signal}, i.e. pressure $p_{\indexMic}$ at position $\x_{\indexMic} \in\mathbb{R}^3$ (microphone) to its sound source $\wScalar_{\indexSpeak}$ at $\x_{\indexSpeak} \in\mathbb{R}^3$ (speaker):
\begin{linenomath}\begin{equation}
\label{eq:superpos}
    \pScalar_{\indexMic}(t) = \sum_{\indexSpeak} \g_{\indexMicSpeak}\wScalar_{\indexSpeak}(t)\,.
\end{equation}\end{linenomath}
It can be expressed as follows:
\begin{linenomath}\begin{equation}
    \g_{\indexMicSpeak} = \ampMod_{\indexMicSpeak} \text{exp}\left(i \phaseShift_{\indexMicSpeak} \right)\,, \quad 
    \phaseShift_{\indexMicSpeak} = \phaseMod_{\indexMicSpeak} + \waveNumber \Vert\x_{\indexMic} - \x_{\indexSpeak} \Vert\,.
    \label{eq:meth_g}
\end{equation}\end{linenomath}
Here, $\ampMod_{\indexMicSpeak}$ is a function modulating the amplitude, $\waveNumber$ the wavenumber, $\phaseShift_{\indexMicSpeak}$ the phase shift and $\phaseMod_{\indexMicSpeak}$ the phase modulation. Apart from the phase shift caused by the wave having to travel a physical distance $s_{\indexMicSpeak}=\Vert\x_{\indexMic} - \x_{\indexSpeak} \Vert$, we consider the phase modulation to contain several kinds of possibly nonlinear effects, like dispersion.
Given that we neglect wind for now, the amplitude modulation can be dependent on  the frequency of the speaker (e.g.~if the source is not a point source but a Gaussian source \philipp{\cite{SteinStraubeSesterhennWeinzierlLemke2019b}}) and the phase modulation can also be dependent on the frequency.

The consideration of wind leads to more complicated acoustic transfer functions. Since wind stretches and compresses sound waves, it is useful to introduce the wave number modulation factor $\waveNumberMod_{\indexMicSpeak}\left(\Ma\right)$ into \Cref{eq:meth_g} which is dependent on the ratio between wind speed and speed of sound $\speedSound$, known as the Mach number $\Ma$. The wind also leads to an in- or decrease in amplitude and influences the phase shift. Therefore $\ampMod_{\indexMicSpeak}\left(\f,\Ma\right)$ and $\phaseMod_{\indexMicSpeak}\left(\f,\Ma\right)$ are dependent on the Mach number.

%The acoustic transfer function $g_{\indexMicSpeak}$ can be thus expressed as follows \philipp{brauchen wir die formel hier wirklich?, ich würde die acoustic transfer function hier nicht nochmal schreiben, da sie dann unten zum 3ten mal auftaucht. Sage z.b.
The acoustic transfer function $g_{\indexMicSpeak}(\f,\Ma)\colon \mathbb{R}^2\to \mathbb{C}$ can be thus expressed as a function of the frequency and Mach number
\begin{linenomath}\begin{equation}
    \begin{split}
    \g_{\indexMicSpeak}(f,&Ma) =
    \ampMod_{\indexMicSpeak}\left(\f,\Ma\right)\\
    &\cdot\text{exp}\left(i\left(\phaseMod_{\indexMicSpeak}\left(\f,\Ma\right) + \waveNumberMod_{\indexMicSpeak}\left(\Ma\right)ks_{nm}\right)\right)\,,
    \end{split}
\end{equation}\end{linenomath}
 where the nonlinear functions $\ampMod_{\indexMicSpeak}\left(\f,\Ma\right),\allowbreak \phaseMod_{\indexMicSpeak}\left(\f,\Ma\right)$ and $ \waveNumberMod_{\indexMicSpeak}\left(\Ma\right)$ need to be determined from data or measurements. 
Since nonlinear effects can vary between different spatial positions, the three functions 
are dependent on the locations $\x_{\indexMic}$ and $\x_{\indexSpeak}$. Hence, in this manuscript we use separate neural networks to approximate $\ampMod_{\indexMicSpeak},\phaseMod_{\indexMicSpeak}$ and $\waveNumberMod_{\indexMicSpeak}$. 
As an alternative approach, the neural network could be used for all speaker and microphone combinations. However, this would lead to an even more complex function to be approximated, with poor approximation quality and less insight into the system.%$\ampMod_{\indexMicSpeak}\left(\f,\Ma\right),\phaseMod_{\indexMicSpeak}\left(\f,\Ma\right),\waveNumberMod_{\indexMicSpeak}\left(\Ma\right)$ 

Note, that the functions have at most two inputs and are thus low-dimensional. In this case, another possibility would be to interpolate between the measured values. This is not done here, since it would pose the assumption, that the outputs have to be a smooth function of the inputs, which doesn't have to be the case for complicated systems. 

%%%%%%%%%%%%%%%%%%%%%%%%%%%%%%%%%%%%%%%%%%%%%%%%%%%%%%%%%%%%
% Philipp changed order:
%%%%%%%%%%%%%%%%%%%%%%%%%%%%%%%%%%%%%%%%%%%%%%%%%%%%%%%%%%%%

\subsection{Training Procedure} \label{subsec:train}
Finally, we introduce the generalised acoustic transfer function $\G^\text{NN} = (\g^\text{NN}_{\indexMicSpeak})_{\indexSpeak=1\ldots\numberSpeak,\indexMic=1\ldots\numberMic}$ between microphone at $\x_{\indexMic}$ and speaker at $\x_{\indexSpeak}$: 
% \begin{linenomath}\begin{equation}
%     g_{kj}(f,M) = H_{a,jk}(f,M)\cdot\text{exp}\left(i\left(H_{\varphi,jk}(f,M) + H_{k,jk}(M)ks_{kj}\right)\right)
%     \label{eq:meth_acousticTransferFunctionFromNN}
% \end{equation}\end{linenomath}
\begin{linenomath}\begin{equation}
\begin{split}
    \g^\text{NN}_{\indexMicSpeak}(\f,&\Ma) =    \ampMod^{\NN}_{\indexMicSpeak}(\f,\Ma)\\
    &\cdot\text{exp}\left(i\left(\phaseMod^{\NN}_{\indexMicSpeak}(\f,\Ma) + \waveNumberMod^{\NN}_{\indexMicSpeak}(\Ma)k\s_{\indexMicSpeak}\right)\right)\,.
\end{split}
\label{eq:meth_acousticTransferFunctionFromNN}
\end{equation}\end{linenomath}
As explained above, $\G^\text{NN}$ is able to include nonlinear effects using the learnable functions $\ampMod^{\NN}_{\indexMicSpeak}\left(\f,\Ma\right)$, $\phaseMod^{\NN}_{\indexMicSpeak}\left(\f,\Ma\right)$ and $\waveNumberMod^{\NN}_{\indexMicSpeak}\left(\Ma\right)$, which are approximated by fully connected neural networks, as indicated by the NN in their superscript.  

Note, that the neural networks get $\Ma$ as single input for the wind speed, where in reality the wind is a continuous vector field. However, we assume a constant wind profile, which is scaled by the Mach number. It is important to note here that this restriction does not have to be imposed necessarily, as the neural networks can be trained on any physically meaningful wind profile. This is shown in  \Cref{sec:Results}. Moreover, the approach can be generalised for multiple inputs, describing the wind profile. One possible application could be the parametrization of the wind profile in a Fourier series, which is left open for future research.

\begin{figure}[t]
	\setlength\figureheight{0.25\textwidth}
	\setlength\figurewidth{0.25\textwidth}
	\centering
    % This file was created by matlab2tikz.
%
\begin{tikzpicture}

\begin{axis}[%
width=\figurewidth,
height=\figureheight,
at={(0\figurewidth,0\figureheight)},
scale only axis,
xmin=-1.5000000000,
xmax=1.0000000000,
xlabel style={font=\color{white!15!black}},
xlabel={$x$ [m]},
ymin=-1.5000000000,
ymax=1.0000000000,
xmajorgrids,
ymajorgrids,
ylabel style={font=\color{white!15!black}},
ylabel={$y$ [m]},
axis background/.style={fill=white},
axis x line*=bottom,
axis y line*=left,
scaled ticks=false, xticklabel style={/pgf/number format/fixed},yticklabel style={/pgf/number format/fixed}
]
\addplot[only marks, mark=o, mark options={}, mark size=1.7678pt, draw=black, forget plot] table[row sep=crcr]{%
x	y\\
0.5000000000	0.0000000000\\
0.4619397663	0.1913417162\\
0.3535533906	0.3535533906\\
0.1913417162	0.4619397663\\
0.0000000000	0.5000000000\\
-0.1913417162	0.4619397663\\
-0.3535533906	0.3535533906\\
-0.4619397663	0.1913417162\\
-0.5000000000	0.0000000000\\
-0.4619397663	-0.1913417162\\
-0.3535533906	-0.3535533906\\
-0.1913417162	-0.4619397663\\
-0.0000000000	-0.5000000000\\
0.1913417162	-0.4619397663\\
0.3535533906	-0.3535533906\\
0.4619397663	-0.1913417162\\
};
\addplot[only marks, mark=asterisk, mark options={}, mark size=1.3693pt, draw=black, forget plot] table[row sep=crcr]{%
x	y\\
0.1500000000	-0.1500000000\\
0.1500000000	-0.0750000000\\
0.1500000000	0.0000000000\\
0.1500000000	0.0750000000\\
0.1500000000	0.1500000000\\
0.0750000000	0.1500000000\\
0.0000000000	0.1500000000\\
-0.0750000000	0.1500000000\\
-0.1500000000	0.1500000000\\
-0.1500000000	0.0750000000\\
-0.1500000000	0.0000000000\\
-0.1500000000	-0.0750000000\\
-0.1500000000	-0.1500000000\\
-0.0750000000	-0.1500000000\\
0.0000000000	-0.1500000000\\
0.0750000000	-0.1500000000\\
};
\addplot[only marks, mark=asterisk, mark options={}, mark size=1.3693pt, draw=black, forget plot] table[row sep=crcr]{%
x	y\\
-0.8500000000	-1.1500000000\\
-0.8500000000	-1.0750000000\\
-0.8500000000	-1.0000000000\\
-0.8500000000	-0.9250000000\\
-0.8500000000	-0.8500000000\\
-0.9250000000	-0.8500000000\\
-1.0000000000	-0.8500000000\\
-1.0750000000	-0.8500000000\\
-1.1500000000	-0.8500000000\\
-1.1500000000	-0.9250000000\\
-1.1500000000	-1.0000000000\\
-1.1500000000	-1.0750000000\\
-1.1500000000	-1.1500000000\\
-1.0750000000	-1.1500000000\\
-1.0000000000	-1.1500000000\\
-0.9250000000	-1.1500000000\\
};

\addplot[area legend, draw=black, fill=red, forget plot]
table[row sep=crcr] {%
x	y\\
0.1500000000	-0.1500000000\\
0.1500000000	-0.0750000000\\
0.1500000000	0.0000000000\\
0.1500000000	0.0750000000\\
0.1500000000	0.1500000000\\
0.0750000000	0.1500000000\\
0.0000000000	0.1500000000\\
-0.0750000000	0.1500000000\\
-0.1500000000	0.1500000000\\
-0.1500000000	0.0750000000\\
-0.1500000000	0.0000000000\\
-0.1500000000	-0.0750000000\\
-0.1500000000	-0.1500000000\\
-0.0750000000	-0.1500000000\\
0.0000000000	-0.1500000000\\
0.0750000000	-0.1500000000\\
}--cycle;

\addplot[area legend, draw=black, fill=blue, forget plot]
table[row sep=crcr] {%
x	y\\
-0.8500000000	-1.1500000000\\
-0.8500000000	-1.0750000000\\
-0.8500000000	-1.0000000000\\
-0.8500000000	-0.9250000000\\
-0.8500000000	-0.8500000000\\
-0.9250000000	-0.8500000000\\
-1.0000000000	-0.8500000000\\
-1.0750000000	-0.8500000000\\
-1.1500000000	-0.8500000000\\
-1.1500000000	-0.9250000000\\
-1.1500000000	-1.0000000000\\
-1.1500000000	-1.0750000000\\
-1.1500000000	-1.1500000000\\
-1.0750000000	-1.1500000000\\
-1.0000000000	-1.1500000000\\
-0.9250000000	-1.1500000000\\
}--cycle;
\end{axis}
\end{tikzpicture}%
%
	\caption{Simulation setup of \Cref{sec:case_without_wind}. The red square is the bright zone and the blue square is the dark zone. Circles ($\circ$) are indicators for speakers, asterisks ($*$) for microphones.}
	\label{fig:benchNoWind_geometry}
\end{figure}
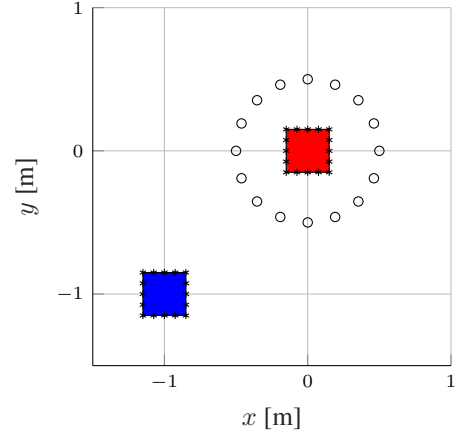

In this paper, the networks are trained on simulation data, but the presented approach is also suitable for measured data. In the following, we describe how the training data for $\ampMod^{\NN}_{\indexMicSpeak}$, $\phaseMod^{\NN}_{\indexMicSpeak}$ and $\waveNumberMod^{\NN}_{\indexMicSpeak}$ is collected from measurements of the pressure at the microphones. We assume a microphone/speaker setup similar to the one illustrated in \Cref{fig:benchNoWind_geometry} with multiple sound sources (speakers) and measurement locations (microphone).

For a given $(Ma,f)$ configuration, we measure the impulse-response correlation between every pair of speaker and microphone. Therefore, a mono frequent signal $\wScalar_{\indexSpeak}\in\mathbb{R}$
\begin{linenomath}\begin{equation}
    \wScalar_{\indexSpeak}\left(t\right) = \hat{w_0} \text{sin}\left(2\pi\f t + \phaseSpeaker_{0}\right),
\end{equation}\end{linenomath}
is produced by a single speaker at position $\x_{\indexSpeak}$. Since both, amplitude $\hat{w_0}$ and phase shift $\phasePressure_{0}$ of the speaker signal can be arbitrarily chosen, we can keep them constant for all microphone and speaker combinations. The pressure signal $\pScalar_{\indexMic}\in\mathbb{R}$
\begin{linenomath}\begin{equation}
    \pScalar_{\indexMic}\left(t\right) = \ampPressure_{\indexMicSpeak} \text{sin}\left(2\pi\f t + \phasePressure_{\indexMicSpeak} \right),
\end{equation}\end{linenomath}
is measured at microphone $\x_{\indexMic}$.  %To simplify the handling of the results it is convenient to keep $\wScalar_0$, meaning the amplitude and phase shift of the driving signal for all speakers, the same during the whole training data generation.

\paragraph{Amplitude and Phase Shift of the Pressure Signal}
\cite{lange_fourier_2020} propose to perform a fast Fourier transformation (FFT) to retrieve an initial guess of the amplitude, phase shift and frequency of the signal and to refine the initial guess using a gradient descent method. 
Since the frequency is known a priori in this case, here the FFT is not required to get a good initial guess. The initial guess is retrieved by the following expression:
\begin{linenomath}\begin{equation}
   \ampPressure_{\indexMicSpeak}^0 = \max_t\left(\vert\pScalar_{\indexMic}(t) \vert\right)    \,,
\end{equation}\end{linenomath}
\begin{linenomath}\begin{equation}
   \phasePressure_{\indexMicSpeak}^{0}=
   \begin{cases}
    \text{asin}\left(\pScalar_{\indexMic}\left(t=0\right)/\ampPressure_{\indexMicSpeak}^{0}\right) & \text{ if } \frac{\partial \pScalar_{\indexMic}\left(t=0\right)}{\partial t}  \geq 0 \\
    \text{asin}\left(-\pScalar_{\indexMic}\left(t=0\right)/\ampPressure_{\indexMicSpeak}^{0}\right)+\pi & \text{ if } \frac{\partial \pScalar_{\indexMic}\left(t=0\right)}{\partial t}  < 0
    \end{cases}
\end{equation}\end{linenomath}
Thereafter the guess can be improved using a gradient descent method.

The training data for one speaker to microphone combination can be calculated from the amplitude $\ampPressure_{\indexMicSpeak}$ and phase shift $\phasePressure_{\indexMicSpeak}$ of the pressure at the microphone for all measured/simulated frequencies and wind speeds. 

\paragraph{Amplitude Modulation}
We want the acoustic transfer function to be the response of the pressure at the microphones from a unit signal (amplitude of 1 and no phase shift) of the speaker. However, restraining to measuring the acoustic transfer function only using a unit signal is impractical. Therefore, the amplitude modulation is calculated as follows:%In order to obtain the acoustic transfer function as a response of the pressure at the microphones from a unit signal (amplitude of 1 and no phase shift) of the speaker, the amplitude modulation is computed using \philipp{wenn es doch unit signal ist brauchen wir $w_0$ nicht?). würde ich weglassen! andernfalls $phi_0=0$ und $w_0=1$ schreiben}
\begin{linenomath}\begin{equation}
    \ampMod_{\indexMicSpeak} = \ampPressure_{\indexMicSpeak}/\wScalar_0.
\end{equation}\end{linenomath}
The neural network has two inputs (frequency and Mach number) and one output.

\paragraph{Wavenumber Modulation Factor}
The wavenumber modulation factor can be iteratively calculated with $\waveNumberMod^{\iterationStep = 0} = 1$ as initial guess with the following scheme: 
\begin{linenomath}\begin{equation}
    \waveNumberMod^{\iterationStep+1}_{\indexMicSpeak} = \waveNumberMod^{\iterationStep}_{\indexMicSpeak} - \gamma \frac{\partial}{\partial \waveNumberMod_{\indexMicSpeak}}
    \frac{1}{\f^+ - \f^-}\int_{\f^-}^{\f^+}\frac{\partial  \phaseMod_{\indexMicSpeak}^{\iterationStep}}{\partial \f} \text{d}\f
\end{equation}\end{linenomath}
\begin{linenomath}\begin{equation}
    \phaseMod_{\indexMicSpeak}^{\iterationStep+1} = \phasePressure_{\indexMicSpeak} - \phaseSpeaker_{0} - \waveNumberMod_{\indexMicSpeak}^{\iterationStep+1}\waveNumber\s_{\indexMicSpeak}.
\end{equation}\end{linenomath}
Here, $\f^+$ is the upper limit of considered frequencies, $\f^-$ the respective lower limit and $\gamma$ is a step size of the iteration process. The goal of the iteration process is that the phase modulation is constant with respect to the Mach number or, if this proves to be unfeasible for example due to dispersion, to minimise the mean of $\frac{\partial}{\partial \f}\phaseMod_{\indexMicSpeak}$.

\begin{figure}[hb!]
\centering
    	\setlength\figureheight{\textwidth}
    	\setlength\figurewidth{\textwidth}

        \input{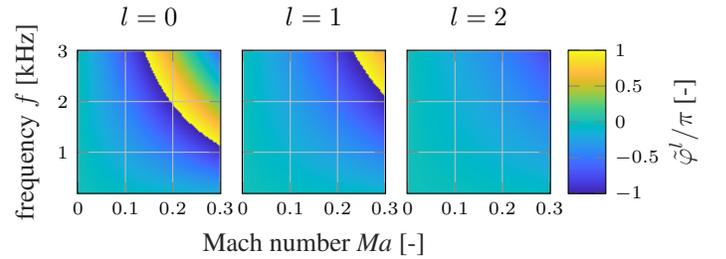}%}

		\caption{First iteration steps of the iteration scheme to calculate the wavenumber modulation factor of the speaker and microphone indicated by crosses in \Cref{fig:benchWind_geometry}. The initial phase modulation $\tilde{\phi}^0$ is represented in the left colour plot ($l=0$).}
		\label{fig:iterationProcessWaveNumberCorr}
\end{figure}

The integral is intentionally not solved since the phase angle is brought into an interval of $[-\pi, \pi]$ and therefore it may be discontinuous (see \Cref{fig:iterationProcessWaveNumberCorr} and \ref{sec:appIterateWaveModFac}). The numeric scheme that calculates the derivative must be able to handle this discontinuity. If the sound source is not a point source, then using the exact distance of the speaker to the microphone can lead to a wave number correction factor $\waveNumberMod_{\indexMicSpeak}\neq1$ for $\Ma = 0$. The neural network has one input (Mach number) and one output.

\paragraph{Phase Modulation}
The phase modulation can be calculated if the wavenumber modulation factor is known:
\begin{linenomath}\begin{equation}
    \phaseMod_{\indexMicSpeak}=\phasePressure_{\indexMicSpeak} - \phaseSpeaker_{0} - \waveNumberMod_{\indexMicSpeak}\waveNumber\s_{\indexMicSpeak}.
\end{equation}\end{linenomath}
Here, the phase shift of the speaker signal is included to achieve that the acoustic transfer function is the response of a unit signal. To handle a possible discontinuity of the phase modulation, stemming from the periodicity of the harmonic excitation, the network is not trained on the value of the phase modulation but on $\text{sin}\left(\phaseMod_{\indexMicSpeak}\right)$ and $\text{cos}\left(\phaseMod_{\indexMicSpeak}\right)$. In order to get the phase modulation, the two dimensional arcustangens function atan2 is used (see \ref{sec:appDiscontinuityPhaseAngle} for detailed explanation). The neural network thus has two inputs (frequency and Mach number) and two outputs $\phaseMod^{\NN}_{\indexMicSpeak} = \text{atan2}\left(\phaseMod^{\NN}_{1,\indexMicSpeak}, \phaseMod^{\NN}_{2,\indexMicSpeak}\right)$.

Note that scaling the output to a constant interval of $[-1,1]$ is done to ensure independence of the loss from the setup. The frequency values are scaled into the interval $[0,1]$ to improve the performance of the neural network.

The amplitude modulation and phase modulation neural networks have two hidden layers with 15 and 10 neurons. The wave number modulation network consists of two hidden layers with 10 and 5 neurons. The activation function of all networks' hidden layers is the hyperbolic tangent sigmoid function (tansig), where the output layers do not have an activation (purelin). The neural networks are trained using MATLAB's implementation of the Levenberg-Marquardt-Algorithm (trainlm) and the training is interrupted when the mean square error loss (MSE) sinks below $10^{-7}$.\footnote{see provided code at {\color{blue}\url{https://github.com/henrysallandt/Supervised-Learning-for-Multi-Zone-Sound-Field-Reproduction-under-Harsh-Environmental-Conditions/}}} The number of neurons might need to be adapted when handling more complicated cases.

\subsection{Multizone Sound Field Reproduction}\label{subsec:multizone_sound_field_repro}
With the superposition principle \Cref{eq:superpos} and a given transfer function $\G=(\g_{\indexMicSpeak})_{\indexMic=1,\dots,\numberMic,\indexSpeak=1,\dots,\numberSpeak}\,{\in \mathbb{C}^{\numberMic\times\numberSpeak}}$
the vector of speaker signals $\w=(\wScalar_{\indexSpeak})_{\indexSpeak=1,\dots,\numberSpeak}\,{\in\mathbb{C}^{\numberSpeak}}$ can be optimised towards a desired pressure $\p_\text{goal}=(p_{\indexMic})_{\indexMic=1,\dots,\numberMic}\,{\in\mathbb{C}^{\numberMic}}$ at the microphone locations 
% \begin{linenomath}\begin{equation}
%     \pScalar_{\indexMic} = \sum_{\indexSpeak} \g_{\indexMicSpeak}\wScalar_{\indexSpeak},
% \end{equation}\end{linenomath}
% \begin{linenomath}\begin{equation}
%     \pScalar_{\indexMic} = \sum_{\indexSpeak} \g_{\indexMicSpeak}\wScalar_{\indexSpeak},
% \end{equation}\end{linenomath}
% where $\wScalar_{\indexSpeak}$ is the speaker weight $\indexSpeak=1\ldots\numberSpeak$. 
by solving the linear system \footnote{There are nonlinearities that affect the propagation of sound (wind, temperature stratification) but since the sound pressure amplitude is small dissipative effects can be neglected. Since the nonlinearities are fixed for all microphones and speakers the superposition principle can be assumed.}
\begin{linenomath}\begin{equation}
    \p_\text{goal} = \G\w\,.
\end{equation}\end{linenomath}
%with $\p=(p_n)_{n=1,\dots,N},\w=(w_m)_{m=1,\dots,M}, \G=(\g_{nm})_{n=1,\dots,N,m=1,\dots,M}$.
For a two zone sound field reproduction problem with one dark (index d) and one bright (index b) zone, as shown in \Cref{fig:benchNoWind_geometry}, two optimisation problems arise \cite{choi_generation_2002}:
\begin{align}
    \min_\w \Vert\p_{\text{goal,b}} - \G_\text{b}\w\Vert_2^2 \quad \text{and} \quad
    \min_\w\Vert\p_{\text{goal,d}} - \G_\text{d}\w\Vert_2^2
\end{align}
If the two optimisation problems are weighted equally, this leads to the following expression:
\begin{linenomath}\begin{equation}
\begin{split}
    \w & = \argmin_{\w} \left\Vert
    \begin{bmatrix}
        \p_{\text{goal,b}}\\
        \p_{\text{goal,d}}
    \end{bmatrix}
    -
    \begin{bmatrix}
        \G_{\text{b}}\\
        \G_{\text{d}}
    \end{bmatrix}
    \w\right\Vert_2^2 \\
    & = \argmin_{\w} \Vert\p_\text{goal}-\hat{\G}\w\Vert_2^2\,,
\end{split}
    \label{eq:meth_optimisationProblem}
\end{equation}\end{linenomath}
 with $\hat{\G}=[\G_\text{b},\G_\text{d}]^\text{T}$.

As proposed by \cite{tikhonov_numerical_1990}, Tikhonov regularisation is used to penalise unfeasible solutions. % with possibly indefinite energy. 
Therefore, we add $\regParam \Vert\w\Vert_2^2,\,\regParam>0$ to the objective function $\J$ that is sought to be minimised:
\begin{linenomath}\begin{equation}
    \J = \Vert\p_\text{goal}-\hat{\G}\w\Vert_2^2 + \regParam\Vert\w\Vert_2^2\,,
\end{equation}\end{linenomath}
resulting in a well posed regression problem.
% \begin{linenomath}\begin{equation}
%     w:\text{argmin } \Vert\mtensor{w}\Vert_2^2
% \end{equation}\end{linenomath}
% which can be weighted to the optimisation problem in \Cref{eq:meth_optimisationProblem} with the regularisation parameter $\lambda$. The corresponding goal function $J$ which needs to be minimised is
% \begin{linenomath}\begin{equation}
%     \label{eq:total_optimization_goal}
%     J = \Vert\mtensor{p}_\text{goal}-\mtensor{G}\mtensor{w}\Vert_2^2 + \lambda\Vert\mtensor{w}\Vert_2^2
% \end{equation}\end{linenomath}
This can be solved directly with
\begin{linenomath}\begin{equation}
    \w = \left[\hat{\G}^\text{H}\hat{\G} + \lambda\mtensor{I}\right]^{-1}\hat{\G}^\mtensor{H}\p_\text{goal}
    \label{eq:meth_matrixEquation}
\end{equation}\end{linenomath}
% The optimised speaker weights need to be multiplied with the excitation of the speaker when recording the acoustic transfer functions to get the optimised speaker excitations \philipp{$\leftarrow$ umformulieren oder weg lassen, verstehe ich nicht}\henry{ich bin ein großer großer Fan von diesem Satz und eher ein Verfechter des $/w_0$ im Amplitude Modulation part der Training Procedure Hier ein Alternativvorschlag:}.

% The acoustic transfer functions are gained using a measured pressure signal of a microphone for a given excitation $\wScalar_0$ from one speaker. The result of \Cref{eq:meth_matrixEquation} are optimised speaker weights $\wScalar_{\indexSpeak} \in \mathbb{C}$. The optimised speaker excitations are obtained by the product of $\wScalar_{\indexSpeak}$ and $\wScalar_{0}$.

\subsection{Optimisation Procedure} \label{subsec:opt}
Our methodology is visualised in the flow chart shown in \Cref{fig:meth_flowChartAlgorithm}. First, a frequency, a wind speed (Mach number) and the goal pressures at the microphones are chosen. Next, the neural networks are evaluated to reproduce the acoustic transfer functions using \Cref{eq:meth_acousticTransferFunctionFromNN} to generate $\hat{\G}$. Finally the \philipp{objective function} can be minimised using \Cref{eq:meth_matrixEquation}. The optimisation generates a set of speaker signals $w_{\indexSpeak}$. This enables the reproduction of sound fields for frequencies and Mach numbers in a given parameter interval.

%%%%%%%%%%%%%%%%%%%%%%%%%%%%%%%%%%%%%%%%%%%%%%%%%%%%%%%%%%%%%%%
%%%% FLOWCHART
%%%%%%%%%%%%%%%%%%%%%%%%%%%%%%%%%%%%%%%%%%%%%%%%%%%%%%%%%%%%%
% Define block styles
\tikzstyle{materia}=[draw, fill=blue!20, text width=6.0em, text centered,
  minimum height=1.5em, drop shadow]
 % \tikzstyle{raute}=[draw, fill=green!20, text width=6.0em,]
\tikzstyle{etape} = [materia, text width=7em, minimum width=8em,
  minimum height=3em, rounded corners]
\tikzstyle{etapeWide} = [materia, text width=10em, minimum width=12em,
  minimum height=3em, rounded corners]
\tikzstyle{texto} = [above, text width=6em, text centered]
\tikzstyle{linepart} = [draw, thick, color=black!50, -latex', dashed]
\tikzstyle{line} = [draw, thick, color=black!50, -latex']
\tikzstyle{lineNoArrow} = [draw, thick, color=black!50]
\tikzstyle{ur}=[draw, text centered, minimum height=0.01em]
\tikzstyle{decision}=[diamond, minimum width=0.5cm, minimum height=0.5cm, text centered, draw=black, fill=green!30]
\tikzstyle{cloud} = [draw, ellipse,fill=red!20, node distance=2.5cm,
    minimum height=2 cm]

% Draw background
\newcommand{\background}[5]{%
  \begin{pgfonlayer}{background}
    % Left-top corner of the background rectangle
    \path (#1.west |- #2.north)+(-0.4,0.4) node (a1) {};
    % Right-bottom corner of the background rectanle
    \path (#3.east |- #4.south)+(+0.4,-0.4) node (a2) {};
    % Draw the background
    \path[rounded corners, draw=black!50, dashed]%fill=grey!20,]
      (a1) rectangle (a2);
      \path (#3.east |- #2.north)+(0,0.25)--(#1.west |- #2.north) node[midway] (#5-n) {};
      \path (#3.east |- #2.south)+(0,-0.35)--(#1.west |- #2.south) node[midway] (#5-s) {};
      \path (#3.east |- #2.north)+(0.7,0)--(#3.east |- #4.south) node[midway] (#5-w) {};
  \end{pgfonlayer}}

\newcommand{\transreceptor}[3]{%
  \path [linepart] (#1.east) -- node [above]
    {\scriptsize #2} (#3);}

\usetikzlibrary{positioning}

\tikzset{%
  every neuron/.style={
    circle,
    draw,
    minimum size=0.4cm
  },
  neuron missing/.style={
    draw=none, 
    scale=1,
    text height=0.222cm,
    execute at begin node=\color{black}$\vdots$
  },
}
\newcommand{\posNNx}[0]{-5.5cm}
\newcommand{\posNNy}[0]{+1.1cm}

\begin{figure}[h!]
	\centering
	\resizebox{\columnwidth}{!}{
\begin{tikzpicture}[scale=0.7,transform shape,node distance=2.8cm]
  %%%%%%%%%
  % Nodes
  %%%%%%%%%
    \node (freqMa) [etape] {frequency $\f$, Mach number $\Ma$};
    \node (phaseModulation) [etape,below of=freqMa] {phase \\modulation $\phaseMod^{\NN}_{\indexMicSpeak}(\f,\Ma)$};
    \node (amplitudeModulation) [etape,left of=phaseModulation, xshift=-1.5cm] {amplitude modulation $\ampMod^{\NN}_{\indexMicSpeak}(\f,\Ma)$};
    \node (waveNumberModulation) [etape,right of=phaseModulation, xshift=1.5cm] {wave number modulation $\waveNumberMod^{\NN}_{\indexMicSpeak}(\Ma)$};
    \node (acousticTF) [yshift=-0.3cm,etape,below of=phaseModulation] {acoustuic transfer function $\g_{\indexMicSpeak}$};
    
    \node (goalFunctional) [etapeWide,below of=acousticTF,yshift=0.5cm] {objective function $\J=\Vert\p_\text{goal}-\hat{\G}\w\Vert_2^2 + \regParam\Vert\w\Vert_2^2$};
    
    \node (microphoneGoals) [etape,left of=goalFunctional,xshift=-1.5cm] {goal pressure at microphones $\pScalar_{\indexMic}$};

    \node (finish) [etape,right of=goalFunctional,xshift=1.5cm] {optimized speaker signals $\wScalar_{\indexSpeak}$};

  %%%%%%%%%%
  % Arrows
  %%%%%%%%%%
    \draw [line] (freqMa) -- node [anchor=west, yshift=0.3cm] {evaluate neural networks} (phaseModulation) coordinate[midway] (midFreqPhase);
    \draw [line] (midFreqPhase) -| (amplitudeModulation);
    \draw [line] (midFreqPhase) -| (waveNumberModulation);
    
    \draw [line] (phaseModulation) -- node [anchor=west, yshift=-0.3cm] {\Cref{eq:meth_acousticTransferFunctionFromNN}}(acousticTF) coordinate[midway] (midPhaseATF);
    \draw [lineNoArrow] (amplitudeModulation) |- (midPhaseATF);
    \draw [lineNoArrow] (waveNumberModulation) |- (midPhaseATF);
    
    \draw [line] (goalFunctional)  -- node [anchor=south,yshift=0.7cm] {$\w:\text{argmin }\J$} (finish);
    
    \draw [line] (microphoneGoals)  -- (goalFunctional);
    \draw [line] (acousticTF)  -- (goalFunctional);
  %%%%%%%%%%
  % Boxes
  %%%%%%%%%%
    \background{amplitudeModulation}{freqMa}{waveNumberModulation}{phaseModulation}{};
    %\background{microphoneGoals}{freqMa}{freqMa}{freqMa}{};
  %%%%%%%%%%%%%%
  % Text fields
  %%%%%%%%%%%%%%
    %\node (text) [above of=freqMa, yshift=-1.2cm] {acoustic transfer function reconstruction using neural networks};
   
  %%%%%%%%%%%%%%
  % Neural Network
  %%%%%%%%%%%%%%
\foreach \m/\l [count=\y] in {1,2}
  \node [every neuron/.try, neuron \m/.try] (input-\m) at (0cm\posNNx,-0.5cm-\y*0.5cm\posNNy) {};

\foreach \m [count=\y] in {1,2,missing,3}
  \node [every neuron/.try, neuron \m/.try ] (hidden1-\m) at (1cm\posNNx,0cm-\y*0.5cm\posNNy) {};
  
\foreach \m [count=\y] in {1,2,missing,3}
  \node [every neuron/.try, neuron \m/.try ] (hidden2-\m) at (2cm\posNNx,0cm-\y*0.5cm\posNNy) {};
  
\foreach \m [count=\y] in {1}
  \node [every neuron/.try, neuron \m/.try ] (output-\m) at (3cm\posNNx,-1.25cm-\y*0cm\posNNy) {};

\foreach \i in {1,2}
  \foreach \j in {1,2,...,3}
    \draw [-] (input-\i) -- (hidden1-\j);
    
\foreach \i in {1,2,...,3}
  \foreach \j in {1,2,...,3}
    \draw [-] (hidden1-\i) -- (hidden2-\j);

\foreach \i in {1,2,...,3}
  \foreach \j in {1}
    \draw [-] (hidden2-\i) -- (output-\j);

\node (MicrophonePic) [below of=amplitudeModulation, yshift=-0.7cm] {\includegraphics[height=1.5cm]{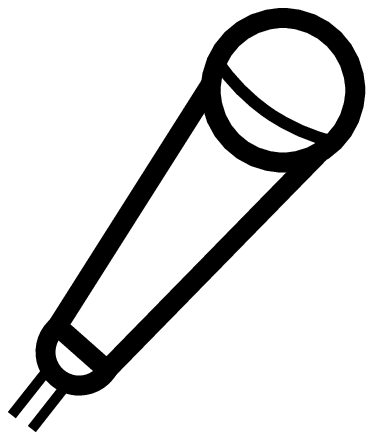}};

\node (SpeakerPic) [below of=waveNumberModulation, yshift=-0.7cm] {\includegraphics[height=1.5cm]{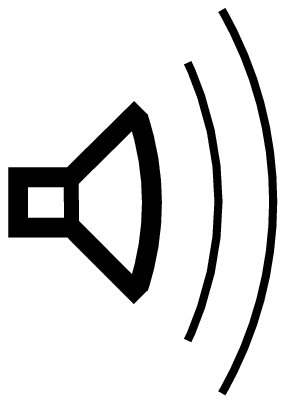}};
\end{tikzpicture}
}
\caption{Flow chart of the implemented algorithm: Choose a frequency and a Mach number, evaluate the neural networks to reconstruct the acoustic transfer functions and solve the optimisation problem to get the optimised speaker signals.}
\label{fig:meth_flowChartAlgorithm}
\end{figure}

\section{Results} \label{sec:Results}

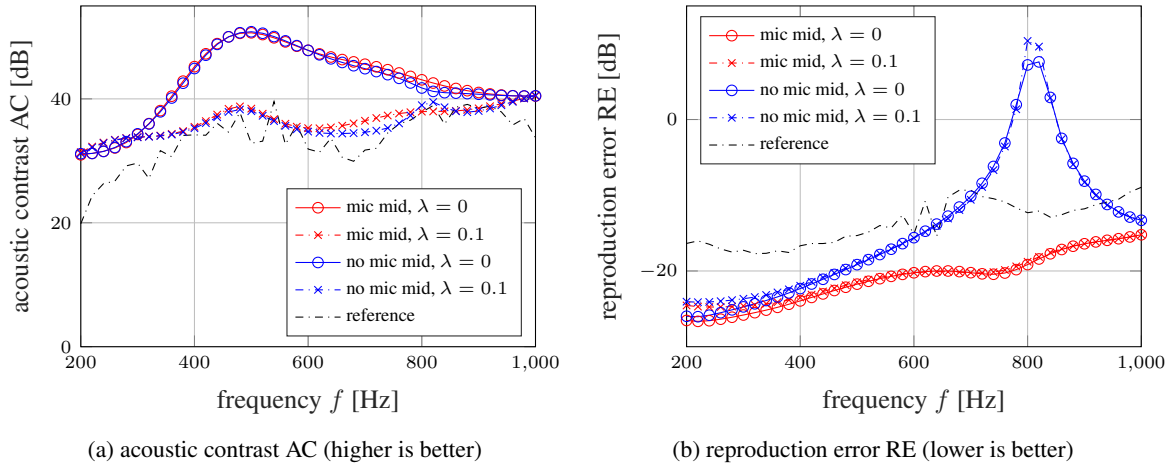
\begin{figure*}[hb!]
    \centering
	\begin{subfigure}[t]{\picWidth}
    	\setlength\figureheight{0.6\textwidth}
    	\setlength\figurewidth{0.8\textwidth}
    	\centering
        % This file was created by matlab2tikz.
%
\begin{tikzpicture}

\begin{axis}[%
width=\figurewidth,
height=\figureheight,
at={(0\figurewidth,0\figureheight)},
scale only axis,
xmin=200.0000000000,
xmax=1000.0000000000,
xlabel style={font=\color{white!15!black}},
xlabel={frequency $\f$ [Hz]},
ymin=0.0000000000,
ymax=55.0000000000,
ylabel style={font=\color{white!15!black}},
ylabel={acoustic contrast $\AC$ [dB]},
axis background/.style={fill=white},
xmajorgrids,
ymajorgrids,
legend style={at={(0.97,0.03)}, anchor=south east, legend cell align=left, align=left, draw=white!15!black},
scaled ticks=false, xticklabel style={/pgf/number format/fixed},yticklabel style={/pgf/number format/fixed}
]
\addplot [color=red, mark=o, mark options={solid, red}]
  table[row sep=crcr]{%
200.0000000000	30.9722178323\\
220.0000000000	31.2017534923\\
240.0000000000	31.5368409186\\
260.0000000000	32.0755927626\\
280.0000000000	32.9516776892\\
300.0000000000	34.2601413337\\
320.0000000000	36.0033357152\\
340.0000000000	38.1027742839\\
360.0000000000	40.4366485868\\
380.0000000000	42.8586504318\\
400.0000000000	45.1999723492\\
420.0000000000	47.2750685735\\
440.0000000000	48.9136849139\\
460.0000000000	50.0153450316\\
480.0000000000	50.5752491682\\
500.0000000000	50.6615338674\\
520.0000000000	50.3813652929\\
540.0000000000	49.8549446785\\
560.0000000000	49.1959893791\\
580.0000000000	48.5000008969\\
600.0000000000	47.8381933326\\
620.0000000000	47.2536060781\\
640.0000000000	46.7595928688\\
660.0000000000	46.3422631308\\
680.0000000000	45.9674258922\\
700.0000000000	45.5918345799\\
720.0000000000	45.1775142752\\
740.0000000000	44.7051250238\\
760.0000000000	44.1798858177\\
780.0000000000	43.6264586921\\
800.0000000000	43.0766252502\\
820.0000000000	42.5578215551\\
840.0000000000	42.0877748366\\
860.0000000000	41.6750292934\\
880.0000000000	41.3222165873\\
900.0000000000	41.0292729374\\
920.0000000000	40.7954655179\\
940.0000000000	40.6203560137\\
960.0000000000	40.5042494625\\
980.0000000000	40.4485249775\\
1000.0000000000	40.4557707726\\
};
\addlegendentry{mic mid, $\lambda=0$}

\addplot [color=red, dashdotted, mark=x, mark options={solid, red}]
  table[row sep=crcr]{%
200.0000000000	31.3940680157\\
220.0000000000	32.2978020522\\
240.0000000000	32.9554400111\\
260.0000000000	33.4028976451\\
280.0000000000	33.6776016494\\
300.0000000000	33.8311415955\\
320.0000000000	33.9367592604\\
340.0000000000	34.0808172504\\
360.0000000000	34.3467575740\\
380.0000000000	34.8024513061\\
400.0000000000	35.4903433321\\
420.0000000000	36.4084109521\\
440.0000000000	37.4625588802\\
460.0000000000	38.3892763485\\
480.0000000000	38.7795560988\\
500.0000000000	38.4295777239\\
520.0000000000	37.6059460033\\
540.0000000000	36.7134992417\\
560.0000000000	35.9890082293\\
580.0000000000	35.5098886485\\
600.0000000000	35.2744397070\\
620.0000000000	35.2502935000\\
640.0000000000	35.3965343472\\
660.0000000000	35.6730423587\\
680.0000000000	36.0427898648\\
700.0000000000	36.4689972163\\
720.0000000000	36.9097228348\\
740.0000000000	37.3155471404\\
760.0000000000	37.6372829372\\
780.0000000000	37.8444216681\\
800.0000000000	37.9422885454\\
820.0000000000	37.9721071819\\
840.0000000000	37.9932284351\\
860.0000000000	38.0615131965\\
880.0000000000	38.2148918023\\
900.0000000000	38.4674690612\\
920.0000000000	38.8100327196\\
940.0000000000	39.2156141872\\
960.0000000000	39.6485252737\\
980.0000000000	40.0730050361\\
1000.0000000000	40.4554435311\\
};
\addlegendentry{mic mid, $\lambda=0.1$}

\addplot [color=blue, mark=o, mark options={solid, blue}]
  table[row sep=crcr]{%
200.0000000000	31.1498622204\\
220.0000000000	31.1944169763\\
240.0000000000	31.5162290125\\
260.0000000000	32.1514766072\\
280.0000000000	33.0945560641\\
300.0000000000	34.3550927864\\
320.0000000000	35.9589793408\\
340.0000000000	37.9081593467\\
360.0000000000	40.1414980954\\
380.0000000000	42.5300551300\\
400.0000000000	44.8985990967\\
420.0000000000	47.0497878150\\
440.0000000000	48.7955989756\\
460.0000000000	50.0108713134\\
480.0000000000	50.6720716441\\
500.0000000000	50.8349468290\\
520.0000000000	50.5929636774\\
540.0000000000	50.0606607262\\
560.0000000000	49.3588122836\\
580.0000000000	48.5907260018\\
600.0000000000	47.8281084679\\
620.0000000000	47.1121849807\\
640.0000000000	46.4614059439\\
660.0000000000	45.8785749352\\
680.0000000000	45.3553432516\\
700.0000000000	44.8740553448\\
720.0000000000	44.4066023840\\
740.0000000000	43.9092603303\\
760.0000000000	43.3165204320\\
780.0000000000	42.5675860325\\
800.0000000000	41.7558597596\\
820.0000000000	41.2056673850\\
840.0000000000	41.0021506167\\
860.0000000000	40.9265099393\\
880.0000000000	40.8477727191\\
900.0000000000	40.7481209353\\
920.0000000000	40.6432019144\\
940.0000000000	40.5511843865\\
960.0000000000	40.4869474260\\
980.0000000000	40.4627983206\\
1000.0000000000	40.4900732979\\
};
\addlegendentry{no mic mid, $\lambda=0$}

\addplot [color=blue, dashdotted, mark=x, mark options={solid, blue}]
  table[row sep=crcr]{%
200.0000000000	31.2088245287\\
220.0000000000	32.1624103836\\
240.0000000000	32.8614486697\\
260.0000000000	33.3388802968\\
280.0000000000	33.6313504738\\
300.0000000000	33.7901261026\\
320.0000000000	33.8880141855\\
340.0000000000	34.0107807819\\
360.0000000000	34.2408964608\\
380.0000000000	34.6445021848\\
400.0000000000	35.2616020496\\
420.0000000000	36.0890953922\\
440.0000000000	37.0412910767\\
460.0000000000	37.8896339275\\
480.0000000000	38.2823057583\\
500.0000000000	38.0151865787\\
520.0000000000	37.2746290814\\
540.0000000000	36.4109186575\\
560.0000000000	35.6550023459\\
580.0000000000	35.0906447502\\
600.0000000000	34.7187740419\\
620.0000000000	34.5061585021\\
640.0000000000	34.4123391934\\
660.0000000000	34.4051410440\\
680.0000000000	34.4696284430\\
700.0000000000	34.6122893774\\
720.0000000000	34.8644592439\\
740.0000000000	35.2944215910\\
760.0000000000	36.0394972637\\
780.0000000000	37.3151475162\\
800.0000000000	38.9775994506\\
820.0000000000	39.4867668822\\
840.0000000000	38.6436742695\\
860.0000000000	37.9982695692\\
880.0000000000	37.8514746153\\
900.0000000000	38.0476828851\\
920.0000000000	38.4505349711\\
940.0000000000	38.9665434220\\
960.0000000000	39.5257662871\\
980.0000000000	40.0735167242\\
1000.0000000000	40.5643568485\\
};
\addlegendentry{no mic mid, $\lambda=0.1$}

\addplot [color=black, dashdotted]
  table[row sep=crcr]{%
200.0000000000	19.8396000000\\
220.0000000000	24.3606000000\\
240.0000000000	26.3177000000\\
260.0000000000	26.7274000000\\
280.0000000000	29.1851000000\\
300.0000000000	29.6402000000\\
320.0000000000	27.1370000000\\
340.0000000000	31.6428000000\\
360.0000000000	30.6567000000\\
380.0000000000	34.1006000000\\
400.0000000000	34.1916000000\\
420.0000000000	34.1309000000\\
440.0000000000	36.0577000000\\
460.0000000000	34.6316000000\\
480.0000000000	37.8023000000\\
500.0000000000	32.8414000000\\
520.0000000000	33.1751000000\\
540.0000000000	39.5622000000\\
560.0000000000	32.6593000000\\
580.0000000000	35.8149000000\\
600.0000000000	31.9007000000\\
620.0000000000	31.3849000000\\
640.0000000000	33.6454000000\\
660.0000000000	30.5657000000\\
680.0000000000	29.9285000000\\
700.0000000000	31.7339000000\\
720.0000000000	32.0221000000\\
740.0000000000	34.4950000000\\
760.0000000000	35.9666000000\\
780.0000000000	36.5583000000\\
800.0000000000	37.8327000000\\
820.0000000000	38.9402000000\\
840.0000000000	35.9211000000\\
860.0000000000	38.6519000000\\
880.0000000000	39.1071000000\\
900.0000000000	38.4699000000\\
920.0000000000	38.8340000000\\
940.0000000000	37.8934000000\\
960.0000000000	35.7846000000\\
980.0000000000	36.8314000000\\
1000.0000000000	33.7213000000\\
};
\addlegendentry{reference}

\end{axis}

\begin{axis}[%
width=1.226994\figurewidth,
height=1.226994\figureheight,
at={(-0.159509\figurewidth,-0.134969\figureheight)},
scale only axis,
xmin=0.0000000000,
xmax=1.0000000000,
ymin=0.0000000000,
ymax=1.0000000000,
axis line style={draw=none},
ticks=none,
axis x line*=bottom,
axis y line*=left,
scaled ticks=false, xticklabel style={/pgf/number format/fixed},yticklabel style={/pgf/number format/fixed}
]
\end{axis}
\end{tikzpicture}%
        \caption{acoustic contrast $\AC$ (higher is better)}
        \label{fig:benchNoWind_acousticContrast}
    \end{subfigure}
~
	\begin{subfigure}[t]{\picWidth}
    	\setlength\figureheight{0.6\textwidth}
    	\setlength\figurewidth{0.8\textwidth}
    	\centering
        % This file was created by matlab2tikz.
%
\begin{tikzpicture}

\begin{axis}[%
width=\figurewidth,
height=\figureheight,
at={(0\figurewidth,0\figureheight)},
scale only axis,
xmin=200.0000000000,
xmax=1000.0000000000,
xlabel style={font=\color{white!15!black}},
xlabel={frequency $\f$ [Hz]},
ymin=-30.0000000000,
ymax=15.0000000000,
ylabel style={font=\color{white!15!black}},
ylabel={reproduction error $\RE$ [dB]},
axis background/.style={fill=white},
xmajorgrids,
ymajorgrids,
legend style={at={(0.03,0.97)}, anchor=north west, legend cell align=left, align=left, draw=white!15!black},
scaled ticks=false, xticklabel style={/pgf/number format/fixed},yticklabel style={/pgf/number format/fixed}
]
\addplot [color=red, mark=o, mark options={solid, red}]
  table[row sep=crcr]{%
200.0000000000	-26.5350013878\\
220.0000000000	-26.6518889160\\
240.0000000000	-26.5684415994\\
260.0000000000	-26.3677974819\\
280.0000000000	-26.1069017268\\
300.0000000000	-25.8149290390\\
320.0000000000	-25.5034731683\\
340.0000000000	-25.1691626941\\
360.0000000000	-24.8006130541\\
380.0000000000	-24.3922244250\\
400.0000000000	-23.9508056611\\
420.0000000000	-23.4913041182\\
440.0000000000	-23.0295628137\\
460.0000000000	-22.5782143955\\
480.0000000000	-22.1460641946\\
500.0000000000	-21.7390759367\\
520.0000000000	-21.3616177938\\
540.0000000000	-21.0175650258\\
560.0000000000	-20.7112535195\\
580.0000000000	-20.4482897413\\
600.0000000000	-20.2361356246\\
620.0000000000	-20.0842446005\\
640.0000000000	-20.0032559750\\
660.0000000000	-20.0022863529\\
680.0000000000	-20.0826688803\\
700.0000000000	-20.2260567233\\
720.0000000000	-20.3771455635\\
740.0000000000	-20.4326570153\\
760.0000000000	-20.2684150630\\
780.0000000000	-19.8218840088\\
800.0000000000	-19.1563416245\\
820.0000000000	-18.4159007886\\
840.0000000000	-17.7294185978\\
860.0000000000	-17.1637914060\\
880.0000000000	-16.7304150810\\
900.0000000000	-16.4077810806\\
920.0000000000	-16.1602913407\\
940.0000000000	-15.9497128407\\
960.0000000000	-15.7410082694\\
980.0000000000	-15.5047469792\\
1000.0000000000	-15.2176513930\\
};
\addlegendentry{mic mid, $\lambda=0$}

\addplot [color=red, dashdotted, mark=x, mark options={solid, red}]
  table[row sep=crcr]{%
200.0000000000	-24.5534443263\\
220.0000000000	-24.7046154476\\
240.0000000000	-24.7598209399\\
260.0000000000	-24.7512932469\\
280.0000000000	-24.7009359282\\
300.0000000000	-24.6175637615\\
320.0000000000	-24.4985265356\\
340.0000000000	-24.3345112618\\
360.0000000000	-24.1156173788\\
380.0000000000	-23.8367541217\\
400.0000000000	-23.5008278269\\
420.0000000000	-23.1190166115\\
440.0000000000	-22.7084901092\\
460.0000000000	-22.2888273003\\
480.0000000000	-21.8786226982\\
500.0000000000	-21.4932603818\\
520.0000000000	-21.1440661463\\
540.0000000000	-20.8385614492\\
560.0000000000	-20.5814149859\\
580.0000000000	-20.3756757686\\
600.0000000000	-20.2238100683\\
620.0000000000	-20.1279895141\\
640.0000000000	-20.0890682928\\
660.0000000000	-20.1037003179\\
680.0000000000	-20.1590553076\\
700.0000000000	-20.2251451358\\
720.0000000000	-20.2477309265\\
740.0000000000	-20.1519560658\\
760.0000000000	-19.8710539957\\
780.0000000000	-19.3934054786\\
800.0000000000	-18.7825149621\\
820.0000000000	-18.1409582910\\
840.0000000000	-17.5558726576\\
860.0000000000	-17.0718502609\\
880.0000000000	-16.6939362805\\
900.0000000000	-16.4024791761\\
920.0000000000	-16.1670474925\\
940.0000000000	-15.9554726003\\
960.0000000000	-15.7384991056\\
980.0000000000	-15.4916559850\\
1000.0000000000	-15.1957011704\\
};
\addlegendentry{mic mid, $\lambda=0.1$}

\addplot [color=blue, mark=o, mark options={solid, blue}]
  table[row sep=crcr]{%
200.0000000000	-25.9809522595\\
220.0000000000	-26.0397163748\\
240.0000000000	-25.8407092557\\
260.0000000000	-25.5111798765\\
280.0000000000	-25.1333802741\\
300.0000000000	-24.7406295517\\
320.0000000000	-24.3338188803\\
340.0000000000	-23.8995647710\\
360.0000000000	-23.4228613478\\
380.0000000000	-22.8964429915\\
400.0000000000	-22.3244947739\\
420.0000000000	-21.7189269054\\
440.0000000000	-21.0928779379\\
460.0000000000	-20.4561374283\\
480.0000000000	-19.8134226556\\
500.0000000000	-19.1645423991\\
520.0000000000	-18.5053982621\\
540.0000000000	-17.8291085569\\
560.0000000000	-17.1268097314\\
580.0000000000	-16.3879198058\\
600.0000000000	-15.5998226743\\
620.0000000000	-14.7469696653\\
640.0000000000	-13.8092729454\\
660.0000000000	-12.7593839426\\
680.0000000000	-11.5579019873\\
700.0000000000	-10.1443293959\\
720.0000000000	-8.4184264253\\
740.0000000000	-6.1974094954\\
760.0000000000	-3.1006928648\\
780.0000000000	1.9591917333\\
800.0000000000	7.2078944689\\
820.0000000000	7.6117651875\\
840.0000000000	2.9165956161\\
860.0000000000	-2.4917323607\\
880.0000000000	-5.7900921531\\
900.0000000000	-8.1352686389\\
920.0000000000	-9.8897504773\\
940.0000000000	-11.2158969431\\
960.0000000000	-12.1975628689\\
980.0000000000	-12.8826997233\\
1000.0000000000	-13.3027977799\\
};
\addlegendentry{no mic mid, $\lambda=0$}

\addplot [color=blue, dashdotted, mark=x, mark options={solid, blue}]
  table[row sep=crcr]{%
200.0000000000	-24.0549682254\\
220.0000000000	-24.1222933484\\
240.0000000000	-24.0910450161\\
260.0000000000	-23.9931328518\\
280.0000000000	-23.8494412879\\
300.0000000000	-23.6675650639\\
320.0000000000	-23.4438914240\\
340.0000000000	-23.1685832511\\
360.0000000000	-22.8314374894\\
380.0000000000	-22.4267139256\\
400.0000000000	-21.9556768098\\
420.0000000000	-21.4265053488\\
440.0000000000	-20.8520877201\\
460.0000000000	-20.2467840301\\
480.0000000000	-19.6233309309\\
500.0000000000	-18.9906539706\\
520.0000000000	-18.3527678655\\
540.0000000000	-17.7085560314\\
560.0000000000	-17.0520980300\\
580.0000000000	-16.3731862462\\
600.0000000000	-15.6576222781\\
620.0000000000	-14.8868445237\\
640.0000000000	-14.0364689171\\
660.0000000000	-13.0733337242\\
680.0000000000	-11.9503387911\\
700.0000000000	-10.5972597116\\
720.0000000000	-8.9028945055\\
740.0000000000	-6.6771344648\\
760.0000000000	-3.5603632588\\
780.0000000000	1.3565003234\\
800.0000000000	10.3752021775\\
820.0000000000	9.6298533275\\
840.0000000000	2.8537475823\\
860.0000000000	-2.5894774471\\
880.0000000000	-5.9370943795\\
900.0000000000	-8.2934690439\\
920.0000000000	-10.0260438288\\
940.0000000000	-11.3125515975\\
960.0000000000	-12.2503993473\\
980.0000000000	-12.8967077300\\
1000.0000000000	-13.2880957954\\
};
\addlegendentry{no mic mid, $\lambda=0.1$}

\addplot [color=black, dashdotted]
  table[row sep=crcr]{%
200.0000000000	-16.3622000000\\
220.0000000000	-16.0273000000\\
240.0000000000	-16.4708000000\\
260.0000000000	-17.0856000000\\
280.0000000000	-17.5057000000\\
300.0000000000	-17.2642000000\\
320.0000000000	-17.7077000000\\
340.0000000000	-17.6686000000\\
360.0000000000	-17.3337000000\\
380.0000000000	-17.5592000000\\
400.0000000000	-16.6716000000\\
420.0000000000	-16.3756000000\\
440.0000000000	-16.3754000000\\
460.0000000000	-16.2196000000\\
480.0000000000	-15.4176000000\\
500.0000000000	-15.3240000000\\
520.0000000000	-14.8646000000\\
540.0000000000	-14.0003000000\\
560.0000000000	-14.3193000000\\
580.0000000000	-12.6144000000\\
600.0000000000	-15.1207000000\\
620.0000000000	-10.7613000000\\
640.0000000000	-15.4394000000\\
660.0000000000	-10.2394000000\\
680.0000000000	-9.3284900000\\
700.0000000000	-9.3438600000\\
720.0000000000	-10.1766000000\\
740.0000000000	-10.4566000000\\
760.0000000000	-10.8767000000\\
780.0000000000	-11.6628000000\\
800.0000000000	-12.3086000000\\
820.0000000000	-12.0827000000\\
840.0000000000	-12.8998000000\\
860.0000000000	-12.5727000000\\
880.0000000000	-12.0043000000\\
900.0000000000	-11.7939000000\\
920.0000000000	-11.2488000000\\
940.0000000000	-10.8906000000\\
960.0000000000	-10.5245000000\\
980.0000000000	-9.5746600000\\
1000.0000000000	-8.9517400000\\
};
\addlegendentry{reference}

\end{axis}
\end{tikzpicture}%

		\caption{reproduction error $\RE$ (lower is better)}
		\label{fig:benchNoWind_reproductionError}
	\end{subfigure}
    \caption{Comparison of the results of the solution with the microphone in the middle and without the microphone in the middle. To show the influence of the regularisation parameter the solution with two different values of the regularisation parameter is shown. For a comparison we include the results of \cite{du_multizone_2021}.}
    \label{fig:benchNoWind_compRefPaper}
\end{figure*}
In the following section, we test the presented methodology. To compare the results, two variables are introduced: First the reproduction error (\RE, lower is better) which measures the quality of the sound field in the bright zone $\S_\text{b}$:
\begin{linenomath}\begin{equation}
    \text{\RE} = 10\text{log}_{10}\frac{1}{\left|\S_\text{b}\right|}\int_{\S_\text{b}}\frac{\left|\pScalar_\text{rep} - \pScalar_\text{goal} \right|^2}
    {\left|\pScalar_\text{rep} \right|^2}\text{d}\S\,.
\end{equation}\end{linenomath}
Second the acoustic contrast (\AC, higher is better) which measures the energy density difference between the dark zone $\S_\text{d}$ and the bright zone $S_\text{b}$:
\begin{linenomath}\begin{equation}
    \text{\AC} = 10\text{log}_{10}\frac{\left|S_\text{d}\right| \int_{S_\text{b}}\left|\pScalar_\text{rep} \right|^2\text{d}S}
    {\left|S_\text{b}\right| \int_{S_\text{d}}\left|\pScalar_\text{rep} \right|^2\text{d}S}\,.
\end{equation}\end{linenomath}
The regularisation introduces an additional optimisation problem which can be competing with the original optimisation problem in \Cref{eq:meth_optimisationProblem}. Therefore a higher regularisation parameter $\regParam$ usually leads to a lower $\AC$ and a higher \RE.
%%%%%%%%%%%%%%%%%%%%%%%%%%%%%%%%%%%%%%%%%%%%%%%%%%%%%%%%%%%%%%%%%%%%%%%%%%%%%%%%%%
\subsection{Synthetic Test Case without Wind}
\begin{figure*}[ht!]
    \centering
	\begin{subfigure}[t]{\picWidth}
    	\setlength\figureheight{0.6\textwidth}
    	\setlength\figurewidth{0.7\textwidth}
    	\centering
        % This file was created by matlab2tikz.
%
\begin{tikzpicture}

\begin{axis}[%
width=0.838677\figurewidth,
height=\figureheight,
at={(0\figurewidth,0\figureheight)},
scale only axis,
point meta min=-40.0000000000,
point meta max=10.0000000000,
axis on top,
xmin=-1.5,
xmax=1,
xlabel style={font=\color{white!15!black}},
xlabel={$x$ [m]},
ymin=-1.5,
ymax=1,
xmajorgrids,
ymajorgrids,
ylabel style={font=\color{white!15!black}},
ylabel={$y$ [m]},
axis background/.style={fill=white},
colormap/hot2,
colorbar,
colorbar style={ylabel style={font=\color{white!15!black}}, ylabel={rel. sound energy density level [dB]}}
]
\addplot [forget plot] graphics [xmin=-2.0100502513, xmax=2.0100502513, ymin=-2.0100502513, ymax=2.0100502513] {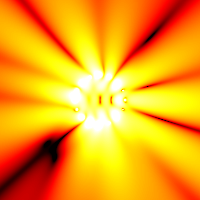};
\addplot[only marks, mark=asterisk, mark options={}, mark size=1.7678pt, draw=red, forget plot] table[row sep=crcr]{%
x	y\\
0.1500000000	-0.1500000000\\
0.1500000000	-0.0750000000\\
0.1500000000	0.0000000000\\
0.1500000000	0.0750000000\\
0.1500000000	0.1500000000\\
0.0750000000	0.1500000000\\
0.0000000000	0.1500000000\\
-0.0750000000	0.1500000000\\
-0.1500000000	0.1500000000\\
-0.1500000000	0.0750000000\\
-0.1500000000	0.0000000000\\
-0.1500000000	-0.0750000000\\
-0.1500000000	-0.1500000000\\
-0.0750000000	-0.1500000000\\
0.0000000000	-0.1500000000\\
0.0750000000	-0.1500000000\\
};
\addplot[only marks, mark=asterisk, mark options={}, mark size=1.7678pt, draw=blue, forget plot] table[row sep=crcr]{%
x	y\\
-0.8500000000	-1.1500000000\\
-0.8500000000	-1.0750000000\\
-0.8500000000	-1.0000000000\\
-0.8500000000	-0.9250000000\\
-0.8500000000	-0.8500000000\\
-0.9250000000	-0.8500000000\\
-1.0000000000	-0.8500000000\\
-1.0750000000	-0.8500000000\\
-1.1500000000	-0.8500000000\\
-1.1500000000	-0.9250000000\\
-1.1500000000	-1.0000000000\\
-1.1500000000	-1.0750000000\\
-1.1500000000	-1.1500000000\\
-1.0750000000	-1.1500000000\\
-1.0000000000	-1.1500000000\\
-0.9250000000	-1.1500000000\\
};
\addplot[only marks, mark=o, mark options={}, mark size=1.7678pt, draw=blue, forget plot] table[row sep=crcr]{%
x	y\\
0.5000000000	0.0000000000\\
0.4619397663	0.1913417162\\
0.3535533906	0.3535533906\\
0.1913417162	0.4619397663\\
0.0000000000	0.5000000000\\
-0.1913417162	0.4619397663\\
-0.3535533906	0.3535533906\\
-0.4619397663	0.1913417162\\
-0.5000000000	0.0000000000\\
-0.4619397663	-0.1913417162\\
-0.3535533906	-0.3535533906\\
-0.1913417162	-0.4619397663\\
-0.0000000000	-0.5000000000\\
0.1913417162	-0.4619397663\\
0.3535533906	-0.3535533906\\
0.4619397663	-0.1913417162\\
};
\end{axis}

\begin{axis}[%
width=1.226994\figurewidth,
height=1.226994\figureheight,
at={(-0.140681\figurewidth,-0.134969\figureheight)},
scale only axis,
point meta min=0.0000000000,
point meta max=1.0000000000,
xmin=0.0000000000,
xmax=1.0000000000,
ymin=0.0000000000,
ymax=1.0000000000,
axis line style={draw=none},
ticks=none,
axis x line*=bottom,
axis y line*=left
]
\end{axis}
\end{tikzpicture}%

		\caption{without microphone in the middle}
		\label{fig:benchNoWind_noMicMiddleEnergyLevel}
	\end{subfigure}
~
	\begin{subfigure}[t]{\picWidth}
    	\setlength\figureheight{0.6\textwidth}
    	\setlength\figurewidth{0.7\textwidth}
    	\centering
        % This file was created by matlab2tikz.
%
\begin{tikzpicture}

\begin{axis}[%
width=0.838677\figurewidth,
height=\figureheight,
at={(0\figurewidth,0\figureheight)},
scale only axis,
point meta min=-40.0000000000,
point meta max=10.0000000000,
axis on top,
xmin=-1.5,
xmax=1,
xlabel style={font=\color{white!15!black}},
xlabel={$x$ [m]},
ymin=-1.5,
ymax=1,
xmajorgrids,
ymajorgrids,
ylabel style={font=\color{white!15!black}},
ylabel={$y$ [m]},
axis background/.style={fill=white},
colormap/hot2,
colorbar,
colorbar style={ylabel style={font=\color{white!15!black}}, ylabel={rel. sound energy density level [dB]}}
]
\addplot [forget plot] graphics [xmin=-2.0100502513, xmax=2.0100502513, ymin=-2.0100502513, ymax=2.0100502513] {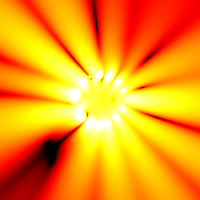};
\addplot[only marks, mark=asterisk, mark options={}, mark size=1.7678pt, draw=red, forget plot] table[row sep=crcr]{%
x	y\\
0.0000000000	0.0000000000\\
0.1500000000	-0.1500000000\\
0.1500000000	-0.0750000000\\
0.1500000000	0.0000000000\\
0.1500000000	0.0750000000\\
0.1500000000	0.1500000000\\
0.0750000000	0.1500000000\\
0.0000000000	0.1500000000\\
-0.0750000000	0.1500000000\\
-0.1500000000	0.1500000000\\
-0.1500000000	0.0750000000\\
-0.1500000000	0.0000000000\\
-0.1500000000	-0.0750000000\\
-0.1500000000	-0.1500000000\\
-0.0750000000	-0.1500000000\\
0.0000000000	-0.1500000000\\
0.0750000000	-0.1500000000\\
};
\addplot[only marks, mark=asterisk, mark options={}, mark size=1.7678pt, draw=blue, forget plot] table[row sep=crcr]{%
x	y\\
-1.0000000000	-1.0000000000\\
-0.8500000000	-1.1500000000\\
-0.8500000000	-1.0750000000\\
-0.8500000000	-1.0000000000\\
-0.8500000000	-0.9250000000\\
-0.8500000000	-0.8500000000\\
-0.9250000000	-0.8500000000\\
-1.0000000000	-0.8500000000\\
-1.0750000000	-0.8500000000\\
-1.1500000000	-0.8500000000\\
-1.1500000000	-0.9250000000\\
-1.1500000000	-1.0000000000\\
-1.1500000000	-1.0750000000\\
-1.1500000000	-1.1500000000\\
-1.0750000000	-1.1500000000\\
-1.0000000000	-1.1500000000\\
-0.9250000000	-1.1500000000\\
};
\addplot[only marks, mark=o, mark options={}, mark size=1.7678pt, draw=blue, forget plot] table[row sep=crcr]{%
x	y\\
0.5000000000	0.0000000000\\
0.4619397663	0.1913417162\\
0.3535533906	0.3535533906\\
0.1913417162	0.4619397663\\
0.0000000000	0.5000000000\\
-0.1913417162	0.4619397663\\
-0.3535533906	0.3535533906\\
-0.4619397663	0.1913417162\\
-0.5000000000	0.0000000000\\
-0.4619397663	-0.1913417162\\
-0.3535533906	-0.3535533906\\
-0.1913417162	-0.4619397663\\
-0.0000000000	-0.5000000000\\
0.1913417162	-0.4619397663\\
0.3535533906	-0.3535533906\\
0.4619397663	-0.1913417162\\
};
\end{axis}

\begin{axis}[%
width=1.226994\figurewidth,
height=1.226994\figureheight,
at={(-0.140681\figurewidth,-0.134969\figureheight)},
scale only axis,
point meta min=0.0000000000,
point meta max=1.0000000000,
xmin=0.0000000000,
xmax=1.0000000000,
ymin=0.0000000000,
ymax=1.0000000000,
axis line style={draw=none},
ticks=none,
axis x line*=bottom,
axis y line*=left
]
\end{axis}
\end{tikzpicture}%
        \caption{with microphone in the middle}
        \label{fig:benchNoWind_micMiddleEnergyLevel}
    \end{subfigure}
    \caption{Comparison of the sound energy density level relative to the bright zone in dB at $f=810\unit{Hz}$ with and without microphone in the middle of the zones. Circles are indicators for speakers, asterisks for microphones.}
    \label{fig:benchNoWind_compMicNoMic}
\end{figure*}
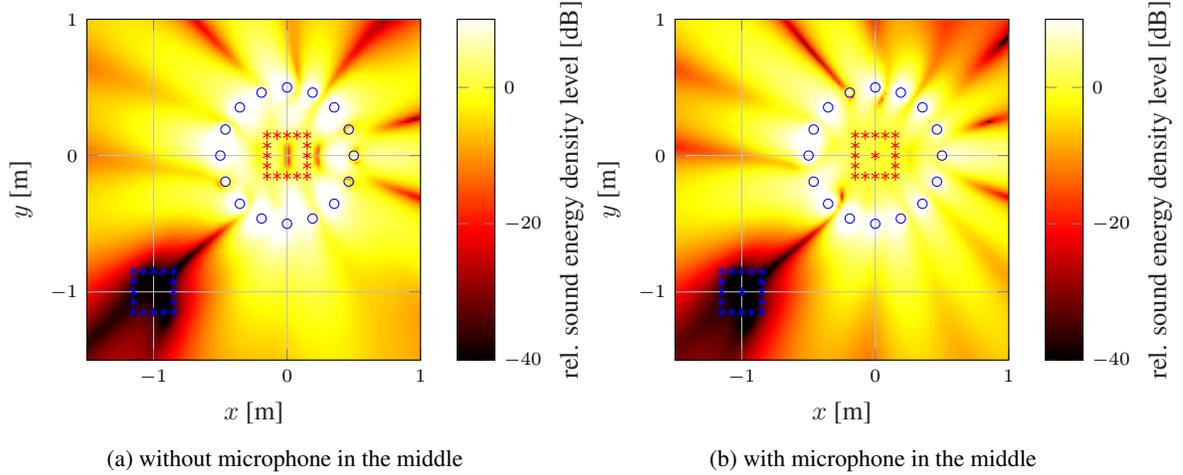
\label{sec:case_without_wind}
Here, we reproduce the results of \cite{du_multizone_2021} to verify the proposed methodology.
\paragraph{Simulation Setup}
The simulation setup is displayed in \Cref{fig:benchNoWind_geometry} and consists of a circular speaker setup with 16 uniformly spaced speakers with a radius of $0.5\unit{m}$. The two sound field reproduction zones are squares with a side length of each $0.3\unit{m}$ placed at $\left(0\unit{m}, 0\unit{m}\right)$ (red) and $\left(-1\unit{m}, -1\unit{m}\right)$ (blue) with each 16 microphones on their boundaries. The red zone is the bright zone where the sound field of a monopole placed at $\left(5\unit{m}, 0\unit{m}\right)$ is aimed to be reproduced. The blue zone is the dark zone where the pressure amplitude is aimed to be minimised. When evaluating the neural networks, the Mach number of the wind is set to zero. The speed of sound is chosen as $343\unit{m/s}$.

The training data for the neural networks is created using the free field Green's function of the 3D wave equation convoluted with a point source.\footnote{for Green's function of the wave equation in 3D space see \ref{sec:appDerivGreen} and set $\Ma=0$. For Green's function of the wave equation in 2D space see \cite{bailly_numerical_2000} and set $\Ma=0$.}
\paragraph{Results}
The training of the neural networks takes approx. $1\unit{h}$ on a standard notebook CPU while the evaluation takes approx. $10\unit{s}$ and the optimisation problem is solved in a split of a second.

\Cref{fig:benchNoWind_compRefPaper} shows the $\RE$ and the $\AC$ in the frequency interval $\left[200,1000\right]\unit{Hz}$. With the geometry shown in \Cref{fig:benchNoWind_geometry} the results of the $\AC$ are equal or better than the reference results of \cite{du_multizone_2021} for both considered regularisation parameters and frequencies which can be seen at \Cref{fig:benchNoWind_acousticContrast}. However, \Cref{fig:benchNoWind_reproductionError} shows that the results of  \cite{du_multizone_2021} cannot be matched: there is some sort of a resonance at the frequency $\f=810\unit{Hz}$ which leads to a small silent region in the bright zone. The resonance is not found in \cite{du_multizone_2021}. This can be explained by their symmetric setup and missing noise\footnote{an examination of noise on the solution can be found in \ref{sec:appNoise}} or the Monte Carlo trials which are mentioned in \cite{du_multizone_2021}. The small silent region inside the bright zone can be prevented by adding a microphone in the middle of the bright zone. To better understand the behaviour of the resonance an additional setup is examined in \ref{sec:appResonance}. Note, that we refrain from using noise in all of our test cases in order to make it deterministic and replicable for benchmarking.

A comparison of \Cref{fig:benchNoWind_noMicMiddleEnergyLevel,fig:benchNoWind_micMiddleEnergyLevel} shows that the solution is comparable regarding the overall trend except that there is no silent region in the bright zone.

Also the $\AC$ which is shown in \Cref{fig:benchNoWind_acousticContrast} is akin for both geometries. Near the resonance, there is a slightly better $\AC$ which is due to the missing silent region in the bright zone which makes the bright zone more silent and therefore the $\AC$ lower.
However, the $\RE$, shown in \Cref{fig:benchNoWind_reproductionError}, is much lower with the microphones in the middle of the zones and quite resistant to changes in the regularisation parameter $\regParam$. The $\RE$ is approximately $10\unit{dB}$ lower compared to \cite{du_multizone_2021}.

Overall, we conclude that the proposed method and its implementation is valid for this application. For all following examples there will be a microphone placed in the middle of both zones.

\subsection{Synthetic Test Case with Wind}
\label{sec:case_with_wind}
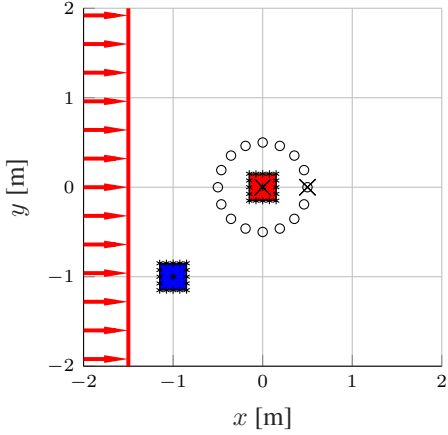
\begin{figure}[hb!]
    \centering
	\setlength\figureheight{0.25\textwidth}
	\setlength\figurewidth{0.25\textwidth}
	\centering
    % This file was created by matlab2tikz.
%
\begin{tikzpicture}

\begin{axis}[%
width=\figurewidth,
height=\figureheight,
at={(0\figurewidth,0\figureheight)},
scale only axis,
xmin=-2.0000000000,
xmax=2.0000000000,
xlabel style={font=\color{white!15!black}},
xlabel={$x$ [m]},
ymin=-2.0000000000,
ymax=2.0000000000,
ylabel style={font=\color{white!15!black}},
ylabel={$y$ [m]},
axis background/.style={fill=white},
axis x line*=bottom,
axis y line*=left,
xmajorgrids,
ymajorgrids,
scaled ticks=false, xticklabel style={/pgf/number format/fixed},yticklabel style={/pgf/number format/fixed}
]
\addplot[only marks, mark=o, mark options={}, mark size=1.7678pt, draw=black, forget plot] table[row sep=crcr]{%
x	y\\
0.5000000000	0.0000000000\\
0.4619397663	0.1913417162\\
0.3535533906	0.3535533906\\
0.1913417162	0.4619397663\\
0.0000000000	0.5000000000\\
-0.1913417162	0.4619397663\\
-0.3535533906	0.3535533906\\
-0.4619397663	0.1913417162\\
-0.5000000000	0.0000000000\\
-0.4619397663	-0.1913417162\\
-0.3535533906	-0.3535533906\\
-0.1913417162	-0.4619397663\\
-0.0000000000	-0.5000000000\\
0.1913417162	-0.4619397663\\
0.3535533906	-0.3535533906\\
0.4619397663	-0.1913417162\\
};
\addplot[only marks, mark=asterisk, mark options={}, mark size=1.3693pt, draw=black, forget plot] table[row sep=crcr]{%
x	y\\
0.0000000000	0.0000000000\\
0.1500000000	-0.1500000000\\
0.1500000000	-0.0750000000\\
0.1500000000	0.0000000000\\
0.1500000000	0.0750000000\\
0.1500000000	0.1500000000\\
0.0750000000	0.1500000000\\
0.0000000000	0.1500000000\\
-0.0750000000	0.1500000000\\
-0.1500000000	0.1500000000\\
-0.1500000000	0.0750000000\\
-0.1500000000	0.0000000000\\
-0.1500000000	-0.0750000000\\
-0.1500000000	-0.1500000000\\
-0.0750000000	-0.1500000000\\
0.0000000000	-0.1500000000\\
0.0750000000	-0.1500000000\\
};
\addplot[only marks, mark=asterisk, mark options={}, mark size=1.3693pt, draw=black, forget plot] table[row sep=crcr]{%
x	y\\
-1.0000000000	-1.0000000000\\
-0.8500000000	-1.1500000000\\
-0.8500000000	-1.0750000000\\
-0.8500000000	-1.0000000000\\
-0.8500000000	-0.9250000000\\
-0.8500000000	-0.8500000000\\
-0.9250000000	-0.8500000000\\
-1.0000000000	-0.8500000000\\
-1.0750000000	-0.8500000000\\
-1.1500000000	-0.8500000000\\
-1.1500000000	-0.9250000000\\
-1.1500000000	-1.0000000000\\
-1.1500000000	-1.0750000000\\
-1.1500000000	-1.1500000000\\
-1.0750000000	-1.1500000000\\
-1.0000000000	-1.1500000000\\
-0.9250000000	-1.1500000000\\
};
\addplot[only marks, mark=x, mark options={}, mark size=4.3301pt, draw=black, forget plot] table[row sep=crcr]{%
x	y\\
0.0000000000	0.0000000000\\
};
\addplot[only marks, mark=x, mark options={}, mark size=4.3301pt, draw=black, forget plot] table[row sep=crcr]{%
x	y\\
0.5000000000	0.0000000000\\
};
\addplot [color=red, line width=1.5pt, forget plot]
  table[row sep=crcr]{%
-1.5000000000	-2.0000000000\\
-1.5000000000	-1.9600000000\\
-1.5000000000	-1.9200000000\\
-1.5000000000	-1.8800000000\\
-1.5000000000	-1.8400000000\\
-1.5000000000	-1.8000000000\\
-1.5000000000	-1.7600000000\\
-1.5000000000	-1.7200000000\\
-1.5000000000	-1.6800000000\\
-1.5000000000	-1.6400000000\\
-1.5000000000	-1.6000000000\\
-1.5000000000	-1.5600000000\\
-1.5000000000	-1.5200000000\\
-1.5000000000	-1.4800000000\\
-1.5000000000	-1.4400000000\\
-1.5000000000	-1.4000000000\\
-1.5000000000	-1.3600000000\\
-1.5000000000	-1.3200000000\\
-1.5000000000	-1.2800000000\\
-1.5000000000	-1.2400000000\\
-1.5000000000	-1.2000000000\\
-1.5000000000	-1.1600000000\\
-1.5000000000	-1.1200000000\\
-1.5000000000	-1.0800000000\\
-1.5000000000	-1.0400000000\\
-1.5000000000	-1.0000000000\\
-1.5000000000	-0.9600000000\\
-1.5000000000	-0.9200000000\\
-1.5000000000	-0.8800000000\\
-1.5000000000	-0.8400000000\\
-1.5000000000	-0.8000000000\\
-1.5000000000	-0.7600000000\\
-1.5000000000	-0.7200000000\\
-1.5000000000	-0.6800000000\\
-1.5000000000	-0.6400000000\\
-1.5000000000	-0.6000000000\\
-1.5000000000	-0.5600000000\\
-1.5000000000	-0.5200000000\\
-1.5000000000	-0.4800000000\\
-1.5000000000	-0.4400000000\\
-1.5000000000	-0.4000000000\\
-1.5000000000	-0.3600000000\\
-1.5000000000	-0.3200000000\\
-1.5000000000	-0.2800000000\\
-1.5000000000	-0.2400000000\\
-1.5000000000	-0.2000000000\\
-1.5000000000	-0.1600000000\\
-1.5000000000	-0.1200000000\\
-1.5000000000	-0.0800000000\\
-1.5000000000	-0.0400000000\\
-1.5000000000	0.0000000000\\
-1.5000000000	0.0400000000\\
-1.5000000000	0.0800000000\\
-1.5000000000	0.1200000000\\
-1.5000000000	0.1600000000\\
-1.5000000000	0.2000000000\\
-1.5000000000	0.2400000000\\
-1.5000000000	0.2800000000\\
-1.5000000000	0.3200000000\\
-1.5000000000	0.3600000000\\
-1.5000000000	0.4000000000\\
-1.5000000000	0.4400000000\\
-1.5000000000	0.4800000000\\
-1.5000000000	0.5200000000\\
-1.5000000000	0.5600000000\\
-1.5000000000	0.6000000000\\
-1.5000000000	0.6400000000\\
-1.5000000000	0.6800000000\\
-1.5000000000	0.7200000000\\
-1.5000000000	0.7600000000\\
-1.5000000000	0.8000000000\\
-1.5000000000	0.8400000000\\
-1.5000000000	0.8800000000\\
-1.5000000000	0.9200000000\\
-1.5000000000	0.9600000000\\
-1.5000000000	1.0000000000\\
-1.5000000000	1.0400000000\\
-1.5000000000	1.0800000000\\
-1.5000000000	1.1200000000\\
-1.5000000000	1.1600000000\\
-1.5000000000	1.2000000000\\
-1.5000000000	1.2400000000\\
-1.5000000000	1.2800000000\\
-1.5000000000	1.3200000000\\
-1.5000000000	1.3600000000\\
-1.5000000000	1.4000000000\\
-1.5000000000	1.4400000000\\
-1.5000000000	1.4800000000\\
-1.5000000000	1.5200000000\\
-1.5000000000	1.5600000000\\
-1.5000000000	1.6000000000\\
-1.5000000000	1.6400000000\\
-1.5000000000	1.6800000000\\
-1.5000000000	1.7200000000\\
-1.5000000000	1.7600000000\\
-1.5000000000	1.8000000000\\
-1.5000000000	1.8400000000\\
-1.5000000000	1.8800000000\\
-1.5000000000	1.9200000000\\
-1.5000000000	1.9600000000\\
-1.5000000000	2.0000000000\\
};
\addplot[-Straight Barb, color=red, line width=1.5pt, point meta={sqrt((\thisrow{u})^2+(\thisrow{v})^2)}, point meta min=0, quiver={u=\thisrow{u}, v=\thisrow{v}, every arrow/.append style={-{Straight Barb[angle'=18.263, scale={1}]}}}]
 table[row sep=crcr] {%
x	y	u	v\\
-2.0000000000	-1.9200000000	0.5000000000	0.0000000000\\
-2.0000000000	-1.6000000000	0.5000000000	0.0000000000\\
-2.0000000000	-1.2800000000	0.5000000000	0.0000000000\\
-2.0000000000	-0.9600000000	0.5000000000	0.0000000000\\
-2.0000000000	-0.6400000000	0.5000000000	0.0000000000\\
-2.0000000000	-0.3200000000	0.5000000000	0.0000000000\\
-2.0000000000	0.0000000000	0.5000000000	0.0000000000\\
-2.0000000000	0.3200000000	0.5000000000	0.0000000000\\
-2.0000000000	0.6400000000	0.5000000000	0.0000000000\\
-2.0000000000	0.9600000000	0.5000000000	0.0000000000\\
-2.0000000000	1.2800000000	0.5000000000	0.0000000000\\
-2.0000000000	1.6000000000	0.5000000000	0.0000000000\\
-2.0000000000	1.9200000000	0.5000000000	0.0000000000\\
};

\addplot[area legend, draw=black, fill=red, forget plot]
table[row sep=crcr] {%
x	y\\
0.1500000000	-0.1500000000\\
0.1500000000	-0.0750000000\\
0.1500000000	0.0000000000\\
0.1500000000	0.0750000000\\
0.1500000000	0.1500000000\\
0.0750000000	0.1500000000\\
0.0000000000	0.1500000000\\
-0.0750000000	0.1500000000\\
-0.1500000000	0.1500000000\\
-0.1500000000	0.0750000000\\
-0.1500000000	0.0000000000\\
-0.1500000000	-0.0750000000\\
-0.1500000000	-0.1500000000\\
-0.0750000000	-0.1500000000\\
0.0000000000	-0.1500000000\\
0.0750000000	-0.1500000000\\
}--cycle;

\addplot[area legend, draw=black, fill=blue, forget plot]
table[row sep=crcr] {%
x	y\\
-0.8500000000	-1.1500000000\\
-0.8500000000	-1.0750000000\\
-0.8500000000	-1.0000000000\\
-0.8500000000	-0.9250000000\\
-0.8500000000	-0.8500000000\\
-0.9250000000	-0.8500000000\\
-1.0000000000	-0.8500000000\\
-1.0750000000	-0.8500000000\\
-1.1500000000	-0.8500000000\\
-1.1500000000	-0.9250000000\\
-1.1500000000	-1.0000000000\\
-1.1500000000	-1.0750000000\\
-1.1500000000	-1.1500000000\\
-1.0750000000	-1.1500000000\\
-1.0000000000	-1.1500000000\\
-0.9250000000	-1.1500000000\\
}--cycle;
\addplot[only marks, mark=x, mark options={}, mark size=4.3301pt, draw=black, forget plot] table[row sep=crcr]{%
x	y\\
0.0000000000	0.0000000000\\
};
\addplot[only marks, mark=x, mark options={}, mark size=4.3301pt, draw=black, forget plot] table[row sep=crcr]{%
x	y\\
0.5000000000	0.0000000000\\
};
\addplot[only marks, mark=x, mark options={}, mark size=4.3301pt, draw=black, forget plot] table[row sep=crcr]{%
x	y\\
0.0000000000	0.0000000000\\
};
\addplot[only marks, mark=x, mark options={}, mark size=4.3301pt, draw=black, forget plot] table[row sep=crcr]{%
x	y\\
0.5000000000	0.0000000000\\
};
\end{axis}

\begin{axis}[%
width=1.226994\figurewidth,
height=1.226994\figureheight,
at={(-0.259969\figurewidth,-0.134969\figureheight)},
scale only axis,
xmin=0.0000000000,
xmax=1.0000000000,
ymin=0.0000000000,
ymax=1.0000000000,
axis line style={draw=none},
ticks=none,
axis x line*=bottom,
axis y line*=left,
scaled ticks=false, xticklabel style={/pgf/number format/fixed},yticklabel style={/pgf/number format/fixed}
]
\end{axis}
\end{tikzpicture}%
	\caption{Geometry of the setup. The red square is the bright zone and the blue square is the dark zone. Circles are indicators for speakers, asterisks for microphones. The uniform wind profile is also shown. The two crosses are indicators for the considered microphone and speaker in \Cref{fig:benchWind_functionsToLearn}.
	}
	\label{fig:benchWind_geometry}
\end{figure}

Next, we introduce wind to the setup, an element that is abstracted from traditional sound field reproduction methods. 
\paragraph{Simulation Setup}
The same simulation setup as in \Cref{sec:case_without_wind} is examined but with the introduction of uniform wind $\left(\Ma\cdot \speedSound, 0, 0\right)\unit{m/s}$ with $\speedSound$ being the speed of sound. The setup is shown in \Cref{fig:benchWind_geometry}. We choose this setup, because the Green's function of of the 3D wave equation is known for uniform wind \footnote{For Green's function of the wave equation with uniform background flow in 3D space see \ref{sec:appDerivGreen}. For Green's function of the wave equation with uniform background flow in 2D space see \cite{bailly_numerical_2000}.}. Hence, this setup serves as a reproducible test case, since numerical experiments can be done easily. 

When training the neural network on measured data, getting a sufficient fine resolution in the frequency range is cheap. It might be costly to measure at different wind speeds and therefore the number of wind speeds is presumably relatively small. To get a feeling of how densely the training data needs to be sampled to get sufficient training data, it is initially created using 15 frequency samples and 3 wind speed samples. The samples are equidistantly distributed. Since this resolution leads to solution quality decreasing effects, the sampling density is increased to the point where these effects disappear (70 frequency samples, 30 wind samples).

In a next step, an investigation of the influence of the MSE loss of the neural networks on the $\AC$ and $\RE$ is performed. The MSE loss training stopping criterion of the neural networks is varied in the range $[10^{-7}, 10^{-1}]$.

To get comparable MSE loss measurements, the trained values of the amplitude modulation and wave number modulation are scaled to the interval $[-1,1]$. The phase modulation training data is not scaled to this interval since it is already scaled due to the sin and cos function.

Finally, the experiments are conducted using the best performing network (MSE loss stopping criterion $10^{-7}$) and sufficiently dense sampled data.

\begin{figure}[ht!]
\centering
	\begin{subfigure}[t]{\picWidth}
    	\setlength\figureheight{0.4\textwidth}
    	\setlength\figurewidth{0.5\textwidth}
    	\centering
        \input{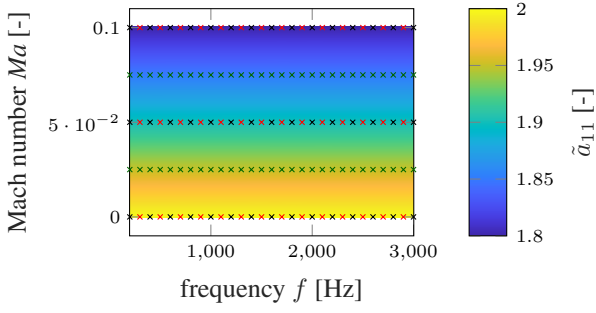}
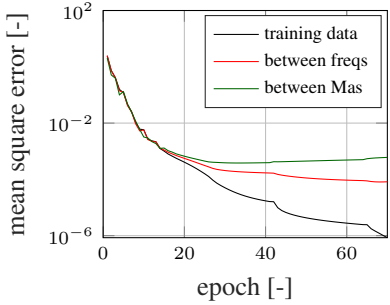
        \caption{amplitude modulation $\ampMod_{11}$ [-]}
        \label{fig:benchWind_ampModulation}
    \end{subfigure}

	\begin{subfigure}[t]{\picWidth}
	    \setlength\figureheight{0.4\textwidth}
	    \setlength\figurewidth{0.5\textwidth}
	    \centering
        % This file was created by matlab2tikz.
%
\definecolor{mycolor1}{rgb}{0,0.4,0}%
\begin{tikzpicture}

\begin{axis}[%
width=\figurewidth,
height=\figureheight,
at={(0\figurewidth,0\figureheight)},
scale only axis,
xmin=0.0000000000,
xmax=70.0000000000,
xlabel style={font=\color{white!15!black}},
xlabel={epoch [-]},
ymode=log,
ymin=0.0000008486,
ymax=100.0000000000,
yminorticks=true,
ylabel style={font=\color{white!15!black}},
ylabel={mean square error [-]},
axis background/.style={fill=white},
xmajorgrids,
ymajorgrids,
legend style={legend cell align=left, align=left, draw=white!15!black},
scaled ticks=false, xticklabel style={/pgf/number format/fixed},yticklabel style={/pgf/number format/fixed}
]
\addplot [color=black]
  table[row sep=crcr]{%
1.0000000000	2.4371308562\\
2.0000000000	0.7342329005\\
3.0000000000	0.4106871957\\
4.0000000000	0.1572117172\\
5.0000000000	0.1196963420\\
6.0000000000	0.0450622790\\
7.0000000000	0.0256964295\\
8.0000000000	0.0105411386\\
9.0000000000	0.0060913887\\
10.0000000000	0.0058138544\\
11.0000000000	0.0026889092\\
12.0000000000	0.0023436109\\
13.0000000000	0.0021707385\\
14.0000000000	0.0012503236\\
15.0000000000	0.0010130287\\
16.0000000000	0.0007981018\\
17.0000000000	0.0006695450\\
18.0000000000	0.0005624518\\
19.0000000000	0.0004801585\\
20.0000000000	0.0004089634\\
21.0000000000	0.0003459433\\
22.0000000000	0.0002896863\\
23.0000000000	0.0002398508\\
24.0000000000	0.0001962016\\
25.0000000000	0.0001586825\\
26.0000000000	0.0001262727\\
27.0000000000	0.0000941584\\
28.0000000000	0.0000734367\\
29.0000000000	0.0000603055\\
30.0000000000	0.0000504290\\
31.0000000000	0.0000427881\\
32.0000000000	0.0000368313\\
33.0000000000	0.0000321637\\
34.0000000000	0.0000284828\\
35.0000000000	0.0000255549\\
36.0000000000	0.0000232010\\
37.0000000000	0.0000212853\\
38.0000000000	0.0000197057\\
39.0000000000	0.0000183856\\
40.0000000000	0.0000172678\\
41.0000000000	0.0000163096\\
42.0000000000	0.0000160851\\
43.0000000000	0.0000098695\\
44.0000000000	0.0000077959\\
45.0000000000	0.0000066906\\
46.0000000000	0.0000059466\\
47.0000000000	0.0000053929\\
48.0000000000	0.0000049585\\
49.0000000000	0.0000046057\\
50.0000000000	0.0000043120\\
51.0000000000	0.0000040629\\
52.0000000000	0.0000038481\\
53.0000000000	0.0000036606\\
54.0000000000	0.0000034950\\
55.0000000000	0.0000033472\\
56.0000000000	0.0000032141\\
57.0000000000	0.0000030934\\
58.0000000000	0.0000029831\\
59.0000000000	0.0000028816\\
60.0000000000	0.0000027878\\
61.0000000000	0.0000027005\\
62.0000000000	0.0000026190\\
63.0000000000	0.0000025426\\
64.0000000000	0.0000024706\\
65.0000000000	0.0000024569\\
66.0000000000	0.0000017661\\
67.0000000000	0.0000014552\\
68.0000000000	0.0000012175\\
69.0000000000	0.0000010129\\
70.0000000000	0.0000008486\\
};
\addlegendentry{training data}

\addplot [color=red]
  table[row sep=crcr]{%
1.0000000000	2.5180533540\\
2.0000000000	0.7252675997\\
3.0000000000	0.4153673547\\
4.0000000000	0.1406987210\\
5.0000000000	0.1214271646\\
6.0000000000	0.0415552138\\
7.0000000000	0.0246994715\\
8.0000000000	0.0093636893\\
9.0000000000	0.0052796448\\
10.0000000000	0.0057326470\\
11.0000000000	0.0026002620\\
12.0000000000	0.0022575040\\
13.0000000000	0.0021441341\\
14.0000000000	0.0012961736\\
15.0000000000	0.0011034665\\
16.0000000000	0.0009231554\\
17.0000000000	0.0008048800\\
18.0000000000	0.0007041201\\
19.0000000000	0.0006255300\\
20.0000000000	0.0005580662\\
21.0000000000	0.0004982507\\
22.0000000000	0.0004442116\\
23.0000000000	0.0003949976\\
24.0000000000	0.0003504009\\
25.0000000000	0.0003113424\\
26.0000000000	0.0002780628\\
27.0000000000	0.0002459099\\
28.0000000000	0.0002268975\\
29.0000000000	0.0002150571\\
30.0000000000	0.0002061569\\
31.0000000000	0.0001991530\\
32.0000000000	0.0001935049\\
33.0000000000	0.0001888666\\
34.0000000000	0.0001849939\\
35.0000000000	0.0001817064\\
36.0000000000	0.0001788688\\
37.0000000000	0.0001763788\\
38.0000000000	0.0001741595\\
39.0000000000	0.0001721530\\
40.0000000000	0.0001703162\\
41.0000000000	0.0001686163\\
42.0000000000	0.0001662677\\
43.0000000000	0.0001505644\\
44.0000000000	0.0001414682\\
45.0000000000	0.0001345299\\
46.0000000000	0.0001287422\\
47.0000000000	0.0001237782\\
48.0000000000	0.0001194742\\
49.0000000000	0.0001157195\\
50.0000000000	0.0001124286\\
51.0000000000	0.0001095314\\
52.0000000000	0.0001069698\\
53.0000000000	0.0001046946\\
54.0000000000	0.0001026649\\
55.0000000000	0.0001008460\\
56.0000000000	0.0000992091\\
57.0000000000	0.0000977298\\
58.0000000000	0.0000963875\\
59.0000000000	0.0000951648\\
60.0000000000	0.0000940471\\
61.0000000000	0.0000930219\\
62.0000000000	0.0000920783\\
63.0000000000	0.0000912074\\
64.0000000000	0.0000904012\\
65.0000000000	0.0000884921\\
66.0000000000	0.0000839981\\
67.0000000000	0.0000817934\\
68.0000000000	0.0000809928\\
69.0000000000	0.0000812443\\
70.0000000000	0.0000822081\\
};
\addlegendentry{between freqs}

\addplot [color=mycolor1]
  table[row sep=crcr]{%
1.0000000000	2.0501991523\\
2.0000000000	0.5128478596\\
3.0000000000	0.3933709427\\
4.0000000000	0.1007641639\\
5.0000000000	0.1309090676\\
6.0000000000	0.0477521943\\
7.0000000000	0.0242274257\\
8.0000000000	0.0117532129\\
9.0000000000	0.0059634731\\
10.0000000000	0.0032052027\\
11.0000000000	0.0029605624\\
12.0000000000	0.0022003800\\
13.0000000000	0.0018653117\\
14.0000000000	0.0012626635\\
15.0000000000	0.0013009682\\
16.0000000000	0.0010435539\\
17.0000000000	0.0009330036\\
18.0000000000	0.0008066713\\
19.0000000000	0.0007315266\\
20.0000000000	0.0006668130\\
21.0000000000	0.0006116556\\
22.0000000000	0.0005634468\\
23.0000000000	0.0005210716\\
24.0000000000	0.0004828614\\
25.0000000000	0.0004471595\\
26.0000000000	0.0004182524\\
27.0000000000	0.0004064121\\
28.0000000000	0.0004006213\\
29.0000000000	0.0003935987\\
30.0000000000	0.0003882383\\
31.0000000000	0.0003847202\\
32.0000000000	0.0003827177\\
33.0000000000	0.0003819010\\
34.0000000000	0.0003819828\\
35.0000000000	0.0003827233\\
36.0000000000	0.0003839279\\
37.0000000000	0.0003854429\\
38.0000000000	0.0003871503\\
39.0000000000	0.0003889616\\
40.0000000000	0.0003908122\\
41.0000000000	0.0003926561\\
42.0000000000	0.0004245693\\
43.0000000000	0.0004219526\\
44.0000000000	0.0004245537\\
45.0000000000	0.0004282204\\
46.0000000000	0.0004320089\\
47.0000000000	0.0004358034\\
48.0000000000	0.0004396237\\
49.0000000000	0.0004434897\\
50.0000000000	0.0004474070\\
51.0000000000	0.0004513708\\
52.0000000000	0.0004553703\\
53.0000000000	0.0004593924\\
54.0000000000	0.0004634229\\
55.0000000000	0.0004674480\\
56.0000000000	0.0004714552\\
57.0000000000	0.0004754330\\
58.0000000000	0.0004793720\\
59.0000000000	0.0004832638\\
60.0000000000	0.0004871017\\
61.0000000000	0.0004908802\\
62.0000000000	0.0004945948\\
63.0000000000	0.0004982422\\
64.0000000000	0.0005018194\\
65.0000000000	0.0005426479\\
66.0000000000	0.0005621943\\
67.0000000000	0.0005767143\\
68.0000000000	0.0005871439\\
69.0000000000	0.0005959608\\
70.0000000000	0.0006047667\\
};
\addlegendentry{between Mas}

\end{axis}

\begin{axis}[%
width=1.226994\figurewidth,
height=1.226994\figureheight,
at={(-0.159509\figurewidth,-0.134969\figureheight)},
scale only axis,
xmin=0.0000000000,
xmax=1.0000000000,
ymin=0.0000000000,
ymax=1.0000000000,
axis line style={draw=none},
ticks=none,
axis x line*=bottom,
axis y line*=left,
scaled ticks=false, xticklabel style={/pgf/number format/fixed},yticklabel style={/pgf/number format/fixed}
]
\end{axis}
\end{tikzpicture}%
        \caption{Performance history of the amplitude modulation network with comparison of the different data sets. The data points are visualised in \Cref{fig:benchWind_ampModulation} in their respective colour.}
    \label{fig:benchWind_perfHistory}
    \end{subfigure}
    \caption{Influence of the sample density on the MSE loss of the amplitude modulation neural network.}%Two non-trivial of the three functions to learn for microphone 1 and speaker 1 (see \Cref{fig:benchWind_geometry} for location). The phase modulation $\phaseMod_{11}$ is everywhere 0 and therefore not displayed. Training data visualisation for error influence examination (see \Cref{fig:benchWind_perfHistory}).}
    \label{fig:benchWind_functionsToLearn}
\end{figure}

%%%%%%%%%%%%%%%%%%%%%%%%%%%%%%%%%%%%%%%%%%%%%%%%%%%%%%%%%%%%%%%%%%%%%%%%%%%%%%%%%%
\paragraph{Results}
\Cref{fig:benchWind_ampModulation} shows the function of the amplitude modulation that needs to be learned by the neural networks. The functions of the phase modulation and wave number modulation are displayed in \Cref{fig:app_functionsToLearn} in \begin{samepage}\ref{sec:appOtherFunctions}\end{samepage}. The distribution of the training data is displayed as well. Black crosses indicate a training data point. Red and purple crosses indicate different test points. The MSE loss of the different data sets are displayed in \Cref{fig:benchWind_perfHistory}. \Cref{fig:benchWind_perfHistory} displays The MSE loss of the amplitude modulation network over the training epochs.

The training data MSE loss decreases down to $10^{-6}$ while the MSE loss of the data points in the gaps of the training data in frequency direction decreases down to $10^{-4}$. Since the sampling density in wind speed direction is initially low (3 wind speeds), the MSE loss in the gaps of the training data in wind speed direction decreases only to $10^{-3}$.

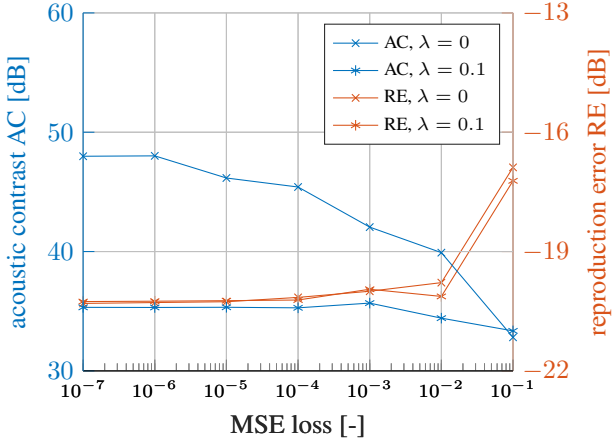
\begin{figure}[ht!]
	\setlength\figureheight{0.25\textwidth}
	\setlength\figurewidth{0.3\textwidth}
	\centering
    % This file was created by matlab2tikz.
%
\definecolor{mycolor1}{rgb}{0.00000,0.44700,0.74100}%
\definecolor{mycolor2}{rgb}{0.85000,0.32500,0.09800}%
\begin{tikzpicture}

\begin{axis}[%
width=\figurewidth,
height=\figureheight,
at={(0\figurewidth,0\figureheight)},
scale only axis,
xmode=log,
xmin=0.0000001000,
xmax=0.1000000000,
xminorticks=true,
xlabel style={font=\color{white!15!black}},
xlabel={MSE loss [-]},
every outer y axis line/.append style={mycolor2},
every y tick label/.append style={font=\color{mycolor2}},
every y tick/.append style={mycolor2},
ymin=-22.000000000,
ymax=-13.0000000000,
ytick={   -13, -16,  -19, -22},
ylabel style={font=\color{mycolor2}},
ylabel={reproduction error $\RE$ [dB]},
axis background/.style={fill=white},
axis x line*=bottom,
axis y line*=right,
xmajorgrids,
ymajorgrids,
scaled ticks=false, xticklabel style={/pgf/number format/fixed},yticklabel style={/pgf/number format/fixed}
]
\addplot [color=mycolor1, mark=x, mark options={solid, mycolor1}]
  table[row sep=crcr]{%
0.0000001000	47.9861307222\\
0.0000010000	48.0193015308\\
0.0000100000	46.1634829473\\
0.0001000000	45.4080691595\\
0.0010000000	42.0483555097\\
0.0100000000	39.9078007813\\
0.1000000000	32.8109020643\\
};

% \addplot [color=mycolor1, mark=o, mark options={solid, mycolor1}]
%   table[row sep=crcr]{%
% 0.0000001000	42.5199940653\\
% 0.0000010000	42.4578331745\\
% 0.0000100000	42.5226147104\\
% 0.0001000000	42.4610860608\\
% 0.0010000000	42.4986222769\\
% 0.0100000000	38.1653686286\\
% 0.1000000000	32.9003336432\\
% };

\addplot [color=mycolor1, mark=asterisk, mark options={solid, mycolor1}]
  table[row sep=crcr]{%
0.0000001000	35.3219236833\\
0.0000010000	35.3131614366\\
0.0000100000	35.3286537777\\
0.0001000000	35.2829229369\\
0.0010000000	35.6791511491\\
0.0100000000	34.4179068120\\
0.1000000000	33.3501202647\\
};

\addplot [color=mycolor2, mark=x, mark options={solid, mycolor2}]
  table[row sep=crcr]{%
0.0000001000	-20.3042045483\\
0.0000010000	-20.2822343697\\
0.0000100000	-20.2616864351\\
0.0001000000	-20.1529311412\\
0.0010000000	-19.9978177184\\
0.0100000000	-19.7815905346\\
0.1000000000	-16.8835860077\\
};

% \addplot [color=mycolor2, mark=o, mark options={solid, mycolor2}]
%   table[row sep=crcr]{%
% 0.0000001000	-20.3148202482\\
% 0.0000010000	-20.3056995566\\
% 0.0000100000	-20.2868293210\\
% 0.0001000000	-20.2985038895\\
% 0.0010000000	-19.9975803460\\
% 0.0100000000	-20.0157355437\\
% 0.1000000000	-16.9323783121\\
% };
% \addlegendentry{$\RE$, $\regParam = 0.01$}

\addplot [color=mycolor2, mark=asterisk, mark options={solid, mycolor2}]
  table[row sep=crcr]{%
0.0000001000	-20.2577284890\\
0.0000010000	-20.2485097004\\
0.0000100000	-20.2353133636\\
0.0001000000	-20.2183340279\\
0.0010000000	-19.9478841227\\
0.0100000000	-20.1281752126\\
0.1000000000	-17.2219521023\\
};

\end{axis}

\begin{axis}[%
width=\figurewidth,
height=\figureheight,
at={(0\figurewidth,0\figureheight)},
scale only axis,
xmode=log,
xmin=0.0000001000,
xmax=0.1000000000,
xminorticks=true,
xlabel style={font=\color{white!15!black}},
xlabel={MSE loss [-]},
every outer y axis line/.append style={mycolor1},
every y tick label/.append style={font=\color{mycolor1}},
every y tick/.append style={mycolor1},
ymin=30,
ymax=60,
ylabel style={font=\color{mycolor1}},
ylabel={acoustic contrast $\AC$ [dB]},
%axis background/.style={fill=white},
axis x line*=bottom,
axis y line*=left,
xmajorgrids,
ymajorgrids,
legend style={at={(0.97,0.97)}, anchor=north east, legend cell align=left, align=left, draw=white!15!black},
scaled ticks=false, xticklabel style={/pgf/number format/fixed},yticklabel style={/pgf/number format/fixed}
]
\addplot [color=mycolor1, mark=x, mark options={solid, mycolor1}]
  table[row sep=crcr]{%
0.0000001000	47.9861307222\\
0.0000010000	48.0193015308\\
0.0000100000	46.1634829473\\
0.0001000000	45.4080691595\\
0.0010000000	42.0483555097\\
0.0100000000	39.9078007813\\
0.1000000000	32.8109020643\\
};
\addlegendentry{$\AC$, $\regParam = 0$}

% \addplot [color=mycolor1, mark=o, mark options={solid, mycolor1}]
%   table[row sep=crcr]{%
% 0.0000001000	42.5199940653\\
% 0.0000010000	42.4578331745\\
% 0.0000100000	42.5226147104\\
% 0.0001000000	42.4610860608\\
% 0.0010000000	42.4986222769\\
% 0.0100000000	38.1653686286\\
% 0.1000000000	32.9003336432\\
% };

\addplot [color=mycolor1, mark=asterisk, mark options={solid, mycolor1}]
  table[row sep=crcr]{%
0.0000001000	35.3219236833\\
0.0000010000	35.3131614366\\
0.0000100000	35.3286537777\\
0.0001000000	35.2829229369\\
0.0010000000	35.6791511491\\
0.0100000000	34.4179068120\\
0.1000000000	33.3501202647\\
};
\addlegendentry{$\AC$, $\regParam = 0.1$}

\addplot [color=mycolor2, mark=x, mark options={solid, mycolor2}]
  table[row sep=crcr]{%
0.0000001000	-20.3042045483\\
0.0000010000	-20.2822343697\\
0.0000100000	-20.2616864351\\
0.0001000000	-20.1529311412\\
0.0010000000	-19.9978177184\\
0.0100000000	-19.7815905346\\
0.1000000000	-16.8835860077\\
};
\addlegendentry{$\RE$, $\regParam = 0$}

% \addplot [color=mycolor2, mark=o, mark options={solid, mycolor2}]
%   table[row sep=crcr]{%
% 0.0000001000	-20.3148202482\\
% 0.0000010000	-20.3056995566\\
% 0.0000100000	-20.2868293210\\
% 0.0001000000	-20.2985038895\\
% 0.0010000000	-19.9975803460\\
% 0.0100000000	-20.0157355437\\
% 0.1000000000	-16.9323783121\\
% };
% \addlegendentry{$\RE$, $\regParam = 0.01$}

\addplot [color=mycolor2, mark=asterisk, mark options={solid, mycolor2}]
  table[row sep=crcr]{%
0.0000001000	-20.2577284890\\
0.0000010000	-20.2485097004\\
0.0000100000	-20.2353133636\\
0.0001000000	-20.2183340279\\
0.0010000000	-19.9478841227\\
0.0100000000	-20.1281752126\\
0.1000000000	-17.2219521023\\
};
\addlegendentry{$\RE$, $\regParam = 0.1$}

\end{axis}

\end{tikzpicture}%
    \caption{Influence of the threshold MSE loss while training the neural networks on the $\AC$ and $\RE$ with $\f=600\unit{Hz}$ and $\Ma=0$}
    \label{fig:benchWind_errorInfluence}
\end{figure}
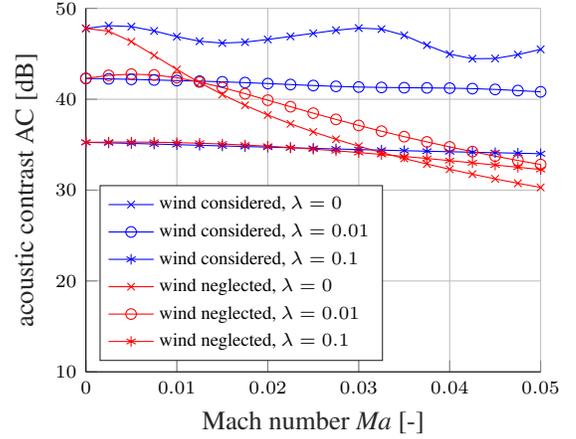
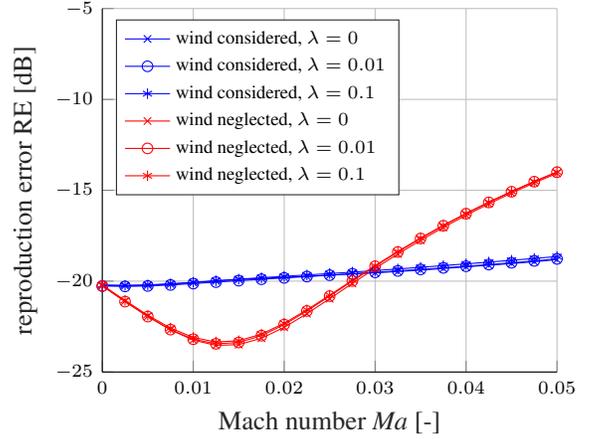
\begin{figure}[ht!]
\centering
	\begin{subfigure}[t]{\picWidth}
    	\setlength\figureheight{0.64\textwidth}
    	\setlength\figurewidth{0.8\textwidth}
    	\centering
        % This file was created by matlab2tikz.
%
\begin{tikzpicture}

\begin{axis}[%
width=\figurewidth,
height=\figureheight,
at={(0\figurewidth,0\figureheight)},
scale only axis,
xmin=0.0000000000,
xmax=0.0500000000,
xlabel style={font=\color{white!15!black}},
xlabel={Mach number $\Ma$ [-]},
ymin=10.0000000000,
ymax=50.0000000000,
ylabel style={font=\color{white!15!black}},
ylabel={acoustic contrast $\AC$ [dB]},
axis background/.style={fill=white},
axis x line*=bottom,
axis y line*=left,
xmajorgrids,
ymajorgrids,
legend style={at={(0.03,0.03)}, anchor=south west, legend cell align=left, align=left, draw=white!15!black},
scaled ticks=false, xticklabel style={/pgf/number format/fixed},yticklabel style={/pgf/number format/fixed}
]
\addplot [color=blue, mark=x, mark options={solid, blue}]
  table[row sep=crcr]{%
0.0000000000	47.7971714406\\
0.0025000000	48.1009060062\\
0.0050000000	47.9935682963\\
0.0075000000	47.5166731946\\
0.0100000000	46.8866623260\\
0.0125000000	46.3805699167\\
0.0150000000	46.1772769339\\
0.0175000000	46.2843847680\\
0.0200000000	46.5748355813\\
0.0225000000	46.9143509066\\
0.0250000000	47.2568700014\\
0.0275000000	47.5911971814\\
0.0300000000	47.8271566609\\
0.0325000000	47.7213236621\\
0.0350000000	47.0329832793\\
0.0375000000	45.9458534974\\
0.0400000000	44.9698385955\\
0.0425000000	44.4571674637\\
0.0450000000	44.4803038979\\
0.0475000000	44.9061170579\\
0.0500000000	45.4882546461\\
};
\addlegendentry{wind considered, $\lambda = 0$}

\addplot [color=blue, mark=o, mark options={solid, blue}]
  table[row sep=crcr]{%
0.0000000000	42.2860172323\\
0.0025000000	42.2442015614\\
0.0050000000	42.1842658767\\
0.0075000000	42.1188561068\\
0.0100000000	42.0541164216\\
0.0125000000	41.9886307824\\
0.0150000000	41.9156890965\\
0.0175000000	41.8286557879\\
0.0200000000	41.7265165301\\
0.0225000000	41.6158898529\\
0.0250000000	41.5083370758\\
0.0275000000	41.4153278071\\
0.0300000000	41.3442134283\\
0.0325000000	41.2966279034\\
0.0350000000	41.2682649260\\
0.0375000000	41.2485438964\\
0.0400000000	41.2210937833\\
0.0425000000	41.1680056073\\
0.0450000000	41.0781312140\\
0.0475000000	40.9544713806\\
0.0500000000	40.8146918876\\
};
\addlegendentry{wind considered, $\lambda = 0.01$}

\addplot [color=blue, mark=asterisk, mark options={solid, blue}]
  table[row sep=crcr]{%
0.0000000000	35.2553828902\\
0.0025000000	35.2011272096\\
0.0050000000	35.1384772928\\
0.0075000000	35.0706512146\\
0.0100000000	35.0005702120\\
0.0125000000	34.9304225465\\
0.0150000000	34.8614253914\\
0.0175000000	34.7938545916\\
0.0200000000	34.7273119797\\
0.0225000000	34.6611309827\\
0.0250000000	34.5948031429\\
0.0275000000	34.5283149486\\
0.0300000000	34.4622852491\\
0.0325000000	34.3978104068\\
0.0350000000	34.3360178292\\
0.0375000000	34.2774995711\\
0.0400000000	34.2219290941\\
0.0425000000	34.1681151505\\
0.0450000000	34.1145126105\\
0.0475000000	34.0599414538\\
0.0500000000	34.0041284524\\
};
\addlegendentry{wind considered, $\lambda = 0.1$}

\addplot [color=red, mark=x, mark options={solid, red}]
  table[row sep=crcr]{%
0.0000000000	47.7971714406\\
0.0025000000	47.4883360167\\
0.0050000000	46.3371616319\\
0.0075000000	44.8222475071\\
0.0100000000	43.2759239979\\
0.0125000000	41.8304761805\\
0.0150000000	40.5179206664\\
0.0175000000	39.3339416548\\
0.0200000000	38.2639430327\\
0.0225000000	37.2921362923\\
0.0250000000	36.4043214216\\
0.0275000000	35.5884754344\\
0.0300000000	34.8346319460\\
0.0325000000	34.1345735518\\
0.0350000000	33.4815132235\\
0.0375000000	32.8698172685\\
0.0400000000	32.2947791122\\
0.0425000000	31.7524391955\\
0.0450000000	31.2394428882\\
0.0475000000	30.7529285283\\
0.0500000000	30.2904389473\\
};
\addlegendentry{wind neglected, $\lambda = 0$}

\addplot [color=red, mark=o, mark options={solid, red}]
  table[row sep=crcr]{%
0.0000000000	42.2860172323\\
0.0025000000	42.6261772622\\
0.0050000000	42.7634038006\\
0.0075000000	42.6749408717\\
0.0100000000	42.3712729715\\
0.0125000000	41.8914027243\\
0.0150000000	41.2873723060\\
0.0175000000	40.6090694921\\
0.0200000000	39.8960467790\\
0.0225000000	39.1759670740\\
0.0250000000	38.4664697863\\
0.0275000000	37.7779073066\\
0.0300000000	37.1157632275\\
0.0325000000	36.4824401357\\
0.0350000000	35.8784708125\\
0.0375000000	35.3033001634\\
0.0400000000	34.7557753782\\
0.0425000000	34.2344467752\\
0.0450000000	33.7377488547\\
0.0475000000	33.2641066491\\
0.0500000000	32.8119959334\\
};
\addlegendentry{wind neglected, $\lambda = 0.01$}

\addplot [color=red, mark=asterisk, mark options={solid, red}]
  table[row sep=crcr]{%
0.0000000000	35.2553828902\\
0.0025000000	35.2832238773\\
0.0050000000	35.2879457457\\
0.0075000000	35.2690398140\\
0.0100000000	35.2263848900\\
0.0125000000	35.1602542353\\
0.0150000000	35.0713011621\\
0.0175000000	34.9605248731\\
0.0200000000	34.8292202378\\
0.0225000000	34.6789166265\\
0.0250000000	34.5113114907\\
0.0275000000	34.3282041153\\
0.0300000000	34.1314340555\\
0.0325000000	33.9228274873\\
0.0350000000	33.7041533204\\
0.0375000000	33.4770896804\\
0.0400000000	33.2432003821\\
0.0425000000	33.0039203521\\
0.0450000000	32.7605485888\\
0.0475000000	32.5142471229\\
0.0500000000	32.2660444905\\
};
\addlegendentry{wind neglected, $\lambda = 0.1$}

\end{axis}

\begin{axis}[%
width=1.226994\figurewidth,
height=1.226994\figureheight,
at={(-0.159509\figurewidth,-0.134969\figureheight)},
scale only axis,
xmin=0.0000000000,
xmax=1.0000000000,
ymin=0.0000000000,
ymax=1.0000000000,
axis line style={draw=none},
ticks=none,
axis x line*=bottom,
axis y line*=left,
scaled ticks=false, xticklabel style={/pgf/number format/fixed},yticklabel style={/pgf/number format/fixed}
]
\end{axis}
\end{tikzpicture}%
        \caption{acoustic contrast $\AC$ [dB] (higher is better)}
        \label{fig:benchWind_acousticContrastOverWind}
    \end{subfigure}
	\begin{subfigure}[t]{\picWidth}
    	\setlength\figureheight{0.64\textwidth}
    	\setlength\figurewidth{0.8\textwidth}
    	\centering
        % This file was created by matlab2tikz.
%
\begin{tikzpicture}

\begin{axis}[%
width=\figurewidth,
height=\figureheight,
at={(0\figurewidth,0\figureheight)},
scale only axis,
xmin=0.0000000000,
xmax=0.0500000000,
xlabel style={font=\color{white!15!black}},
xlabel={Mach number $\Ma$ [-]},
ymin=-25.0000000000,
ymax=-5.0000000000,
ylabel style={font=\color{white!15!black}},
ylabel={reproduction error $\RE$ [dB]},
axis background/.style={fill=white},
axis x line*=bottom,
axis y line*=left,
xmajorgrids,
ymajorgrids,
legend style={at={(0.03,0.97)}, anchor=north west, legend cell align=left, align=left, draw=white!15!black},
scaled ticks=false, xticklabel style={/pgf/number format/fixed},yticklabel style={/pgf/number format/fixed}
]
\addplot [color=blue, mark=x, mark options={solid, blue}]
  table[row sep=crcr]{%
0.0000000000	-20.2260577864\\
0.0025000000	-20.2479296816\\
0.0050000000	-20.2233953511\\
0.0075000000	-20.1631684165\\
0.0100000000	-20.0797595434\\
0.0125000000	-19.9897551423\\
0.0150000000	-19.9091202966\\
0.0175000000	-19.8447838889\\
0.0200000000	-19.7908316180\\
0.0225000000	-19.7350324875\\
0.0250000000	-19.6713527119\\
0.0275000000	-19.6042516334\\
0.0300000000	-19.5388504248\\
0.0325000000	-19.4711793531\\
0.0350000000	-19.3931607598\\
0.0375000000	-19.3044987025\\
0.0400000000	-19.2129552442\\
0.0425000000	-19.1236172960\\
0.0450000000	-19.0327500053\\
0.0475000000	-18.9315383698\\
0.0500000000	-18.8139651139\\
};
\addlegendentry{wind considered, $\lambda = 0$}

\addplot [color=blue, mark=o, mark options={solid, blue}]
  table[row sep=crcr]{%
0.0000000000	-20.2682548075\\
0.0025000000	-20.2899717990\\
0.0050000000	-20.2655527068\\
0.0075000000	-20.2094089221\\
0.0100000000	-20.1351880864\\
0.0125000000	-20.0540363361\\
0.0150000000	-19.9732526223\\
0.0175000000	-19.8957823472\\
0.0200000000	-19.8210681030\\
0.0225000000	-19.7470714012\\
0.0250000000	-19.6722069887\\
0.0275000000	-19.5959094281\\
0.0300000000	-19.5179681771\\
0.0325000000	-19.4379120848\\
0.0350000000	-19.3551609939\\
0.0375000000	-19.2693896693\\
0.0400000000	-19.1803465530\\
0.0425000000	-19.0873008687\\
0.0450000000	-18.9889695363\\
0.0475000000	-18.8843536544\\
0.0500000000	-18.7739519598\\
};
\addlegendentry{wind considered, $\lambda = 0.01$}

\addplot [color=blue, mark=asterisk, mark options={solid, blue}]
  table[row sep=crcr]{%
0.0000000000	-20.2136760328\\
0.0025000000	-20.2339026001\\
0.0050000000	-20.2057351447\\
0.0075000000	-20.1441323752\\
0.0100000000	-20.0637835403\\
0.0125000000	-19.9768791721\\
0.0150000000	-19.8914610783\\
0.0175000000	-19.8109019061\\
0.0200000000	-19.7346831418\\
0.0225000000	-19.6600831678\\
0.0250000000	-19.5840003556\\
0.0275000000	-19.5042136430\\
0.0300000000	-19.4198427638\\
0.0325000000	-19.3311920015\\
0.0350000000	-19.2392653310\\
0.0375000000	-19.1451616992\\
0.0400000000	-19.0495373125\\
0.0425000000	-18.9523491007\\
0.0450000000	-18.8530125597\\
0.0475000000	-18.7508913837\\
0.0500000000	-18.6458302761\\
};
\addlegendentry{wind considered, $\lambda = 0.1$}

\addplot [color=red, mark=x, mark options={solid, red}]
  table[row sep=crcr]{%
0.0000000000	-20.2260577864\\
0.0025000000	-21.0969012925\\
0.0050000000	-21.9359782762\\
0.0075000000	-22.6801466808\\
0.0100000000	-23.2428700522\\
0.0125000000	-23.5333306624\\
0.0150000000	-23.4955308395\\
0.0175000000	-23.1409339719\\
0.0200000000	-22.5412890454\\
0.0225000000	-21.7895753934\\
0.0250000000	-20.9658720199\\
0.0275000000	-20.1251765991\\
0.0300000000	-19.3000343420\\
0.0325000000	-18.5074404831\\
0.0350000000	-17.7549109006\\
0.0375000000	-17.0446085573\\
0.0400000000	-16.3758536029\\
0.0425000000	-15.7465662057\\
0.0450000000	-15.1540676204\\
0.0475000000	-14.5955118309\\
0.0500000000	-14.0681092272\\
};
\addlegendentry{wind neglected, $\lambda = 0$}

\addplot [color=red, mark=o, mark options={solid, red}]
  table[row sep=crcr]{%
0.0000000000	-20.2682548075\\
0.0025000000	-21.1281152145\\
0.0050000000	-21.9491986753\\
0.0075000000	-22.6667334162\\
0.0100000000	-23.1947652866\\
0.0125000000	-23.4471957184\\
0.0150000000	-23.3761109169\\
0.0175000000	-22.9997359721\\
0.0200000000	-22.3912763459\\
0.0225000000	-21.6407474367\\
0.0250000000	-20.8241084158\\
0.0275000000	-19.9931235404\\
0.0300000000	-19.1783763419\\
0.0325000000	-18.3958552490\\
0.0350000000	-17.6526392436\\
0.0375000000	-16.9507550218\\
0.0400000000	-16.2895308721\\
0.0425000000	-15.6669568760\\
0.0450000000	-15.0804448235\\
0.0475000000	-14.5272405406\\
0.0500000000	-14.0046392914\\
};
\addlegendentry{wind neglected, $\lambda = 0.01$}

\addplot [color=red, mark=asterisk, mark options={solid, red}]
  table[row sep=crcr]{%
0.0000000000	-20.2136760328\\
0.0025000000	-21.0636343448\\
0.0050000000	-21.8735917721\\
0.0075000000	-22.5800000545\\
0.0100000000	-23.0992583854\\
0.0125000000	-23.3481429229\\
0.0150000000	-23.2803451550\\
0.0175000000	-22.9130236588\\
0.0200000000	-22.3164768297\\
0.0225000000	-21.5779757659\\
0.0250000000	-20.7719399521\\
0.0275000000	-19.9496560285\\
0.0300000000	-19.1417972687\\
0.0325000000	-18.3646442471\\
0.0350000000	-17.6255911607\\
0.0375000000	-16.9269374847\\
0.0400000000	-16.2682281734\\
0.0425000000	-15.6476191256\\
0.0450000000	-15.0626471642\\
0.0475000000	-14.5106520777\\
0.0500000000	-13.9889999393\\
};
\addlegendentry{wind neglected, $\lambda = 0.1$}

\end{axis}

\begin{axis}[%
width=1.226994\figurewidth,
height=1.226994\figureheight,
at={(-0.159509\figurewidth,-0.134969\figureheight)},
scale only axis,
xmin=0.0000000000,
xmax=1.0000000000,
ymin=0.0000000000,
ymax=1.0000000000,
axis line style={draw=none},
ticks=none,
axis x line*=bottom,
axis y line*=left,
scaled ticks=false, xticklabel style={/pgf/number format/fixed},yticklabel style={/pgf/number format/fixed}
]
\end{axis}
\end{tikzpicture}%
        \caption{reproduction error $\RE$ [dB] (lower is better)}
        \label{fig:benchWind_reproductionErrorOverWind}
    \end{subfigure}
    \caption{Comparison of the solution dependent on the Mach number and the regularisation parameter with neglected wind (red) and considered wind (blue) at $\f=600\unit{Hz}$.}
    \label{fig:benchWind_solutionOverWind}
\end{figure}

The influence of the MSE loss on the $\AC$ and $\RE$ is shown in \Cref{fig:benchWind_errorInfluence}. The $\AC$ of the unregularised solution is starting to decrease at a MSE loss of $10^{-6}$ and continuously decreases from $48\unit{dB}$ down to $32.8\unit{dB}$. However, a regularisation of $\regParam=0.1$ leads to a constant $\AC$ of $35.3\unit{dB}$ until a MSE loss of $10^{-3}$. While the $\AC$ decreases for rising MSE loss down to $33.4\unit{dB}$, the $\RE$ is relatively resistant to changes of the MSE loss. Up to a MSE loss of $10^{-4}$, the $\RE$ is constant $-20.2\unit{dB}$ and then slowly starts increasing. At the MSE loss of $10^{-1}$ the $\RE$ jumps up to $-17\unit{dB}$.

\begin{figure}[ht!]
	\centering
	\setlength\figureheight{0.27\textwidth}
	\setlength\figurewidth{0.32\textwidth}
	%\resizebox{\linewidth}{!}{
    % This file was created by matlab2tikz.
%
\begin{tikzpicture}

\begin{axis}[%
width=\figurewidth,
height=\figureheight,
at={(0\figurewidth,0\figureheight)},
scale only axis,
xmin=0.0000000000,
xmax=0.0500000000,
xlabel style={font=\color{white!15!black}},
xlabel={Mach number $\Ma$ [-]},
ymin=-30.0000000000,
ymax=10.0000000000,
ylabel style={font=\color{white!15!black}},
ylabel={boundary reproduction error $\REBoundary$ [dB]},
axis background/.style={fill=white},
axis x line*=bottom,
axis y line*=left,
xmajorgrids,
ymajorgrids,
legend style={at={(0.03,0.97)}, anchor=north west, legend cell align=left, align=left, draw=white!15!black},
scaled ticks=false, xticklabel style={/pgf/number format/fixed},yticklabel style={/pgf/number format/fixed}
]
\addplot [color=blue, mark=x, mark options={solid, blue}]
  table[row sep=crcr]{%
0.0000000000	-28.5268015982\\
0.0025000000	-28.5094061597\\
0.0050000000	-28.4958230059\\
0.0075000000	-28.4843078576\\
0.0100000000	-28.4709756401\\
0.0125000000	-28.4522315177\\
0.0150000000	-28.4260174935\\
0.0175000000	-28.3919409114\\
0.0200000000	-28.3508368703\\
0.0225000000	-28.3042396279\\
0.0250000000	-28.2538039188\\
0.0275000000	-28.2006436717\\
0.0300000000	-28.1449449571\\
0.0325000000	-28.0862743946\\
0.0350000000	-28.0243955264\\
0.0375000000	-27.9598995290\\
0.0400000000	-27.8942020934\\
0.0425000000	-27.8290380660\\
0.0450000000	-27.7658176154\\
0.0475000000	-27.7050724797\\
0.0500000000	-27.6461283439\\
};
\addlegendentry{wind considered, $\lambda = 0$}

\addplot [color=blue, mark=o, mark options={solid, blue}]
  table[row sep=crcr]{%
0.0000000000	-28.0555134218\\
0.0025000000	-28.0119256197\\
0.0050000000	-27.9818808376\\
0.0075000000	-27.9597142354\\
0.0100000000	-27.9396283367\\
0.0125000000	-27.9167338910\\
0.0150000000	-27.8876289651\\
0.0175000000	-27.8505758592\\
0.0200000000	-27.8053724516\\
0.0225000000	-27.7530244609\\
0.0250000000	-27.6953100838\\
0.0275000000	-27.6343101094\\
0.0300000000	-27.5719683343\\
0.0325000000	-27.5097427189\\
0.0350000000	-27.4483951788\\
0.0375000000	-27.3879441922\\
0.0400000000	-27.3277740420\\
0.0425000000	-27.2668633183\\
0.0450000000	-27.2040707083\\
0.0475000000	-27.1384075703\\
0.0500000000	-27.0692394135\\
};
\addlegendentry{wind considered, $\lambda = 0.01$}

\addplot [color=blue, mark=asterisk, mark options={solid, blue}]
  table[row sep=crcr]{%
0.0000000000	-27.3345875908\\
0.0025000000	-27.2823881738\\
0.0050000000	-27.2419299210\\
0.0075000000	-27.2083170480\\
0.0100000000	-27.1765712462\\
0.0125000000	-27.1424981952\\
0.0150000000	-27.1031593501\\
0.0175000000	-27.0570114613\\
0.0200000000	-27.0038022494\\
0.0225000000	-26.9443108852\\
0.0250000000	-26.8800026727\\
0.0275000000	-26.8126489187\\
0.0300000000	-26.7439581292\\
0.0325000000	-26.6752682337\\
0.0350000000	-26.6073467509\\
0.0375000000	-26.5403280686\\
0.0400000000	-26.4737875505\\
0.0425000000	-26.4069217734\\
0.0450000000	-26.3387832653\\
0.0475000000	-26.2685123169\\
0.0500000000	-26.1955176168\\
};
\addlegendentry{wind considered, $\lambda = 0.1$}

\addplot [color=red, mark=x, mark options={solid, red}]
  table[row sep=crcr]{%
0.0000000000	-28.5268015982\\
0.0025000000	-27.3369998958\\
0.0050000000	-25.8198186540\\
0.0075000000	-24.3009695171\\
0.0100000000	-22.8990656005\\
0.0125000000	-21.6375977114\\
0.0150000000	-20.5079258954\\
0.0175000000	-19.4933439767\\
0.0200000000	-18.5770233517\\
0.0225000000	-17.7442395659\\
0.0250000000	-16.9826990830\\
0.0275000000	-16.2822932512\\
0.0300000000	-15.6347286909\\
0.0325000000	-15.0331771807\\
0.0350000000	-14.4719821652\\
0.0375000000	-13.9464237561\\
0.0400000000	-13.4525338694\\
0.0425000000	-12.9869515332\\
0.0450000000	-12.5468095066\\
0.0475000000	-12.1296450411\\
0.0500000000	-11.7333292193\\
};
\addlegendentry{wind neglected, $\lambda = 0$}

\addplot [color=red, mark=o, mark options={solid, red}]
  table[row sep=crcr]{%
0.0000000000	-28.0555134218\\
0.0025000000	-26.9105052843\\
0.0050000000	-25.4725036753\\
0.0075000000	-24.0253155240\\
0.0100000000	-22.6788456328\\
0.0125000000	-21.4587963946\\
0.0150000000	-20.3602186651\\
0.0175000000	-19.3693899153\\
0.0200000000	-18.4715912576\\
0.0225000000	-17.6535454352\\
0.0250000000	-16.9039539178\\
0.0275000000	-16.2134018124\\
0.0300000000	-15.5740882269\\
0.0325000000	-14.9795411615\\
0.0350000000	-14.4243664412\\
0.0375000000	-13.9040407570\\
0.0400000000	-13.4147456366\\
0.0425000000	-12.9532356220\\
0.0450000000	-12.5167338355\\
0.0475000000	-12.1028490629\\
0.0500000000	-11.7095096335\\
};
\addlegendentry{wind neglected, $\lambda = 0.01$}

\addplot [color=red, mark=asterisk, mark options={solid, red}]
  table[row sep=crcr]{%
0.0000000000	-27.3345875908\\
0.0025000000	-26.3281489018\\
0.0050000000	-25.0359070244\\
0.0075000000	-23.7022335403\\
0.0100000000	-22.4366484324\\
0.0125000000	-21.2734840671\\
0.0150000000	-20.2154714591\\
0.0175000000	-19.2542177306\\
0.0200000000	-18.3784889559\\
0.0225000000	-17.5772745266\\
0.0250000000	-16.8407705784\\
0.0275000000	-16.1605681626\\
0.0300000000	-15.5295585571\\
0.0325000000	-14.9417565043\\
0.0350000000	-14.3921172158\\
0.0375000000	-13.8763729954\\
0.0400000000	-13.3908956813\\
0.0425000000	-12.9325838000\\
0.0450000000	-12.4987710235\\
0.0475000000	-12.0871521742\\
0.0500000000	-11.6957233797\\
};
\addlegendentry{wind neglected, $\lambda = 0.1$}

\end{axis}
\end{tikzpicture}%
%}
	\caption{Comparison of the reproduction error $\REBoundary$ at the boundary of the bright zone for neglected and considered wind effects.}
	\label{fig:benchWind_reproductionErrorOnBoundary}
\end{figure}
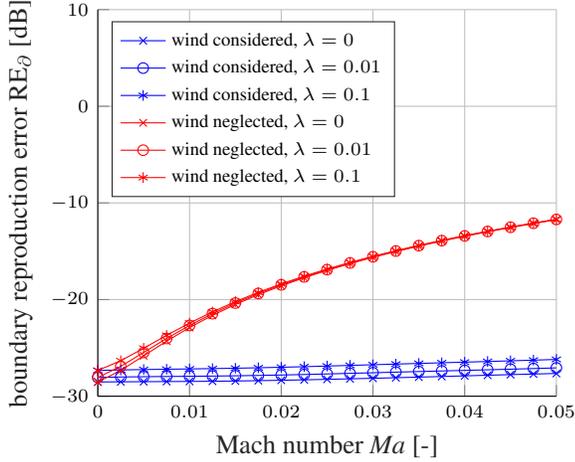
When comparing the $\AC$ in
\Cref{fig:benchWind_acousticContrastOverWind} for neglected and considered wind, one can see that the latter leads to a higher AC. There is a strong dependency on the regularisation parameter. A high regularisation parameter decreases the $\AC$ for $\Ma=0$, but it rises the solution's robustness against small errors in the input. This robustness leads to the fact that the unregularised solution performs worse in terms of the $\AC$ than the solution with the regulisation parameter $\regParam=0.01$ and the Mach number $\Ma>0.012$. The unregularised performance undercuts the performance of the even higher regularised solution with $\regParam=0.1$ and $\Ma>0.032$.

For the $\AC$ with considered wind, a slight, general downwards trend is visible for all regularisation parameters. This is probably due to the fact that the wind stretches and compresses the sound waves dependent on the direction of the sound wave front to the wind. Since for a given microphone not all speakers are located in the same direction, the sound waves get stretched and compressed differently. Therefore, with rising Mach numbers $\Ma$, it is harder to cancel out all sound waves in the dark zone.

At first sight the $\RE$ in \philipp{\Cref{fig:benchWind_reproductionErrorOverWind}} shows an unexpected behaviour. For negelected wind, the RE  decreases with small Mach numbers $\Ma<0.03$ and even outperforms the optimisation with considered wind in this regime. However, this is only an effect caused by the microphone setup and the definition of the RE. For example if we compute the $\RE$, closer to the areas where the pressure is optimised (positions of the microphones), this effect vanishes, as can be seen from
\Cref{fig:benchWind_reproductionErrorOnBoundary}. \philipp{It} shows the $\RE$ computed at the boundary of the bright zone (denoted by $\REBoundary$). It is visible that neglecting the wind leads to an increase in the $\REBoundary$, even for small Mach numbers. 
%This shows that the lower $\RE$ for small Mach numbers with neglected wind effect is due to the microphone distribution and not due to errors in the acoustic transfer function recreation.

Similarly to \Cref{sec:case_without_wind}, the $\RE$ and also the $\REBoundary$ is quite resistant to changes in the regularisation parameter in the studied Mach number range. Furthermore, a slow increasing $\RE$ and $\REBoundary$ is visible with rising Mach numbers, which is probably due to the same reason as the decreasing $\AC$ with higher Mach numbers.

\subsection{Sound Field Reproduction under Harsh Environmental Conditions}
\label{sec:flexingMuscles}

\begin{figure}[t]
	\setlength\figureheight{0.25\textwidth}
	\setlength\figurewidth{0.25\textwidth}
	\centering
    \input{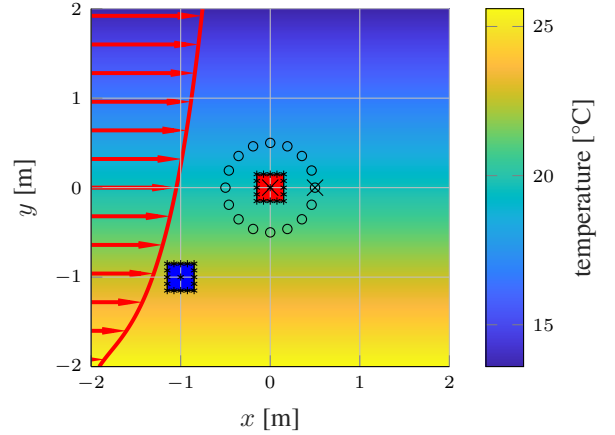}
	\caption{Geometry of the setup. The red square is the bright zone and the blue square is the dark zone. Circles are indicators for speakers, asterisks for microphones. The temperature can be read from the colour scale ($3\unit{°C/m}$ temperature gradient case). The logarithmic wind profile is shown where the wind speed corresponds to the wind speed at $y=0\unit{m}$.}
	\label{fig:muscles_geometry}
\end{figure}

In order to show that the proposed methodology works under harsh environmental conditions, we examine a more complex setting based on simulations of 2D Euler equations.
\paragraph{Simulation Setup}
The distribution of the microphones and the speakers remains unchanged (see \Cref{fig:muscles_geometry}). The underlying wind profile is logarithmic $\left(0.4673\Ma\cdot\speedSound\,\text{ln}\left((y + 2.3\unit{m})/0.3\unit{m} \right), 0\right)\unit{m/s}$ as proposed in \cite{richards_appropriate_1993} for atmospheric boundary layers with $\speedSound$ being the speed of sound. The wind speed corresponds to the wind speed at $y=0\unit{m}$ and it is set to $5\unit{Bft}$ ($9.35\unit{m/s}$). In addition to a nonuniform wind profile, a temperature gradient of $3\unit{°C/m}$ and $30\unit{°C/m}$ is introduced (see \Cref{fig:muscles_geometry}) where at $y=0\unit{m}$ the speed of sound is $343\unit{m/s}$. With this setup, no analytical sound field is known.

\begin{figure}[t]
	\setlength\figureheight{0.15\textwidth}
	\setlength\figurewidth{0.25\textwidth}
	\centering
    % This file was created by matlab2tikz.
%
\begin{tikzpicture}

\begin{axis}[%
width=\figurewidth,
height=\figureheight,
at={(0\figurewidth,0\figureheight)},
scale only axis,
point meta min=0.2325889917,
point meta max=0.7623955014,
axis on top,
xmin=286.3636363636,
xmax=3013.6363636364,
xlabel style={font=\color{white!15!black}},
xlabel={frequency $\f$ [Hz]},
ymin=-0.0013157895,
ymax=0.0513157895,
xmajorgrids,
ymajorgrids,
ylabel style={font=\color{white!15!black}},
ylabel={Mach number $\Ma$ [-]},
axis background/.style={fill=white},
colormap={mymap}{[1pt] rgb(0pt)=(0.2422,0.1504,0.6603); rgb(1pt)=(0.2444,0.1534,0.6728); rgb(2pt)=(0.2464,0.1569,0.6847); rgb(3pt)=(0.2484,0.1607,0.6961); rgb(4pt)=(0.2503,0.1648,0.7071); rgb(5pt)=(0.2522,0.1689,0.7179); rgb(6pt)=(0.254,0.1732,0.7286); rgb(7pt)=(0.2558,0.1773,0.7393); rgb(8pt)=(0.2576,0.1814,0.7501); rgb(9pt)=(0.2594,0.1854,0.761); rgb(11pt)=(0.2628,0.1932,0.7828); rgb(12pt)=(0.2645,0.1972,0.7937); rgb(13pt)=(0.2661,0.2011,0.8043); rgb(14pt)=(0.2676,0.2052,0.8148); rgb(15pt)=(0.2691,0.2094,0.8249); rgb(16pt)=(0.2704,0.2138,0.8346); rgb(17pt)=(0.2717,0.2184,0.8439); rgb(18pt)=(0.2729,0.2231,0.8528); rgb(19pt)=(0.274,0.228,0.8612); rgb(20pt)=(0.2749,0.233,0.8692); rgb(21pt)=(0.2758,0.2382,0.8767); rgb(22pt)=(0.2766,0.2435,0.884); rgb(23pt)=(0.2774,0.2489,0.8908); rgb(24pt)=(0.2781,0.2543,0.8973); rgb(25pt)=(0.2788,0.2598,0.9035); rgb(26pt)=(0.2794,0.2653,0.9094); rgb(27pt)=(0.2798,0.2708,0.915); rgb(28pt)=(0.2802,0.2764,0.9204); rgb(29pt)=(0.2806,0.2819,0.9255); rgb(30pt)=(0.2809,0.2875,0.9305); rgb(31pt)=(0.2811,0.293,0.9352); rgb(32pt)=(0.2813,0.2985,0.9397); rgb(33pt)=(0.2814,0.304,0.9441); rgb(34pt)=(0.2814,0.3095,0.9483); rgb(35pt)=(0.2813,0.315,0.9524); rgb(36pt)=(0.2811,0.3204,0.9563); rgb(37pt)=(0.2809,0.3259,0.96); rgb(38pt)=(0.2807,0.3313,0.9636); rgb(39pt)=(0.2803,0.3367,0.967); rgb(40pt)=(0.2798,0.3421,0.9702); rgb(41pt)=(0.2791,0.3475,0.9733); rgb(42pt)=(0.2784,0.3529,0.9763); rgb(43pt)=(0.2776,0.3583,0.9791); rgb(44pt)=(0.2766,0.3638,0.9817); rgb(45pt)=(0.2754,0.3693,0.984); rgb(46pt)=(0.2741,0.3748,0.9862); rgb(47pt)=(0.2726,0.3804,0.9881); rgb(48pt)=(0.271,0.386,0.9898); rgb(49pt)=(0.2691,0.3916,0.9912); rgb(50pt)=(0.267,0.3973,0.9924); rgb(51pt)=(0.2647,0.403,0.9935); rgb(52pt)=(0.2621,0.4088,0.9946); rgb(53pt)=(0.2591,0.4145,0.9955); rgb(54pt)=(0.2556,0.4203,0.9965); rgb(55pt)=(0.2517,0.4261,0.9974); rgb(56pt)=(0.2473,0.4319,0.9983); rgb(57pt)=(0.2424,0.4378,0.9991); rgb(58pt)=(0.2369,0.4437,0.9996); rgb(59pt)=(0.2311,0.4497,0.9995); rgb(60pt)=(0.225,0.4559,0.9985); rgb(61pt)=(0.2189,0.462,0.9968); rgb(62pt)=(0.2128,0.4682,0.9948); rgb(63pt)=(0.2066,0.4743,0.9926); rgb(64pt)=(0.2006,0.4803,0.9906); rgb(65pt)=(0.195,0.4861,0.9887); rgb(66pt)=(0.1903,0.4919,0.9867); rgb(67pt)=(0.1869,0.4975,0.9844); rgb(68pt)=(0.1847,0.503,0.9819); rgb(69pt)=(0.1831,0.5084,0.9793); rgb(70pt)=(0.1818,0.5138,0.9766); rgb(71pt)=(0.1806,0.5191,0.9738); rgb(72pt)=(0.1795,0.5244,0.9709); rgb(73pt)=(0.1785,0.5296,0.9677); rgb(74pt)=(0.1778,0.5349,0.9641); rgb(75pt)=(0.1773,0.5401,0.9602); rgb(76pt)=(0.1768,0.5452,0.956); rgb(77pt)=(0.1764,0.5504,0.9516); rgb(78pt)=(0.1755,0.5554,0.9473); rgb(79pt)=(0.174,0.5605,0.9432); rgb(80pt)=(0.1716,0.5655,0.9393); rgb(81pt)=(0.1686,0.5705,0.9357); rgb(82pt)=(0.1649,0.5755,0.9323); rgb(83pt)=(0.161,0.5805,0.9289); rgb(84pt)=(0.1573,0.5854,0.9254); rgb(85pt)=(0.154,0.5902,0.9218); rgb(86pt)=(0.1513,0.595,0.9182); rgb(87pt)=(0.1492,0.5997,0.9147); rgb(88pt)=(0.1475,0.6043,0.9113); rgb(89pt)=(0.1461,0.6089,0.908); rgb(90pt)=(0.1446,0.6135,0.905); rgb(91pt)=(0.1429,0.618,0.9022); rgb(92pt)=(0.1408,0.6226,0.8998); rgb(93pt)=(0.1383,0.6272,0.8975); rgb(94pt)=(0.1354,0.6317,0.8953); rgb(95pt)=(0.1321,0.6363,0.8932); rgb(96pt)=(0.1288,0.6408,0.891); rgb(97pt)=(0.1253,0.6453,0.8887); rgb(98pt)=(0.1219,0.6497,0.8862); rgb(99pt)=(0.1185,0.6541,0.8834); rgb(100pt)=(0.1152,0.6584,0.8804); rgb(101pt)=(0.1119,0.6627,0.877); rgb(102pt)=(0.1085,0.6669,0.8734); rgb(103pt)=(0.1048,0.671,0.8695); rgb(104pt)=(0.1009,0.675,0.8653); rgb(105pt)=(0.0964,0.6789,0.8609); rgb(106pt)=(0.0914,0.6828,0.8562); rgb(107pt)=(0.0855,0.6865,0.8513); rgb(108pt)=(0.0789,0.6902,0.8462); rgb(109pt)=(0.0713,0.6938,0.8409); rgb(110pt)=(0.0628,0.6972,0.8355); rgb(111pt)=(0.0535,0.7006,0.8299); rgb(112pt)=(0.0433,0.7039,0.8242); rgb(113pt)=(0.0328,0.7071,0.8183); rgb(114pt)=(0.0234,0.7103,0.8124); rgb(115pt)=(0.0155,0.7133,0.8064); rgb(116pt)=(0.0091,0.7163,0.8003); rgb(117pt)=(0.0046,0.7192,0.7941); rgb(118pt)=(0.0019,0.722,0.7878); rgb(119pt)=(0.0009,0.7248,0.7815); rgb(120pt)=(0.0018,0.7275,0.7752); rgb(121pt)=(0.0046,0.7301,0.7688); rgb(122pt)=(0.0094,0.7327,0.7623); rgb(123pt)=(0.0162,0.7352,0.7558); rgb(124pt)=(0.0253,0.7376,0.7492); rgb(125pt)=(0.0369,0.74,0.7426); rgb(126pt)=(0.0504,0.7423,0.7359); rgb(127pt)=(0.0638,0.7446,0.7292); rgb(128pt)=(0.077,0.7468,0.7224); rgb(129pt)=(0.0899,0.7489,0.7156); rgb(130pt)=(0.1023,0.751,0.7088); rgb(131pt)=(0.1141,0.7531,0.7019); rgb(132pt)=(0.1252,0.7552,0.695); rgb(133pt)=(0.1354,0.7572,0.6881); rgb(134pt)=(0.1448,0.7593,0.6812); rgb(135pt)=(0.1532,0.7614,0.6741); rgb(136pt)=(0.1609,0.7635,0.6671); rgb(137pt)=(0.1678,0.7656,0.6599); rgb(138pt)=(0.1741,0.7678,0.6527); rgb(139pt)=(0.1799,0.7699,0.6454); rgb(140pt)=(0.1853,0.7721,0.6379); rgb(141pt)=(0.1905,0.7743,0.6303); rgb(142pt)=(0.1954,0.7765,0.6225); rgb(143pt)=(0.2003,0.7787,0.6146); rgb(144pt)=(0.2061,0.7808,0.6065); rgb(145pt)=(0.2118,0.7828,0.5983); rgb(146pt)=(0.2178,0.7849,0.5899); rgb(147pt)=(0.2244,0.7869,0.5813); rgb(148pt)=(0.2318,0.7887,0.5725); rgb(149pt)=(0.2401,0.7905,0.5636); rgb(150pt)=(0.2491,0.7922,0.5546); rgb(151pt)=(0.2589,0.7937,0.5454); rgb(152pt)=(0.2695,0.7951,0.536); rgb(153pt)=(0.2809,0.7964,0.5266); rgb(154pt)=(0.2929,0.7975,0.517); rgb(155pt)=(0.3052,0.7985,0.5074); rgb(156pt)=(0.3176,0.7994,0.4975); rgb(157pt)=(0.3301,0.8002,0.4876); rgb(158pt)=(0.3424,0.8009,0.4774); rgb(159pt)=(0.3548,0.8016,0.4669); rgb(160pt)=(0.3671,0.8021,0.4563); rgb(161pt)=(0.3795,0.8026,0.4454); rgb(162pt)=(0.3921,0.8029,0.4344); rgb(163pt)=(0.405,0.8031,0.4233); rgb(164pt)=(0.4184,0.803,0.4122); rgb(165pt)=(0.4322,0.8028,0.4013); rgb(166pt)=(0.4463,0.8024,0.3904); rgb(167pt)=(0.4608,0.8018,0.3797); rgb(168pt)=(0.4753,0.8011,0.3691); rgb(169pt)=(0.4899,0.8002,0.3586); rgb(170pt)=(0.5044,0.7993,0.348); rgb(171pt)=(0.5187,0.7982,0.3374); rgb(172pt)=(0.5329,0.797,0.3267); rgb(173pt)=(0.547,0.7957,0.3159); rgb(175pt)=(0.5748,0.7929,0.2941); rgb(176pt)=(0.5886,0.7913,0.2833); rgb(177pt)=(0.6024,0.7896,0.2726); rgb(178pt)=(0.6161,0.7878,0.2622); rgb(179pt)=(0.6297,0.7859,0.2521); rgb(180pt)=(0.6433,0.7839,0.2423); rgb(181pt)=(0.6567,0.7818,0.2329); rgb(182pt)=(0.6701,0.7796,0.2239); rgb(183pt)=(0.6833,0.7773,0.2155); rgb(184pt)=(0.6963,0.775,0.2075); rgb(185pt)=(0.7091,0.7727,0.1998); rgb(186pt)=(0.7218,0.7703,0.1924); rgb(187pt)=(0.7344,0.7679,0.1852); rgb(188pt)=(0.7468,0.7654,0.1782); rgb(189pt)=(0.759,0.7629,0.1717); rgb(190pt)=(0.771,0.7604,0.1658); rgb(191pt)=(0.7829,0.7579,0.1608); rgb(192pt)=(0.7945,0.7554,0.157); rgb(193pt)=(0.806,0.7529,0.1546); rgb(194pt)=(0.8172,0.7505,0.1535); rgb(195pt)=(0.8281,0.7481,0.1536); rgb(196pt)=(0.8389,0.7457,0.1546); rgb(197pt)=(0.8495,0.7435,0.1564); rgb(198pt)=(0.86,0.7413,0.1587); rgb(199pt)=(0.8703,0.7392,0.1615); rgb(200pt)=(0.8804,0.7372,0.165); rgb(201pt)=(0.8903,0.7353,0.1695); rgb(202pt)=(0.9,0.7336,0.1749); rgb(203pt)=(0.9093,0.7321,0.1815); rgb(204pt)=(0.9184,0.7308,0.189); rgb(205pt)=(0.9272,0.7298,0.1973); rgb(206pt)=(0.9357,0.729,0.2061); rgb(207pt)=(0.944,0.7285,0.2151); rgb(208pt)=(0.9523,0.7284,0.2237); rgb(209pt)=(0.9606,0.7285,0.2312); rgb(210pt)=(0.9689,0.7292,0.2373); rgb(211pt)=(0.977,0.7304,0.2418); rgb(212pt)=(0.9842,0.733,0.2446); rgb(213pt)=(0.99,0.7365,0.2429); rgb(214pt)=(0.9946,0.7407,0.2394); rgb(215pt)=(0.9966,0.7458,0.2351); rgb(216pt)=(0.9971,0.7513,0.2309); rgb(217pt)=(0.9972,0.7569,0.2267); rgb(218pt)=(0.9971,0.7626,0.2224); rgb(219pt)=(0.9969,0.7683,0.2181); rgb(220pt)=(0.9966,0.774,0.2138); rgb(221pt)=(0.9962,0.7798,0.2095); rgb(222pt)=(0.9957,0.7856,0.2053); rgb(223pt)=(0.9949,0.7915,0.2012); rgb(224pt)=(0.9938,0.7974,0.1974); rgb(225pt)=(0.9923,0.8034,0.1939); rgb(226pt)=(0.9906,0.8095,0.1906); rgb(227pt)=(0.9885,0.8156,0.1875); rgb(228pt)=(0.9861,0.8218,0.1846); rgb(229pt)=(0.9835,0.828,0.1817); rgb(230pt)=(0.9807,0.8342,0.1787); rgb(231pt)=(0.9778,0.8404,0.1757); rgb(232pt)=(0.9748,0.8467,0.1726); rgb(233pt)=(0.972,0.8529,0.1695); rgb(234pt)=(0.9694,0.8591,0.1665); rgb(235pt)=(0.9671,0.8654,0.1636); rgb(236pt)=(0.9651,0.8716,0.1608); rgb(237pt)=(0.9634,0.8778,0.1582); rgb(238pt)=(0.9619,0.884,0.1557); rgb(239pt)=(0.9608,0.8902,0.1532); rgb(240pt)=(0.9601,0.8963,0.1507); rgb(241pt)=(0.9596,0.9023,0.148); rgb(242pt)=(0.9595,0.9084,0.145); rgb(243pt)=(0.9597,0.9143,0.1418); rgb(244pt)=(0.9601,0.9203,0.1382); rgb(245pt)=(0.9608,0.9262,0.1344); rgb(246pt)=(0.9618,0.932,0.1304); rgb(247pt)=(0.9629,0.9379,0.1261); rgb(248pt)=(0.9642,0.9437,0.1216); rgb(249pt)=(0.9657,0.9494,0.1168); rgb(250pt)=(0.9674,0.9552,0.1116); rgb(251pt)=(0.9692,0.9609,0.1061); rgb(252pt)=(0.9711,0.9667,0.1001); rgb(253pt)=(0.973,0.9724,0.0938); rgb(254pt)=(0.9749,0.9782,0.0872); rgb(255pt)=(0.9769,0.9839,0.0805)},
colorbar,
colorbar style={ylabel style={font=\color{white!15!black}},ylabel=$\ampMod_{11}$ [-]}
]
\addplot [forget plot] graphics [xmin=286.3636363636, xmax=3013.6363636364, ymin=-0.0013157895, ymax=0.0513157895] {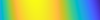};
\end{axis}
\end{tikzpicture}%
	\caption{Amplitude modulation $\ampMod_{11}$ [-] of the acoustic transfer function between the speaker and microphone indicated with a cross in \Cref{fig:muscles_geometry} which shows the frequency dependency of the amplitude modulation due to the source not being a point source.}
	\label{fig:muscles_ampModulation}
\end{figure}

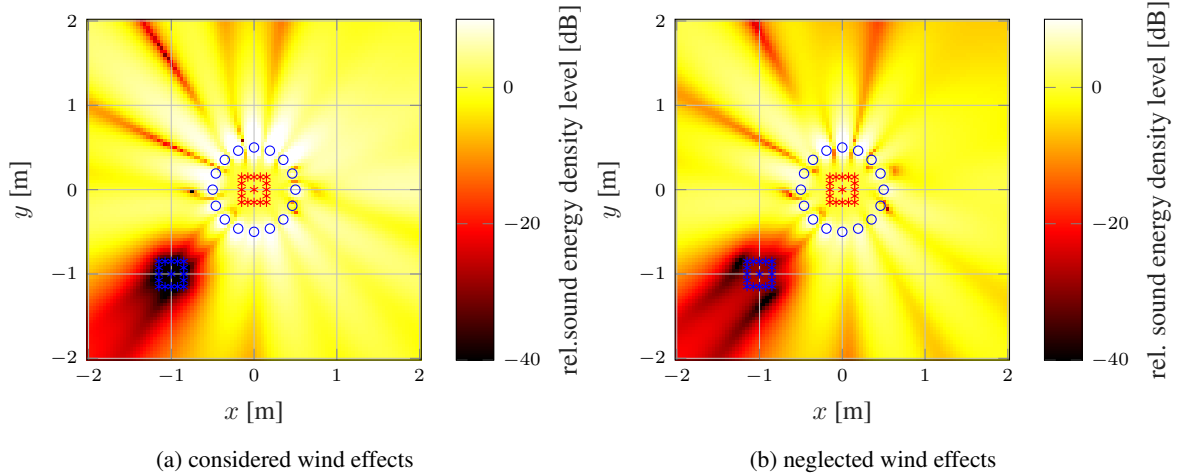
\begin{figure*}[ht!]
    \centering
	\begin{subfigure}[t]{\picWidth}
    	\setlength\figureheight{0.6\textwidth}
    	\setlength\figurewidth{0.7\textwidth}
    	\centering
        % This file was created by matlab2tikz.
%
\begin{tikzpicture}

\begin{axis}[%
width=0.84052\figurewidth,
height=\figureheight,
at={(0\figurewidth,0\figureheight)},
scale only axis,
point meta min=-40.0000000000,
point meta max=10.0000000000,
axis on top,
xmin=-2.0202020202,
xmax=2.0202020202,
xlabel style={font=\color{white!15!black}},
xlabel={$x$ [m]},
ymin=-2.0202020202,
ymax=2.0202020202,
xmajorgrids,
ymajorgrids,
ylabel style={font=\color{white!15!black}},
ylabel={$y$ [m]},
axis background/.style={fill=white},
colormap/hot2,
colorbar,
colorbar style={ylabel style={font=\color{white!15!black}},ylabel=rel.sound energy density level [dB]}
]
\addplot [forget plot] graphics [xmin=-2.0202020202, xmax=2.0202020202, ymin=-2.0202020202, ymax=2.0202020202] {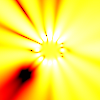};
\addplot[only marks, mark=asterisk, mark options={}, mark size=1.7678pt, draw=red, forget plot] table[row sep=crcr]{%
x	y\\
0.0000000000	0.0000000000\\
0.1500000000	-0.1500000000\\
0.1500000000	-0.0750000000\\
0.1500000000	0.0000000000\\
0.1500000000	0.0750000000\\
0.1500000000	0.1500000000\\
0.0750000000	0.1500000000\\
0.0000000000	0.1500000000\\
-0.0750000000	0.1500000000\\
-0.1500000000	0.1500000000\\
-0.1500000000	0.0750000000\\
-0.1500000000	0.0000000000\\
-0.1500000000	-0.0750000000\\
-0.1500000000	-0.1500000000\\
-0.0750000000	-0.1500000000\\
0.0000000000	-0.1500000000\\
0.0750000000	-0.1500000000\\
};
\addplot[only marks, mark=asterisk, mark options={}, mark size=1.7678pt, draw=blue, forget plot] table[row sep=crcr]{%
x	y\\
-1.0000000000	-1.0000000000\\
-0.8500000000	-1.1500000000\\
-0.8500000000	-1.0750000000\\
-0.8500000000	-1.0000000000\\
-0.8500000000	-0.9250000000\\
-0.8500000000	-0.8500000000\\
-0.9250000000	-0.8500000000\\
-1.0000000000	-0.8500000000\\
-1.0750000000	-0.8500000000\\
-1.1500000000	-0.8500000000\\
-1.1500000000	-0.9250000000\\
-1.1500000000	-1.0000000000\\
-1.1500000000	-1.0750000000\\
-1.1500000000	-1.1500000000\\
-1.0750000000	-1.1500000000\\
-1.0000000000	-1.1500000000\\
-0.9250000000	-1.1500000000\\
};
\addplot[only marks, mark=o, mark options={}, mark size=1.7678pt, draw=blue, forget plot] table[row sep=crcr]{%
x	y\\
0.5000000000	0.0000000000\\
0.4619397663	0.1913417162\\
0.3535533906	0.3535533906\\
0.1913417162	0.4619397663\\
0.0000000000	0.5000000000\\
-0.1913417162	0.4619397663\\
-0.3535533906	0.3535533906\\
-0.4619397663	0.1913417162\\
-0.5000000000	0.0000000000\\
-0.4619397663	-0.1913417162\\
-0.3535533906	-0.3535533906\\
-0.1913417162	-0.4619397663\\
-0.0000000000	-0.5000000000\\
0.1913417162	-0.4619397663\\
0.3535533906	-0.3535533906\\
0.4619397663	-0.1913417162\\
};
\end{axis}

\begin{axis}[%
width=1.226994\figurewidth,
height=1.226994\figureheight,
at={(-0.14099\figurewidth,-0.134969\figureheight)},
scale only axis,
point meta min=0.0000000000,
point meta max=1.0000000000,
xmin=0.0000000000,
xmax=1.0000000000,
ymin=0.0000000000,
ymax=1.0000000000,
axis line style={draw=none},
ticks=none,
axis x line*=bottom,
axis y line*=left
]
\end{axis}
\end{tikzpicture}%
		\caption{considered wind effects}
		\label{fig:muscles_windConsideredEnergyLevel}
	\end{subfigure}
~
	\begin{subfigure}[t]{\picWidth}
    	\setlength\figureheight{0.6\textwidth}
    	\setlength\figurewidth{0.7\textwidth}
    	\centering
        % This file was created by matlab2tikz.
%
\begin{tikzpicture}

\begin{axis}[%
width=0.84052\figurewidth,
height=\figureheight,
at={(0\figurewidth,0\figureheight)},
scale only axis,
point meta min=-40.0000000000,
point meta max=10.0000000000,
axis on top,
xmin=-2.0202020202,
xmax=2.0202020202,
xlabel style={font=\color{white!15!black}},
xlabel={$x$ [m]},
ymin=-2.0202020202,
ymax=2.0202020202,
xmajorgrids,
ymajorgrids,
ylabel style={font=\color{white!15!black}},
ylabel={$y$ [m]},
axis background/.style={fill=white},
colormap/hot2,
colorbar,
colorbar style={ylabel style={font=\color{white!15!black}},ylabel=rel. sound energy density level [dB]}
]
\addplot [forget plot] graphics [xmin=-2.0202020202, xmax=2.0202020202, ymin=-2.0202020202, ymax=2.0202020202] {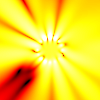};
\addplot[only marks, mark=asterisk, mark options={}, mark size=1.7678pt, draw=red, forget plot] table[row sep=crcr]{%
x	y\\
0.0000000000	0.0000000000\\
0.1500000000	-0.1500000000\\
0.1500000000	-0.0750000000\\
0.1500000000	0.0000000000\\
0.1500000000	0.0750000000\\
0.1500000000	0.1500000000\\
0.0750000000	0.1500000000\\
0.0000000000	0.1500000000\\
-0.0750000000	0.1500000000\\
-0.1500000000	0.1500000000\\
-0.1500000000	0.0750000000\\
-0.1500000000	0.0000000000\\
-0.1500000000	-0.0750000000\\
-0.1500000000	-0.1500000000\\
-0.0750000000	-0.1500000000\\
0.0000000000	-0.1500000000\\
0.0750000000	-0.1500000000\\
};
\addplot[only marks, mark=asterisk, mark options={}, mark size=1.7678pt, draw=blue, forget plot] table[row sep=crcr]{%
x	y\\
-1.0000000000	-1.0000000000\\
-0.8500000000	-1.1500000000\\
-0.8500000000	-1.0750000000\\
-0.8500000000	-1.0000000000\\
-0.8500000000	-0.9250000000\\
-0.8500000000	-0.8500000000\\
-0.9250000000	-0.8500000000\\
-1.0000000000	-0.8500000000\\
-1.0750000000	-0.8500000000\\
-1.1500000000	-0.8500000000\\
-1.1500000000	-0.9250000000\\
-1.1500000000	-1.0000000000\\
-1.1500000000	-1.0750000000\\
-1.1500000000	-1.1500000000\\
-1.0750000000	-1.1500000000\\
-1.0000000000	-1.1500000000\\
-0.9250000000	-1.1500000000\\
};
\addplot[only marks, mark=o, mark options={}, mark size=1.7678pt, draw=blue, forget plot] table[row sep=crcr]{%
x	y\\
0.5000000000	0.0000000000\\
0.4619397663	0.1913417162\\
0.3535533906	0.3535533906\\
0.1913417162	0.4619397663\\
0.0000000000	0.5000000000\\
-0.1913417162	0.4619397663\\
-0.3535533906	0.3535533906\\
-0.4619397663	0.1913417162\\
-0.5000000000	0.0000000000\\
-0.4619397663	-0.1913417162\\
-0.3535533906	-0.3535533906\\
-0.1913417162	-0.4619397663\\
-0.0000000000	-0.5000000000\\
0.1913417162	-0.4619397663\\
0.3535533906	-0.3535533906\\
0.4619397663	-0.1913417162\\
};
\end{axis}
\end{tikzpicture}%
        \caption{neglected wind effects}
        \label{fig:muscles_windNeglectedEnergyLevel}
    \end{subfigure}
    \caption{Comparison of the sound energy density level [dB] relative to the bright zone at frequency $f=600\unit{Hz}$ and regularisation parameter $\regParam=0$ with considered and neglected wind effect at Mach number $\Ma=0.0275$ ($5\unit{Bft}$) and temperature stratification at $3\unit{°C/m}$ gradient.}
    \label{fig:muscles_comparisonNeglectNotNeglect}
\end{figure*}
\renewcommand{\arraystretch}{1.2}
\begin{table*}[hb!]
    \centering
    \begin{tabular}{@{}l l l@{}}
        \toprule
         wind status& temperature gradient considered & temperature gradient neglected \\
         \midrule               
         wind considered  &  $\AC = 45.1\unit{dB}$, $\RE = -33.5\unit{dB}$ & $\AC = 44.4\unit{dB}$, $\RE = -33.9\unit{dB}$ \\
         wind neglected &  $\AC = 28.6\unit{dB}$, $\RE = -16.4\unit{dB}$ & $\AC = 28.7\unit{dB}$, $\RE = -16.5\unit{dB}$\\
         \bottomrule
    \end{tabular}
    \caption{Comparison of the wind profiles at $\Ma=0.0275$ (5Bft) (at $y=0\unit{m}$) and the temperature gradient of $3\unit{°C/m}$ on the $\RE$ and $\AC$.}
\label{tab:muscles_results3}
\end{table*}

\renewcommand{\arraystretch}{1.2}
\begin{table*}[hb!]
    \centering
    \begin{tabular}{@{}l l l@{}}
        \toprule
         wind status& temperature gradient considered & temperature gradient neglected \\
         \midrule               
         wind considered  &  $\AC = 46.5\unit{dB}$, $\RE = -33.5\unit{dB}$ & $\AC = 32.7\unit{dB}$, $\RE = -29.6\unit{dB}$ \\
         wind neglected &  $\AC = 28.8\unit{dB}$, $\RE = -16.5\unit{dB}$ & $\AC = 26.1\unit{dB}$, $\RE = -16.4\unit{dB}$\\
         \bottomrule
    \end{tabular}
    \caption{Comparison of the wind profiles at $\Ma=0.0275$ ($5\unit{Bft}$) (at $y=0\unit{m}$) and the temperature gradient of $30\unit{°C/m}$ on the $\RE$ and $\AC$.}
\label{tab:muscles_results30}
\end{table*}
The training data for the networks is created using a high order finite difference solver of the Euler equations in 2D.\footnote{
See \ref{sec:appParamsEuler} for relevant parameters of the simulation. See \cite{lemke_adjoint_2015} for time and space discretisation schemes and boundary conditions.} The reduction from 3D to 2D was done to speed up the process of training data creation. Since there is no qualitative difference to sound propagation from 2D to 3D, this does not reduce the meaning of the showcase.\footnote{In 3D the amplitude of the pressure reduces inverse proportional to the distance $\s$ to the sound source. In 2D it reduces inverse proportional to the square root of the distance $\sqrt{\s}$ to the sound source.} Since a finite difference solver is used, the speakers cannot be modelled as point sources but rather as a source with the Gaussian function as spacial distribution. This leads to a frequency dependency of the amplitude modulation $\ampMod$ of the acoustic transfer functions (see \Cref{fig:muscles_ampModulation}) which can easily be learned by the neural networks. 
%The training data for one setup consists simulations with varying frequency of the speaker and varying wind speed. 7 different frequencies and 3 different wind speeds (including wind speed 0m/s) were simulated. This shows that a small number of simulations needed to be simulated.

\paragraph{Results}
\Cref{fig:muscles_comparisonNeglectNotNeglect} shows the sound energy density level relative to the bright zone with considered and neglected wind and temperature ($3\unit{°C/m}$ gradient case) influence. It is easily visible that the consideration of both influences leads to a higher $\AC$.

An overview of the results when neglecting or considering wind and/or the temperature gradient of $3\unit{°C/m}$ are shown in \Cref{tab:muscles_results3}. As in the previous cases, abstracting from wind leads to a drop of the $\AC$  by $16\unit{dB}$. Also the $\RE$ is by $17\unit{dB}$ higher compared to the model that includes wind dynamics. 

The consideration of the temperature gradient of $3\unit{°C/m}$ seems to have a very small impact for this case since both $\AC$ and $\RE$ do not differ by more than $1\unit{dB}$. It is expected that the $\RE$ is not strongly affected by the temperature gradient since the temperatures are not significantly different from the temperatures in the middle of the domain. However, the dark zone is by $3\unit{K}$ warmer than the bright zone which leads to an increase of 0.7\% of the speed of sound. With the regularisation parameter $\regParam=0$ the wind with the Mach number $\Ma<0.007$ already had an effect on the $\AC$ in the previous case. The influence of this effect can be bigger in real applications where the distance between the speakers and the microphones is much larger and therefore smaller temperature gradients can have a higher effect on the solution. 

To show that the temperature can influence the results, a second setup with a high temperature gradient of $30\unit{°C/m}$ is introduced. The results are displayed in \Cref{tab:muscles_results30} and it is visible that neglecting the temperature gradient leads to worse performance of the multi zone sound field reproduction. Interestingly, when ignoring wind, the absence of the temperature gradient in the model leads only to small additional errors in the $\AC$ ($2.7\unit{dB}$) and negligible differences in the $\RE$ ($0.1\unit{dB}$). 
Both setups imply that the proposed methodology works for setups where no analytical solution of the sound propagation is available.

\section{Discussion}
\label{sec:discussion}
% \begin{itemize}
%     \item proposed methodology works
%     \item with stronger physical effects the difference between the results with the proposed methodology and with the neglected methodology get stronger
%     \item this can be used to either estimate whether considering the physical effect is necessary or how big the error of neglecting it is
%     \item regularisation plays an important role because a bigger regularisation leads to a smaller difference between considering wind and neglecting wind
%     \item methodology is initially relatively computationally expensive when basing on simulated results since the nonlinear effects need to be simulated (i.e. with solving the Euler equations) but can quite easily be done when using measurements. When the neural networks are trained, getting the acoustic transfer functions is not computationally expensive.
%     \item Methodology can be expanded to also include linearly scaled temperature distributions or i.e. reflections as long as enough training data is available
% \end{itemize}
This manuscript provides a proof of concept for a new multi zone sound field reproduction approach that considers harsh environmental conditions.
Our method is able to reproduce results from \cite{du_multizone_2021} and capable of considering several physical effects that influence sound propagation \philipp{e.g.~}wind, temperature stratification, reflections and dispersion.

The results show that the proposed methodology works well for setups with wind and temperature stratification. The magnitude of the physical effect correlates with the difference in quality of the solution between abstracting from wind and temperature or explicitly account for these environmental influences.

Furthermore the results show that the regularisation parameter plays an important role in the decision whether to consider physical effects or not. A stronger regularisation leads to larger costs for the control variables (speaker signal). Therefore, the optimisation procedure will tend to neglect additional physical effects, like wind.

When using simulations for the creation of the training data, the methodology can be initially computationally intensive since the nonlinear effects of wind, temperature fluctuations and other possible effects need to be simulated which leads to solving the Euler equations in 3D. Using measurements can simplify the creation of the training data. One should keep in mind that only the creation of the training data is computationally intensive. After the neural networks have been trained, they can be rapidly evaluated and used inside the optimisation problem. In this way we can instantly generate an optimised speaker signal, which is adapted to the measured wind and temperature values.

It is important to note that training neural networks to replicate nonlinear physical effects on the transfer functions can be used by all methods that are based on transfer functions.

Future studies can possibly explore studying and generalising the framework for complex geometries like rooms. Another point of interest is to test the framework in physical experiments instead of numerical experiments. The framework can also be used to serve as an initial guess for sound field reproduction with adjoint methods to decrease its computational intensity. 
%The methodology works for all physical models where the superposition principle is valid. \philipp{hast du ja schon im absatz davor gesagt.}
%Therefore wave field reproduction of other wave equations like the shallow water equations or the Maxwell equations can be as well handled with this methodology.

\addcontentsline{toc}{section}{Declaration of competing interest}
\section*{Declaration of competing interest}
The authors declare that they have no known competing financial interests or personal relationships that could have appeared to influence the work reported in this paper.

\addcontentsline{toc}{section}{Author Contribution Statement (CRediT)}
\section*{Author Contribution Statement (CRediT)}
\textbf{Henry Sallandt:} Conceptualisation, Methodology, Software, Investigation, Writing - Original Draft, Visualisation\\
\textbf{Philipp Krah:} Conceptualisation, Methodology, Writing - Original Draft, Supervision, Writing - Review \& Editing\\
\textbf{Mathias Lemke:} Conceptualisation, Software (\philipp{CAA} solver), Supervision, Writing - Review \& Editing\\ 

\addcontentsline{toc}{section}{Acknowledgement}
\section*{Acknowledgement}
\philipp{The authors acknowledge financial support by the Deutsche Forschungsgemeinschaft (DFG) within the project LE 3888/2.}
Philipp Krah gratefully acknowledges the support of the Deutsche Forschungsgemeinschaft (DFG) as part of GRK2433 DAEDALUS.

\newpage
\bibliographystyle{elsarticle-harv}
\bibliography{bibliography.bib}

\newpage
\appendix
\section{Discontinuity in Phase Angle}
\label{sec:appDiscontinuityPhaseAngle}
\begin{figure}
	\setlength\figureheight{0.25\textwidth}
	\setlength\figurewidth{0.4\textwidth}
	\centering
    \input{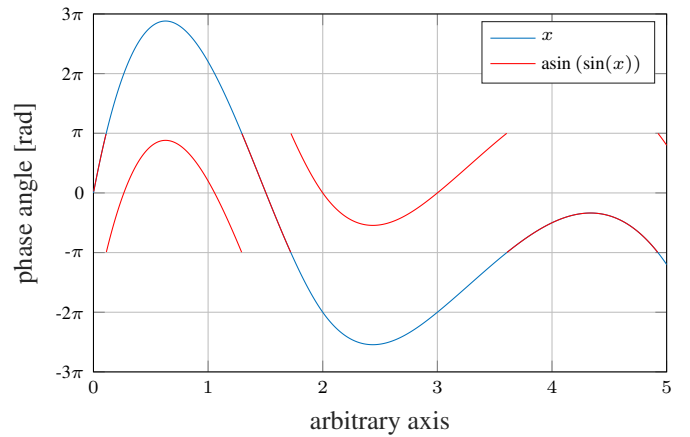}
	\caption{Discontinuity problem: The periodicity of the sin(x) function leads to a discontinuity for $\text{asin}(\text{sin}(x))$ if  $x$ surpasses $[-\pi, \pi]$.}
	\label{fig:app_discontinuity}
\end{figure}

% \begin{itemize}
%     \item neural networks have problems with discontinuous functions
%     \item there are special neural networks for discontinuous functions but it is easier to dodge the discontinuities if possible
%     \item phase angle is discontinuous or does not have a maximum or minimum
%     \item phase angle represented learnt by two variables: real part and imaginary part of the Euler's formula
% \end{itemize}
For a $2\pi$-periodic function $f(x)=y$ it holds $f(x) = f(2\pi n + x)$ with $n\in\mathbb{Z}$. Therefore the inverse function $f^{-1}(y) = x$ only returns values in the interval $[-\pi, \pi]$. This leads to a discontinuity when $x$ surpasses the previously stated interval $[-\pi, \pi]$ which is displayed in \Cref{fig:app_discontinuity}.

The first option would be to detect the discontinuities by looking at the finite difference quotient and then stitching the discontinuities together to get the original function. This leads to the problem that the original function is not bounded and also the stitching process can be complicated when having only few sample points of the function.

The second option is to plug the values into a function that leads to an injective but continuous mapping and train the neural network on the result(s). After evaluating the neural networks the inverse mapping needs to be evaluated to get the original $x$. The chosen function leads to two results $v_1$ and $v_2$ that need to be learned:
\begin{linenomath}\begin{equation}
    v_1 = \text{cos}\left(x\right)
\end{equation}\end{linenomath}
\begin{linenomath}\begin{equation}
    v_2 = \text{sin}\left(x\right)
\end{equation}\end{linenomath}

The inverse mapping is the four quadrant inverse tangens $x=\text{atan2}\left(v_2, v_1\right)$.
%%%%%%%%%%%%%%%%%%%%%%%%%%%%%%%%%%%%%%%%%%%%%%%%%%%%%%%%%%%%%%%%%%%%%%%%%%%%%%%%%%%%%%%%%%%%%%%%
\section{Derivation of Green's Function of the 3D Wave Equation with Uniform Background Flow}
\label{sec:appDerivGreen}
We now seek Green's Function $\greenFunction\left(x,y,z,t\right)$ of the d'Alembert operator (hence 3D wave equation) in a free field with a uniform subsonic mean flow of the form $(\backgroundFlow,0,0)$. This is done similar to \cite{bailly_numerical_2000}, but in 3D. The wave equation of the system is the following:
\begin{linenomath}\begin{equation}
    \frac{\text{d}^2\greenFunction}{\text{d}t^2} - \speedSound^2\nabla^2\greenFunction = \delta\left(x,y,z\right)\text{exp}\left(-i\omega t\right)\,,
\end{equation}\end{linenomath}
where $\frac{\text{d}}{\text{d}t} = \frac{\partial}{\partial t} + \backgroundFlow \frac{\partial}{\partial x}$. With the introduction of a coordinate system moving with the mean flow $\tilde{t} = t$, $\tilde{x} = x-\backgroundFlow t$, $\tilde{y} = y$, $\tilde{z} = z$ the wave equation becomes
\begin{linenomath}\begin{equation}
    \frac{\partial^2\tilde{\greenFunction}}{\partial\tilde{t}^2} - \speedSound^2\tilde{\nabla}^2\tilde{\greenFunction} = \delta\left(\tilde{x}+\backgroundFlow\tilde{t},\tilde{y},\tilde{z}\right)\text{exp}\left(-i\omega \tilde{t}\right),
\end{equation}\end{linenomath}
with $\tilde{\greenFunction}\left(\tilde{x},\tilde{y},\tilde{z},\tilde{t}\right) = \greenFunction\left(x,y,z,t\right)$ and $\delta$ being the $\delta$ distribution. Green's function can be calculated in the primitive variables as:
\begin{linenomath}\begin{equation}
\begin{split}
    \greenFunction\left(x,y,z,t\right)& =  \frac{\text{exp}\left(-i\omega t\right)}{4\pi}\bigintssss\frac{\text{exp}\left(i\omega \tau\right)\theta\left(\tau\right)}{\sqrt{\left(x - \backgroundFlow\tau\right)^2 + y^2 + z^2}}\\
    &\cdot\delta\left(\tau - \sqrt{\frac{\left(x - \backgroundFlow\tau\right)^2 + y^2 + z^2}{\speedSound^2}}\right)\text{d}\tau.
\end{split}
\label{eq:app_greenFunctionIntegral}
\end{equation}\end{linenomath}
Here, $\theta$ is the Heaviside function. The sampling property of the delta distribution can be used to solve the integral by finding the delta distribution's argument root.
\begin{linenomath}\begin{equation}
    0 = \tau - \sqrt{\frac{\left(x - \Ma \: \speedSound\tau\right)^2 + y^2 + z^2}{\speedSound^2}}
\end{equation}\end{linenomath}
The root of the argument can be calculated as
\begin{linenomath}\begin{equation}
    \tau^{1/2} = -\frac{\left(x\Ma\pm\sqrt{x^2 + \left(1-\Ma^2\right)\left(y^2 + z^2\right)}\right)}{\left(1-\Ma^2\right)\speedSound}.
\end{equation}\end{linenomath}
Both values $\tau^1$ and $\tau^2$ need to be checked if they are bigger than 0 because of the Heaviside function. Every positive value is plugged into the integral's argument in \Cref{eq:app_greenFunctionIntegral} and added together to get Green's function. The source term of the speaker is of the form $q=\hat{q}\left(x,y,z\right)\text{exp}\left(-i\omega t\right)$. Therefore the space convolution product of $\hat{q}$ and $\greenFunction$ gives the pressure distribution.
%%%%%%%%%%%%%%%%%%%%%%%%%%%%%%%%%%%%%%%%%%%%%%%%%%%%%%%%%%%%%%%%%%%%%%%%%%%%%%%%%%%%%%%%%%%%%%%%
\section{Iteration Process of Calculating the Wave Number Modulation Factor}
\label{sec:appIterateWaveModFac}
The goal of this calculation is that the phase modulation is constant with respect to the frequency (or more general: the mean derivative with respect to the frequency is minimised). The wind leads to a contraction or an extension of the sound waves. Since in this application a relatively small Mach number $\Ma \ll 1$ is examined the initial wave number modulation of $\waveNumberMod^{\iterationStep=0}=1$ can be assumed. The resulting phase angle can be calculated and if the wave number modulation was correctly assumed this would result in a constant phase modulation with respect to the frequency.

\Cref{fig:iterationProcessWaveNumberCorr} shows the phase modulation for the first 3 iterations. \Cref{fig:app_waveNumberModulation} visualises the respective wavenumber modulation factors for the first 3 iterations. It is visible that for low Mach numbers the assumption of a wave number modulation $\waveNumberMod=1$ is quite accurate. With higher Mach numbers the contraction or extension of the sound waves becomes stronger. To get the correct wave number modulation $\tilde{k}$ the absolute value of the partial derivative of the wrong corrected phase angle $\tilde{\varphi}$ needs to be minimised. This can be achieved with a gradient descent method.

\begin{figure}
	\centering
	\setlength\figureheight{0.2\textwidth}
	\setlength\figurewidth{0.3\textwidth}

	%\resizebox{\linewidth}{!}{
    % This file was created by matlab2tikz.
%
\definecolor{mycolor1}{rgb}{0.00000,0.44700,0.74100}%
\definecolor{mycolor2}{rgb}{0.85000,0.32500,0.09800}%
\definecolor{mycolor3}{rgb}{0.92900,0.69400,0.12500}%
\begin{tikzpicture}

\begin{axis}[%
width=\figurewidth,
height=\figureheight,
at={(0\figurewidth,0\figureheight)},
scale only axis,
xmin=0.0000000000,
xmax=0.3000000000,
xlabel style={font=\color{white!15!black}},
xlabel={Mach number $\Ma$ [-]},
ymin=0.97500000000,
ymax=1.3000000000,
xmajorgrids,
ymajorgrids,
ylabel style={font=\color{white!15!black}},
ylabel={$\waveNumberMod^{\iterationStep}$ [-]},
axis background/.style={fill=white},
legend style={at={(0.03,0.97)}, anchor=north west, legend cell align=left, align=left, draw=white!15!black},
scaled ticks=false, xticklabel style={/pgf/number format/fixed},yticklabel style={/pgf/number format/fixed}
]
\addplot [color=mycolor1]
  table[row sep=crcr]{%
0.0000000000	1.0000000000\\
0.0030303030	1.0000000000\\
0.0060606061	1.0000000000\\
0.0090909091	1.0000000000\\
0.0121212121	1.0000000000\\
0.0151515152	1.0000000000\\
0.0181818182	1.0000000000\\
0.0212121212	1.0000000000\\
0.0242424242	1.0000000000\\
0.0272727273	1.0000000000\\
0.0303030303	1.0000000000\\
0.0333333333	1.0000000000\\
0.0363636364	1.0000000000\\
0.0393939394	1.0000000000\\
0.0424242424	1.0000000000\\
0.0454545455	1.0000000000\\
0.0484848485	1.0000000000\\
0.0515151515	1.0000000000\\
0.0545454545	1.0000000000\\
0.0575757576	1.0000000000\\
0.0606060606	1.0000000000\\
0.0636363636	1.0000000000\\
0.0666666667	1.0000000000\\
0.0696969697	1.0000000000\\
0.0727272727	1.0000000000\\
0.0757575758	1.0000000000\\
0.0787878788	1.0000000000\\
0.0818181818	1.0000000000\\
0.0848484848	1.0000000000\\
0.0878787879	1.0000000000\\
0.0909090909	1.0000000000\\
0.0939393939	1.0000000000\\
0.0969696970	1.0000000000\\
0.1000000000	1.0000000000\\
0.1030303030	1.0000000000\\
0.1060606061	1.0000000000\\
0.1090909091	1.0000000000\\
0.1121212121	1.0000000000\\
0.1151515152	1.0000000000\\
0.1181818182	1.0000000000\\
0.1212121212	1.0000000000\\
0.1242424242	1.0000000000\\
0.1272727273	1.0000000000\\
0.1303030303	1.0000000000\\
0.1333333333	1.0000000000\\
0.1363636364	1.0000000000\\
0.1393939394	1.0000000000\\
0.1424242424	1.0000000000\\
0.1454545455	1.0000000000\\
0.1484848485	1.0000000000\\
0.1515151515	1.0000000000\\
0.1545454545	1.0000000000\\
0.1575757576	1.0000000000\\
0.1606060606	1.0000000000\\
0.1636363636	1.0000000000\\
0.1666666667	1.0000000000\\
0.1696969697	1.0000000000\\
0.1727272727	1.0000000000\\
0.1757575758	1.0000000000\\
0.1787878788	1.0000000000\\
0.1818181818	1.0000000000\\
0.1848484848	1.0000000000\\
0.1878787879	1.0000000000\\
0.1909090909	1.0000000000\\
0.1939393939	1.0000000000\\
0.1969696970	1.0000000000\\
0.2000000000	1.0000000000\\
0.2030303030	1.0000000000\\
0.2060606061	1.0000000000\\
0.2090909091	1.0000000000\\
0.2121212121	1.0000000000\\
0.2151515152	1.0000000000\\
0.2181818182	1.0000000000\\
0.2212121212	1.0000000000\\
0.2242424242	1.0000000000\\
0.2272727273	1.0000000000\\
0.2303030303	1.0000000000\\
0.2333333333	1.0000000000\\
0.2363636364	1.0000000000\\
0.2393939394	1.0000000000\\
0.2424242424	1.0000000000\\
0.2454545455	1.0000000000\\
0.2484848485	1.0000000000\\
0.2515151515	1.0000000000\\
0.2545454545	1.0000000000\\
0.2575757576	1.0000000000\\
0.2606060606	1.0000000000\\
0.2636363636	1.0000000000\\
0.2666666667	1.0000000000\\
0.2696969697	1.0000000000\\
0.2727272727	1.0000000000\\
0.2757575758	1.0000000000\\
0.2787878788	1.0000000000\\
0.2818181818	1.0000000000\\
0.2848484848	1.0000000000\\
0.2878787879	1.0000000000\\
0.2909090909	1.0000000000\\
0.2939393939	1.0000000000\\
0.2969696970	1.0000000000\\
0.3000000000	1.0000000000\\
};
\addlegendentry{$l=0$}

\addplot [color=mycolor2]
  table[row sep=crcr]{%
0.0000000000	1.0000000000\\
0.0030303030	1.0012794123\\
0.0060606061	1.0025666534\\
0.0090909091	1.0038617947\\
0.0121212121	1.0051649088\\
0.0151515152	1.0064760689\\
0.0181818182	1.0077953493\\
0.0212121212	1.0091228252\\
0.0242424242	1.0104585727\\
0.0272727273	1.0118026688\\
0.0303030303	1.0131551915\\
0.0333333333	1.0145162199\\
0.0363636364	1.0158858339\\
0.0393939394	1.0172641145\\
0.0424242424	1.0186511438\\
0.0454545455	1.0200470047\\
0.0484848485	1.0214517815\\
0.0515151515	1.0228655592\\
0.0545454545	1.0242884243\\
0.0575757576	1.0257204641\\
0.0606060606	1.0271617670\\
0.0636363636	1.0286124229\\
0.0666666667	1.0300725225\\
0.0696969697	1.0315421578\\
0.0727272727	1.0330214221\\
0.0757575758	1.0345104098\\
0.0787878788	1.0360092166\\
0.0818181818	1.0375179395\\
0.0848484848	1.0390366767\\
0.0878787879	1.0405655277\\
0.0909090909	1.0421045934\\
0.0939393939	1.0436539760\\
0.0969696970	1.0452137791\\
0.1000000000	1.0467841078\\
0.1030303030	1.0483650683\\
0.1060606061	1.0499567686\\
0.1090909091	1.0515593180\\
0.1121212121	1.0531728272\\
0.1151515152	1.0547974087\\
0.1181818182	1.0564331762\\
0.1212121212	1.0580802452\\
0.1242424242	1.0597387328\\
0.1272727273	1.0614087576\\
0.1303030303	1.0630904400\\
0.1333333333	1.0647839018\\
0.1363636364	1.0664892670\\
0.1393939394	1.0682066608\\
0.1424242424	1.0699362106\\
0.1454545455	1.0716780454\\
0.1484848485	1.0734322960\\
0.1515151515	1.0751990952\\
0.1545454545	1.0769785776\\
0.1575757576	1.0787708798\\
0.1606060606	1.0805761404\\
0.1636363636	1.0823944999\\
0.1666666667	1.0842261010\\
0.1696969697	1.0860710883\\
0.1727272727	1.0879296086\\
0.1757575758	1.0898018109\\
0.1787878788	1.0916878464\\
0.1818181818	1.0935878685\\
0.1848484848	1.0955020328\\
0.1878787879	1.0974304974\\
0.1909090909	1.0993734226\\
0.1939393939	1.1013309713\\
0.1969696970	1.1033033087\\
0.2000000000	1.1052906025\\
0.2030303030	1.1072930231\\
0.2060606061	1.1093107433\\
0.2090909091	1.1113439386\\
0.2121212121	1.1133927875\\
0.2151515152	1.1154574708\\
0.2181818182	1.1175381723\\
0.2212121212	1.1196350788\\
0.2242424242	1.1217483799\\
0.2272727273	1.1238782680\\
0.2303030303	1.1260249388\\
0.2333333333	1.1281885910\\
0.2363636364	1.1303694264\\
0.2393939394	1.1325676501\\
0.2424242424	1.1347834704\\
0.2454545455	1.1370170990\\
0.2484848485	1.1392687510\\
0.2515151515	1.1415386450\\
0.2545454545	1.1438270032\\
0.2575757576	1.1461340512\\
0.2606060606	1.1484600186\\
0.2636363636	1.1508051386\\
0.2666666667	1.1531696483\\
0.2696969697	1.1555537888\\
0.2727272727	1.1579578050\\
0.2757575758	1.1603819462\\
0.2787878788	1.1628264657\\
0.2818181818	1.1652916210\\
0.2848484848	1.1677776742\\
0.2878787879	1.1702848918\\
0.2909090909	1.1728135446\\
0.2939393939	1.1753639084\\
0.2969696970	1.1779362636\\
0.3000000000	1.1805308954\\
};
\addlegendentry{$l=1$}

\addplot [color=mycolor3]
  table[row sep=crcr]{%
0.0000000000	1.0000000000\\
0.0030303030	1.0019729183\\
0.0060606061	1.0039579088\\
0.0090909091	1.0059550819\\
0.0121212121	1.0079645494\\
0.0151515152	1.0099864243\\
0.0181818182	1.0120208212\\
0.0212121212	1.0140678558\\
0.0242424242	1.0161276457\\
0.0272727273	1.0182003096\\
0.0303030303	1.0202859678\\
0.0333333333	1.0223847421\\
0.0363636364	1.0244967559\\
0.0393939394	1.0266221340\\
0.0424242424	1.0287610029\\
0.0454545455	1.0309134908\\
0.0484848485	1.0330797273\\
0.0515151515	1.0352598438\\
0.0545454545	1.0374539733\\
0.0575757576	1.0396622507\\
0.0606060606	1.0418848125\\
0.0636363636	1.0441217969\\
0.0666666667	1.0463733440\\
0.0696969697	1.0486395956\\
0.0727272727	1.0509206957\\
0.0757575758	1.0532167897\\
0.0787878788	1.0555280253\\
0.0818181818	1.0578545520\\
0.0848484848	1.0601965211\\
0.0878787879	1.0625540864\\
0.0909090909	1.0649274032\\
0.0939393939	1.0673166292\\
0.0969696970	1.0697219242\\
0.1000000000	1.0721434500\\
0.1030303030	1.0745813708\\
0.1060606061	1.0770358528\\
0.1090909091	1.0795070646\\
0.1121212121	1.0819951771\\
0.1151515152	1.0845003635\\
0.1181818182	1.0870227994\\
0.1212121212	1.0895626628\\
0.1242424242	1.0921201342\\
0.1272727273	1.0946953965\\
0.1303030303	1.0972886354\\
0.1333333333	1.0999000389\\
0.1363636364	1.1025297978\\
0.1393939394	1.1051781056\\
0.1424242424	1.1078451585\\
0.1454545455	1.1105311554\\
0.1484848485	1.1132362981\\
0.1515151515	1.1159607914\\
0.1545454545	1.1187048429\\
0.1575757576	1.1214686632\\
0.1606060606	1.1242524660\\
0.1636363636	1.1270564679\\
0.1666666667	1.1298808890\\
0.1696969697	1.1327259523\\
0.1727272727	1.1355918842\\
0.1757575758	1.1384789145\\
0.1787878788	1.1413872762\\
0.1818181818	1.1443172059\\
0.1848484848	1.1472689437\\
0.1878787879	1.1502427332\\
0.1909090909	1.1532388217\\
0.1939393939	1.1562574604\\
0.1969696970	1.1592989039\\
0.2000000000	1.1623634111\\
0.2030303030	1.1654512444\\
0.2060606061	1.1685626705\\
0.2090909091	1.1716979602\\
0.2121212121	1.1748573883\\
0.2151515152	1.1780412339\\
0.2181818182	1.1812497805\\
0.2212121212	1.1844833159\\
0.2242424242	1.1877421325\\
0.2272727273	1.1910265272\\
0.2303030303	1.1943368017\\
0.2333333333	1.1976732623\\
0.2363636364	1.2010362203\\
0.2393939394	1.2044259920\\
0.2424242424	1.2078428985\\
0.2454545455	1.2112872663\\
0.2484848485	1.2147594271\\
0.2515151515	1.2182597179\\
0.2545454545	1.2217884814\\
0.2575757576	1.2253460657\\
0.2606060606	1.2289328245\\
0.2636363636	1.2325491177\\
0.2666666667	1.2361953107\\
0.2696969697	1.2398717754\\
0.2727272727	1.2435788894\\
0.2757575758	1.2473170372\\
0.2787878788	1.2510866093\\
0.2818181818	1.2548880030\\
0.2848484848	1.2587216223\\
0.2878787879	1.2625878779\\
0.2909090909	1.2664871880\\
0.2939393939	1.2704199774\\
0.2969696970	1.2743866786\\
0.3000000000	1.2783877314\\
};
\addlegendentry{$l=2$}

\end{axis}

\begin{axis}[%
width=1.226994\figurewidth,
height=1.226994\figureheight,
at={(-0.159509\figurewidth,-0.134969\figureheight)},
scale only axis,
xmin=0.0000000000,
xmax=1.0000000000,
ymin=0.0000000000,
ymax=1.0000000000,
axis line style={draw=none},
ticks=none,
axis x line*=bottom,
axis y line*=left,
scaled ticks=false, xticklabel style={/pgf/number format/fixed},yticklabel style={/pgf/number format/fixed}
]
\end{axis}
\end{tikzpicture}%
%}

	\caption{evolution of the wave number correction factor $\waveNumberMod^{\iterationStep}/\pi$}
	\label{fig:iterationEvolutionWaveNumberCorr}
\end{figure}
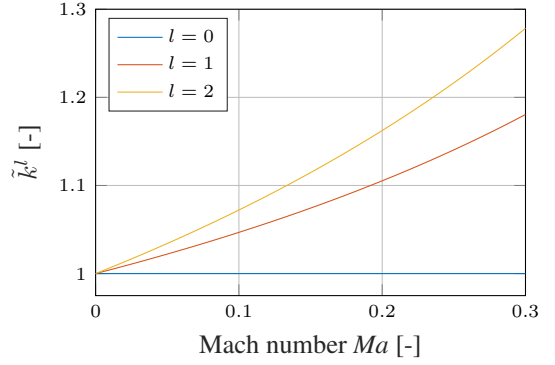

\section{Other Functions to be Learned by the Neural Networks}
\label{sec:appOtherFunctions}
\begin{figure}[ht!]
	\begin{subfigure}[t]{0.48\textwidth}
    	\setlength\figureheight{0.4\textwidth}
    	\setlength\figurewidth{0.5\textwidth}
    	\centering
        % This file was created by matlab2tikz.
%
\begin{tikzpicture}

\begin{axis}[%
width=\figurewidth,
height=\figureheight,
at={(0\figurewidth,0\figureheight)},
scale only axis,
point meta min=-1.0000000000,
point meta max=1.0000000000,
axis on top,
xmin=190.0000000000,
xmax=3010.0000000000,
xlabel style={font=\color{white!15!black}},
xlabel={frequency $\f$ [Hz]},
ymin=-0.0003571429,
ymax=0.1003571429,
xmajorgrids,
ymajorgrids,
ylabel style={font=\color{white!15!black}},
ylabel={Mach number $\Ma$ [-]},
axis background/.style={fill=white},
colormap={mymap}{[1pt] rgb(0pt)=(0.2422,0.1504,0.6603); rgb(1pt)=(0.2444,0.1534,0.6728); rgb(2pt)=(0.2464,0.1569,0.6847); rgb(3pt)=(0.2484,0.1607,0.6961); rgb(4pt)=(0.2503,0.1648,0.7071); rgb(5pt)=(0.2522,0.1689,0.7179); rgb(6pt)=(0.254,0.1732,0.7286); rgb(7pt)=(0.2558,0.1773,0.7393); rgb(8pt)=(0.2576,0.1814,0.7501); rgb(9pt)=(0.2594,0.1854,0.761); rgb(11pt)=(0.2628,0.1932,0.7828); rgb(12pt)=(0.2645,0.1972,0.7937); rgb(13pt)=(0.2661,0.2011,0.8043); rgb(14pt)=(0.2676,0.2052,0.8148); rgb(15pt)=(0.2691,0.2094,0.8249); rgb(16pt)=(0.2704,0.2138,0.8346); rgb(17pt)=(0.2717,0.2184,0.8439); rgb(18pt)=(0.2729,0.2231,0.8528); rgb(19pt)=(0.274,0.228,0.8612); rgb(20pt)=(0.2749,0.233,0.8692); rgb(21pt)=(0.2758,0.2382,0.8767); rgb(22pt)=(0.2766,0.2435,0.884); rgb(23pt)=(0.2774,0.2489,0.8908); rgb(24pt)=(0.2781,0.2543,0.8973); rgb(25pt)=(0.2788,0.2598,0.9035); rgb(26pt)=(0.2794,0.2653,0.9094); rgb(27pt)=(0.2798,0.2708,0.915); rgb(28pt)=(0.2802,0.2764,0.9204); rgb(29pt)=(0.2806,0.2819,0.9255); rgb(30pt)=(0.2809,0.2875,0.9305); rgb(31pt)=(0.2811,0.293,0.9352); rgb(32pt)=(0.2813,0.2985,0.9397); rgb(33pt)=(0.2814,0.304,0.9441); rgb(34pt)=(0.2814,0.3095,0.9483); rgb(35pt)=(0.2813,0.315,0.9524); rgb(36pt)=(0.2811,0.3204,0.9563); rgb(37pt)=(0.2809,0.3259,0.96); rgb(38pt)=(0.2807,0.3313,0.9636); rgb(39pt)=(0.2803,0.3367,0.967); rgb(40pt)=(0.2798,0.3421,0.9702); rgb(41pt)=(0.2791,0.3475,0.9733); rgb(42pt)=(0.2784,0.3529,0.9763); rgb(43pt)=(0.2776,0.3583,0.9791); rgb(44pt)=(0.2766,0.3638,0.9817); rgb(45pt)=(0.2754,0.3693,0.984); rgb(46pt)=(0.2741,0.3748,0.9862); rgb(47pt)=(0.2726,0.3804,0.9881); rgb(48pt)=(0.271,0.386,0.9898); rgb(49pt)=(0.2691,0.3916,0.9912); rgb(50pt)=(0.267,0.3973,0.9924); rgb(51pt)=(0.2647,0.403,0.9935); rgb(52pt)=(0.2621,0.4088,0.9946); rgb(53pt)=(0.2591,0.4145,0.9955); rgb(54pt)=(0.2556,0.4203,0.9965); rgb(55pt)=(0.2517,0.4261,0.9974); rgb(56pt)=(0.2473,0.4319,0.9983); rgb(57pt)=(0.2424,0.4378,0.9991); rgb(58pt)=(0.2369,0.4437,0.9996); rgb(59pt)=(0.2311,0.4497,0.9995); rgb(60pt)=(0.225,0.4559,0.9985); rgb(61pt)=(0.2189,0.462,0.9968); rgb(62pt)=(0.2128,0.4682,0.9948); rgb(63pt)=(0.2066,0.4743,0.9926); rgb(64pt)=(0.2006,0.4803,0.9906); rgb(65pt)=(0.195,0.4861,0.9887); rgb(66pt)=(0.1903,0.4919,0.9867); rgb(67pt)=(0.1869,0.4975,0.9844); rgb(68pt)=(0.1847,0.503,0.9819); rgb(69pt)=(0.1831,0.5084,0.9793); rgb(70pt)=(0.1818,0.5138,0.9766); rgb(71pt)=(0.1806,0.5191,0.9738); rgb(72pt)=(0.1795,0.5244,0.9709); rgb(73pt)=(0.1785,0.5296,0.9677); rgb(74pt)=(0.1778,0.5349,0.9641); rgb(75pt)=(0.1773,0.5401,0.9602); rgb(76pt)=(0.1768,0.5452,0.956); rgb(77pt)=(0.1764,0.5504,0.9516); rgb(78pt)=(0.1755,0.5554,0.9473); rgb(79pt)=(0.174,0.5605,0.9432); rgb(80pt)=(0.1716,0.5655,0.9393); rgb(81pt)=(0.1686,0.5705,0.9357); rgb(82pt)=(0.1649,0.5755,0.9323); rgb(83pt)=(0.161,0.5805,0.9289); rgb(84pt)=(0.1573,0.5854,0.9254); rgb(85pt)=(0.154,0.5902,0.9218); rgb(86pt)=(0.1513,0.595,0.9182); rgb(87pt)=(0.1492,0.5997,0.9147); rgb(88pt)=(0.1475,0.6043,0.9113); rgb(89pt)=(0.1461,0.6089,0.908); rgb(90pt)=(0.1446,0.6135,0.905); rgb(91pt)=(0.1429,0.618,0.9022); rgb(92pt)=(0.1408,0.6226,0.8998); rgb(93pt)=(0.1383,0.6272,0.8975); rgb(94pt)=(0.1354,0.6317,0.8953); rgb(95pt)=(0.1321,0.6363,0.8932); rgb(96pt)=(0.1288,0.6408,0.891); rgb(97pt)=(0.1253,0.6453,0.8887); rgb(98pt)=(0.1219,0.6497,0.8862); rgb(99pt)=(0.1185,0.6541,0.8834); rgb(100pt)=(0.1152,0.6584,0.8804); rgb(101pt)=(0.1119,0.6627,0.877); rgb(102pt)=(0.1085,0.6669,0.8734); rgb(103pt)=(0.1048,0.671,0.8695); rgb(104pt)=(0.1009,0.675,0.8653); rgb(105pt)=(0.0964,0.6789,0.8609); rgb(106pt)=(0.0914,0.6828,0.8562); rgb(107pt)=(0.0855,0.6865,0.8513); rgb(108pt)=(0.0789,0.6902,0.8462); rgb(109pt)=(0.0713,0.6938,0.8409); rgb(110pt)=(0.0628,0.6972,0.8355); rgb(111pt)=(0.0535,0.7006,0.8299); rgb(112pt)=(0.0433,0.7039,0.8242); rgb(113pt)=(0.0328,0.7071,0.8183); rgb(114pt)=(0.0234,0.7103,0.8124); rgb(115pt)=(0.0155,0.7133,0.8064); rgb(116pt)=(0.0091,0.7163,0.8003); rgb(117pt)=(0.0046,0.7192,0.7941); rgb(118pt)=(0.0019,0.722,0.7878); rgb(119pt)=(0.0009,0.7248,0.7815); rgb(120pt)=(0.0018,0.7275,0.7752); rgb(121pt)=(0.0046,0.7301,0.7688); rgb(122pt)=(0.0094,0.7327,0.7623); rgb(123pt)=(0.0162,0.7352,0.7558); rgb(124pt)=(0.0253,0.7376,0.7492); rgb(125pt)=(0.0369,0.74,0.7426); rgb(126pt)=(0.0504,0.7423,0.7359); rgb(127pt)=(0.0638,0.7446,0.7292); rgb(128pt)=(0.077,0.7468,0.7224); rgb(129pt)=(0.0899,0.7489,0.7156); rgb(130pt)=(0.1023,0.751,0.7088); rgb(131pt)=(0.1141,0.7531,0.7019); rgb(132pt)=(0.1252,0.7552,0.695); rgb(133pt)=(0.1354,0.7572,0.6881); rgb(134pt)=(0.1448,0.7593,0.6812); rgb(135pt)=(0.1532,0.7614,0.6741); rgb(136pt)=(0.1609,0.7635,0.6671); rgb(137pt)=(0.1678,0.7656,0.6599); rgb(138pt)=(0.1741,0.7678,0.6527); rgb(139pt)=(0.1799,0.7699,0.6454); rgb(140pt)=(0.1853,0.7721,0.6379); rgb(141pt)=(0.1905,0.7743,0.6303); rgb(142pt)=(0.1954,0.7765,0.6225); rgb(143pt)=(0.2003,0.7787,0.6146); rgb(144pt)=(0.2061,0.7808,0.6065); rgb(145pt)=(0.2118,0.7828,0.5983); rgb(146pt)=(0.2178,0.7849,0.5899); rgb(147pt)=(0.2244,0.7869,0.5813); rgb(148pt)=(0.2318,0.7887,0.5725); rgb(149pt)=(0.2401,0.7905,0.5636); rgb(150pt)=(0.2491,0.7922,0.5546); rgb(151pt)=(0.2589,0.7937,0.5454); rgb(152pt)=(0.2695,0.7951,0.536); rgb(153pt)=(0.2809,0.7964,0.5266); rgb(154pt)=(0.2929,0.7975,0.517); rgb(155pt)=(0.3052,0.7985,0.5074); rgb(156pt)=(0.3176,0.7994,0.4975); rgb(157pt)=(0.3301,0.8002,0.4876); rgb(158pt)=(0.3424,0.8009,0.4774); rgb(159pt)=(0.3548,0.8016,0.4669); rgb(160pt)=(0.3671,0.8021,0.4563); rgb(161pt)=(0.3795,0.8026,0.4454); rgb(162pt)=(0.3921,0.8029,0.4344); rgb(163pt)=(0.405,0.8031,0.4233); rgb(164pt)=(0.4184,0.803,0.4122); rgb(165pt)=(0.4322,0.8028,0.4013); rgb(166pt)=(0.4463,0.8024,0.3904); rgb(167pt)=(0.4608,0.8018,0.3797); rgb(168pt)=(0.4753,0.8011,0.3691); rgb(169pt)=(0.4899,0.8002,0.3586); rgb(170pt)=(0.5044,0.7993,0.348); rgb(171pt)=(0.5187,0.7982,0.3374); rgb(172pt)=(0.5329,0.797,0.3267); rgb(173pt)=(0.547,0.7957,0.3159); rgb(175pt)=(0.5748,0.7929,0.2941); rgb(176pt)=(0.5886,0.7913,0.2833); rgb(177pt)=(0.6024,0.7896,0.2726); rgb(178pt)=(0.6161,0.7878,0.2622); rgb(179pt)=(0.6297,0.7859,0.2521); rgb(180pt)=(0.6433,0.7839,0.2423); rgb(181pt)=(0.6567,0.7818,0.2329); rgb(182pt)=(0.6701,0.7796,0.2239); rgb(183pt)=(0.6833,0.7773,0.2155); rgb(184pt)=(0.6963,0.775,0.2075); rgb(185pt)=(0.7091,0.7727,0.1998); rgb(186pt)=(0.7218,0.7703,0.1924); rgb(187pt)=(0.7344,0.7679,0.1852); rgb(188pt)=(0.7468,0.7654,0.1782); rgb(189pt)=(0.759,0.7629,0.1717); rgb(190pt)=(0.771,0.7604,0.1658); rgb(191pt)=(0.7829,0.7579,0.1608); rgb(192pt)=(0.7945,0.7554,0.157); rgb(193pt)=(0.806,0.7529,0.1546); rgb(194pt)=(0.8172,0.7505,0.1535); rgb(195pt)=(0.8281,0.7481,0.1536); rgb(196pt)=(0.8389,0.7457,0.1546); rgb(197pt)=(0.8495,0.7435,0.1564); rgb(198pt)=(0.86,0.7413,0.1587); rgb(199pt)=(0.8703,0.7392,0.1615); rgb(200pt)=(0.8804,0.7372,0.165); rgb(201pt)=(0.8903,0.7353,0.1695); rgb(202pt)=(0.9,0.7336,0.1749); rgb(203pt)=(0.9093,0.7321,0.1815); rgb(204pt)=(0.9184,0.7308,0.189); rgb(205pt)=(0.9272,0.7298,0.1973); rgb(206pt)=(0.9357,0.729,0.2061); rgb(207pt)=(0.944,0.7285,0.2151); rgb(208pt)=(0.9523,0.7284,0.2237); rgb(209pt)=(0.9606,0.7285,0.2312); rgb(210pt)=(0.9689,0.7292,0.2373); rgb(211pt)=(0.977,0.7304,0.2418); rgb(212pt)=(0.9842,0.733,0.2446); rgb(213pt)=(0.99,0.7365,0.2429); rgb(214pt)=(0.9946,0.7407,0.2394); rgb(215pt)=(0.9966,0.7458,0.2351); rgb(216pt)=(0.9971,0.7513,0.2309); rgb(217pt)=(0.9972,0.7569,0.2267); rgb(218pt)=(0.9971,0.7626,0.2224); rgb(219pt)=(0.9969,0.7683,0.2181); rgb(220pt)=(0.9966,0.774,0.2138); rgb(221pt)=(0.9962,0.7798,0.2095); rgb(222pt)=(0.9957,0.7856,0.2053); rgb(223pt)=(0.9949,0.7915,0.2012); rgb(224pt)=(0.9938,0.7974,0.1974); rgb(225pt)=(0.9923,0.8034,0.1939); rgb(226pt)=(0.9906,0.8095,0.1906); rgb(227pt)=(0.9885,0.8156,0.1875); rgb(228pt)=(0.9861,0.8218,0.1846); rgb(229pt)=(0.9835,0.828,0.1817); rgb(230pt)=(0.9807,0.8342,0.1787); rgb(231pt)=(0.9778,0.8404,0.1757); rgb(232pt)=(0.9748,0.8467,0.1726); rgb(233pt)=(0.972,0.8529,0.1695); rgb(234pt)=(0.9694,0.8591,0.1665); rgb(235pt)=(0.9671,0.8654,0.1636); rgb(236pt)=(0.9651,0.8716,0.1608); rgb(237pt)=(0.9634,0.8778,0.1582); rgb(238pt)=(0.9619,0.884,0.1557); rgb(239pt)=(0.9608,0.8902,0.1532); rgb(240pt)=(0.9601,0.8963,0.1507); rgb(241pt)=(0.9596,0.9023,0.148); rgb(242pt)=(0.9595,0.9084,0.145); rgb(243pt)=(0.9597,0.9143,0.1418); rgb(244pt)=(0.9601,0.9203,0.1382); rgb(245pt)=(0.9608,0.9262,0.1344); rgb(246pt)=(0.9618,0.932,0.1304); rgb(247pt)=(0.9629,0.9379,0.1261); rgb(248pt)=(0.9642,0.9437,0.1216); rgb(249pt)=(0.9657,0.9494,0.1168); rgb(250pt)=(0.9674,0.9552,0.1116); rgb(251pt)=(0.9692,0.9609,0.1061); rgb(252pt)=(0.9711,0.9667,0.1001); rgb(253pt)=(0.973,0.9724,0.0938); rgb(254pt)=(0.9749,0.9782,0.0872); rgb(255pt)=(0.9769,0.9839,0.0805)},
colorbar,
colorbar style={ylabel style={font=\color{white!15!black}}}
]
\addplot [forget plot] graphics [xmin=190.0000000000, xmax=3010.0000000000, ymin=-0.0003571429, ymax=0.1003571429] {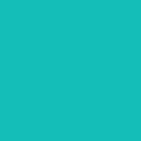};
\end{axis}
\end{tikzpicture}%
        \caption{phase modulation $\phaseMod_{11}$ [-]}
        \label{fig:app_phaseModulation}
    \end{subfigure}
~
	\begin{subfigure}[t]{0.48\textwidth}
    	\setlength\figureheight{0.4\textwidth}
    	\setlength\figurewidth{0.5\textwidth}
    	\centering
        % This file was created by matlab2tikz.
%
\definecolor{mycolor1}{rgb}{0.00000,0.44700,0.74100}%
\begin{tikzpicture}

\begin{axis}[%
width=\figurewidth,
height=\figureheight,
at={(0\figurewidth,0\figureheight)},
scale only axis,
xmin=0.0000000000,
xmax=0.1000000000,
xlabel style={font=\color{white!15!black}},
xlabel={Mach number $\Ma$ [-]},
ymin=1.0000000000,
xmajorgrids,
ymajorgrids,
ymax=1.1200000000,
ylabel style={font=\color{white!15!black}},
ylabel={wave number modulation $\waveNumberMod_{11}$ [-]},
axis background/.style={fill=white},
scaled ticks=false, xticklabel style={/pgf/number format/fixed},yticklabel style={/pgf/number format/fixed}
]
\addplot [color=mycolor1, forget plot]
  table[row sep=crcr]{%
0.0000000000	1.0000000000\\
0.0007142857	1.0007147963\\
0.0014285714	1.0014306151\\
0.0021428571	1.0021474588\\
0.0028571429	1.0028653294\\
0.0035714286	1.0035842293\\
0.0042857143	1.0043041605\\
0.0050000000	1.0050251254\\
0.0057142857	1.0057471262\\
0.0064285714	1.0064701651\\
0.0071428571	1.0071942443\\
0.0078571429	1.0079193662\\
0.0085714286	1.0086455328\\
0.0092857143	1.0093727466\\
0.0100000000	1.0101010097\\
0.0107142857	1.0108303245\\
0.0114285714	1.0115606932\\
0.0121428571	1.0122921181\\
0.0128571429	1.0130246016\\
0.0135714286	1.0137581458\\
0.0142857143	1.0144927531\\
0.0150000000	1.0152284259\\
0.0157142857	1.0159651663\\
0.0164285714	1.0167029769\\
0.0171428571	1.0174418598\\
0.0178571429	1.0181818175\\
0.0185714286	1.0189228523\\
0.0192857143	1.0196649665\\
0.0200000000	1.0204081625\\
0.0207142857	1.0211524427\\
0.0214285714	1.0218978094\\
0.0221428571	1.0226442651\\
0.0228571429	1.0233918120\\
0.0235714286	1.0241404527\\
0.0242857143	1.0248901894\\
0.0250000000	1.0256410247\\
0.0257142857	1.0263929609\\
0.0264285714	1.0271460005\\
0.0271428571	1.0279001458\\
0.0278571429	1.0286553994\\
0.0285714286	1.0294117637\\
0.0292857143	1.0301692410\\
0.0300000000	1.0309278339\\
0.0307142857	1.0316875449\\
0.0314285714	1.0324483764\\
0.0321428571	1.0332103309\\
0.0328571429	1.0339734109\\
0.0335714286	1.0347376189\\
0.0342857143	1.0355029573\\
0.0350000000	1.0362694288\\
0.0357142857	1.0370370357\\
0.0364285714	1.0378057807\\
0.0371428571	1.0385756663\\
0.0378571429	1.0393466950\\
0.0385714286	1.0401188693\\
0.0392857143	1.0408921918\\
0.0400000000	1.0416666652\\
0.0407142857	1.0424422919\\
0.0414285714	1.0432190745\\
0.0421428571	1.0439970156\\
0.0428571429	1.0447761178\\
0.0435714286	1.0455563837\\
0.0442857143	1.0463378160\\
0.0450000000	1.0471204172\\
0.0457142857	1.0479041899\\
0.0464285714	1.0486891368\\
0.0471428571	1.0494752606\\
0.0478571429	1.0502625638\\
0.0485714286	1.0510510492\\
0.0492857143	1.0518407194\\
0.0500000000	1.0526315771\\
0.0507142857	1.0534236249\\
0.0514285714	1.0542168655\\
0.0521428571	1.0550113017\\
0.0528571429	1.0558069362\\
0.0535714286	1.0566037716\\
0.0542857143	1.0574018106\\
0.0550000000	1.0582010561\\
0.0557142857	1.0590015107\\
0.0564285714	1.0598031773\\
0.0571428571	1.0606060584\\
0.0578571429	1.0614101570\\
0.0585714286	1.0622154758\\
0.0592857143	1.0630220175\\
0.0600000000	1.0638297850\\
0.0607142857	1.0646387810\\
0.0614285714	1.0654490083\\
0.0621428571	1.0662604698\\
0.0628571429	1.0670731683\\
0.0635714286	1.0678871067\\
0.0642857143	1.0687022876\\
0.0650000000	1.0695187141\\
0.0657142857	1.0703363889\\
0.0664285714	1.0711553150\\
0.0671428571	1.0719754951\\
0.0678571429	1.0727969323\\
0.0685714286	1.0736196293\\
0.0692857143	1.0744435891\\
0.0700000000	1.0752688145\\
0.0707142857	1.0760953086\\
0.0714285714	1.0769230742\\
0.0721428571	1.0777521142\\
0.0728571429	1.0785824317\\
0.0735714286	1.0794140295\\
0.0742857143	1.0802469107\\
0.0750000000	1.0810810782\\
0.0757142857	1.0819165349\\
0.0764285714	1.0827532840\\
0.0771428571	1.0835913283\\
0.0778571429	1.0844306709\\
0.0785714286	1.0852713148\\
0.0792857143	1.0861132630\\
0.0800000000	1.0869565186\\
0.0807142857	1.0878010847\\
0.0814285714	1.0886469642\\
0.0821428571	1.0894941602\\
0.0828571429	1.0903426759\\
0.0835714286	1.0911925143\\
0.0842857143	1.0920436785\\
0.0850000000	1.0928961715\\
0.0857142857	1.0937499966\\
0.0864285714	1.0946051569\\
0.0871428571	1.0954616554\\
0.0878571429	1.0963194954\\
0.0885714286	1.0971786799\\
0.0892857143	1.0980392122\\
0.0900000000	1.0989010954\\
0.0907142857	1.0997643326\\
0.0914285714	1.1006289272\\
0.0921428571	1.1014948823\\
0.0928571429	1.1023622011\\
0.0935714286	1.1032308868\\
0.0942857143	1.1041009427\\
0.0950000000	1.1049723719\\
0.0957142857	1.1058451779\\
0.0964285714	1.1067193638\\
0.0971428571	1.1075949329\\
0.0978571429	1.1084718884\\
0.0985714286	1.1093502338\\
0.0992857143	1.1102299723\\
0.1000000000	1.1111111071\\
};
\end{axis}
\end{tikzpicture}%
        \caption{wave number modulation $\waveNumberMod_{11}$ [-]}
        \label{fig:app_waveNumberModulation}
    \end{subfigure}
    \caption{Two remaining of the three functions to be learned for microphone 1 and speaker 1 (see \Cref{fig:benchWind_geometry} for location).}
    \label{fig:app_functionsToLearn}
\end{figure}
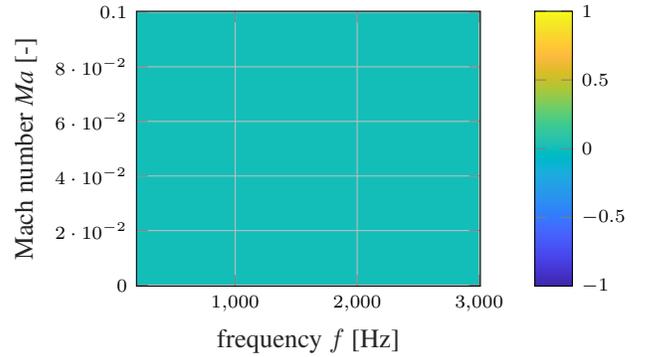
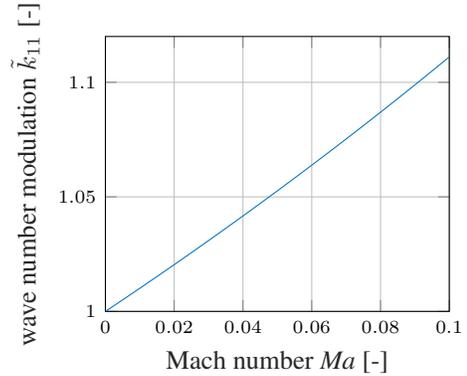

The functions are displayed in \Cref{fig:app_functionsToLearn}.
%%%%%%%%%%%%%%%%%%%%%%%%%%%%%%%%%%%%%%%%%%%%%%%%%%%%%%%%%%%%%%%%%%%%%%%%%%%%%%%%%%%%%%%%%%%%%%%%
\section{Resonance of the Microphone Array}
\label{sec:appResonance}
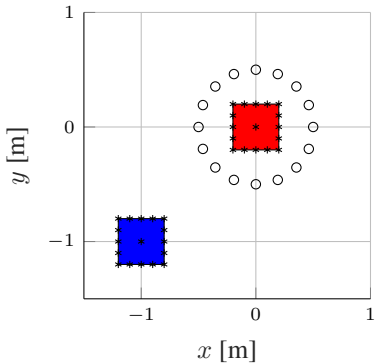
\begin{figure}
	\setlength\figureheight{0.2\textwidth}
	\setlength\figurewidth{0.2\textwidth}
	\centering
    % This file was created by matlab2tikz.
%
\begin{tikzpicture}

\begin{axis}[%
width=\figurewidth,
height=\figureheight,
at={(0\figurewidth,0\figureheight)},
scale only axis,
xmin=-1.5000000000,
xmax=1.0000000000,
xlabel style={font=\color{white!15!black}},
xlabel={$x$ [m]},
ymin=-1.5000000000,
ymax=1.0000000000,
ylabel style={font=\color{white!15!black}},
ylabel={$y$ [m]},
axis background/.style={fill=white},
axis x line*=bottom,
axis y line*=left,
xmajorgrids,
ymajorgrids,
scaled ticks=false, xticklabel style={/pgf/number format/fixed},yticklabel style={/pgf/number format/fixed}
]
\addplot[only marks, mark=o, mark options={}, mark size=1.7678pt, draw=black, forget plot] table[row sep=crcr]{%
x	y\\
0.5000000000	0.0000000000\\
0.4619397663	0.1913417162\\
0.3535533906	0.3535533906\\
0.1913417162	0.4619397663\\
0.0000000000	0.5000000000\\
-0.1913417162	0.4619397663\\
-0.3535533906	0.3535533906\\
-0.4619397663	0.1913417162\\
-0.5000000000	0.0000000000\\
-0.4619397663	-0.1913417162\\
-0.3535533906	-0.3535533906\\
-0.1913417162	-0.4619397663\\
-0.0000000000	-0.5000000000\\
0.1913417162	-0.4619397663\\
0.3535533906	-0.3535533906\\
0.4619397663	-0.1913417162\\
};
\addplot[only marks, mark=asterisk, mark options={}, mark size=1.3693pt, draw=black, forget plot] table[row sep=crcr]{%
x	y\\
0.0000000000	0.0000000000\\
0.2000000000	-0.2000000000\\
0.2000000000	-0.1000000000\\
0.2000000000	0.0000000000\\
0.2000000000	0.1000000000\\
0.2000000000	0.2000000000\\
0.1000000000	0.2000000000\\
-0.0000000000	0.2000000000\\
-0.1000000000	0.2000000000\\
-0.2000000000	0.2000000000\\
-0.2000000000	0.1000000000\\
-0.2000000000	-0.0000000000\\
-0.2000000000	-0.1000000000\\
-0.2000000000	-0.2000000000\\
-0.1000000000	-0.2000000000\\
0.0000000000	-0.2000000000\\
0.1000000000	-0.2000000000\\
};
\addplot[only marks, mark=asterisk, mark options={}, mark size=1.3693pt, draw=black, forget plot] table[row sep=crcr]{%
x	y\\
-1.0000000000	-1.0000000000\\
-0.8000000000	-1.2000000000\\
-0.8000000000	-1.1000000000\\
-0.8000000000	-1.0000000000\\
-0.8000000000	-0.9000000000\\
-0.8000000000	-0.8000000000\\
-0.9000000000	-0.8000000000\\
-1.0000000000	-0.8000000000\\
-1.1000000000	-0.8000000000\\
-1.2000000000	-0.8000000000\\
-1.2000000000	-0.9000000000\\
-1.2000000000	-1.0000000000\\
-1.2000000000	-1.1000000000\\
-1.2000000000	-1.2000000000\\
-1.1000000000	-1.2000000000\\
-1.0000000000	-1.2000000000\\
-0.9000000000	-1.2000000000\\
};

\addplot[area legend, draw=black, fill=red, forget plot]
table[row sep=crcr] {%
x	y\\
0.2000000000	-0.2000000000\\
0.2000000000	-0.1000000000\\
0.2000000000	0.0000000000\\
0.2000000000	0.1000000000\\
0.2000000000	0.2000000000\\
0.1000000000	0.2000000000\\
-0.0000000000	0.2000000000\\
-0.1000000000	0.2000000000\\
-0.2000000000	0.2000000000\\
-0.2000000000	0.1000000000\\
-0.2000000000	-0.0000000000\\
-0.2000000000	-0.1000000000\\
-0.2000000000	-0.2000000000\\
-0.1000000000	-0.2000000000\\
0.0000000000	-0.2000000000\\
0.1000000000	-0.2000000000\\
}--cycle;

\addplot[area legend, draw=black, fill=blue, forget plot]
table[row sep=crcr] {%
x	y\\
-0.8000000000	-1.2000000000\\
-0.8000000000	-1.1000000000\\
-0.8000000000	-1.0000000000\\
-0.8000000000	-0.9000000000\\
-0.8000000000	-0.8000000000\\
-0.9000000000	-0.8000000000\\
-1.0000000000	-0.8000000000\\
-1.1000000000	-0.8000000000\\
-1.2000000000	-0.8000000000\\
-1.2000000000	-0.9000000000\\
-1.2000000000	-1.0000000000\\
-1.2000000000	-1.1000000000\\
-1.2000000000	-1.2000000000\\
-1.1000000000	-1.2000000000\\
-1.0000000000	-1.2000000000\\
-0.9000000000	-1.2000000000\\
}--cycle;
\end{axis}
\end{tikzpicture}%
%
	\caption{Simulation setup of the additional setup. The square side length is $0.4\unit{m}$ instead of $0.3\unit{m}$. The red square is the bright zone and the blue square is the dark zone. Circles ($\circ$) are indicators for speakers, asterisks ($*$) for microphones. Again the setups with a microphone in the middle of the squares and without are compared.}
	\label{fig:app_biggerSquareGeometry}
\end{figure}

To further examine the resonance in the bright zone another setup with a bigger square side length of $0.4\unit{m}$ (square side length of the original setup is $0.3\unit{m}$) is considered and visualised in \Cref{fig:app_biggerSquareGeometry}. The training data is created using Green's function of the wave equation in 3D.

\Cref{fig:app_biggerSquareRE} shows the RE over the previously considered frequency interval $[200,1000]\unit{Hz}$. The resonance that can be seen in \Cref{fig:benchNoWind_noMicMiddleEnergyLevel} and occurs at the frequency $810\unit{Hz}$ for the original setup occurs in the examined setup at $600\unit{Hz}$. This resonance can again be inhibited by introducing the microphone in the middle of the microphone array. A second resonance is visible in \Cref{fig:app_biggerSquareSolution} and occurs at approx. $940\unit{Hz}$. For the original setup this resonance was not inside the considered frequency interval. The placement of the microphone in the middle of the bright zone does not lead to the inhibition of the second resonance since the antinodes of this resonance are not located in the middle of the square.

\begin{figure}
	\centering
	\setlength\figureheight{0.28\textwidth}
	\setlength\figurewidth{0.32\textwidth}
	%\resizebox{\linewidth}{!}{
    % This file was created by matlab2tikz.
%
\begin{tikzpicture}

\begin{axis}[%
width=\figurewidth,
height=\figureheight,
at={(0\figurewidth,0\figureheight)},
scale only axis,
xmin=200.0000000000,
xmax=1000.0000000000,
xlabel style={font=\color{white!15!black}},
xlabel={frequency $f$ [Hz]},
ymin=-30.0000000000,
ymax=30.0000000000,
ylabel style={font=\color{white!15!black}},
ylabel={reproduction error RE [dB]},
axis background/.style={fill=white},
xmajorgrids,
ymajorgrids,
legend style={at={(0.03,0.97)}, anchor=north west, legend cell align=left, align=left, draw=white!15!black},
scaled ticks=false, xticklabel style={/pgf/number format/fixed},yticklabel style={/pgf/number format/fixed}
]
\addplot [color=red, mark=o, mark options={solid, red}]
  table[row sep=crcr]{%
200.0000000000	-20.3097052145\\
220.0000000000	-20.4266731802\\
240.0000000000	-20.4593141786\\
260.0000000000	-20.4008131714\\
280.0000000000	-20.2485075794\\
300.0000000000	-20.0090237653\\
320.0000000000	-19.6963875806\\
340.0000000000	-19.3266720813\\
360.0000000000	-18.9146756248\\
380.0000000000	-18.4744426611\\
400.0000000000	-18.0221719327\\
420.0000000000	-17.5797217363\\
440.0000000000	-17.1773528238\\
460.0000000000	-16.8544098680\\
480.0000000000	-16.6563244518\\
500.0000000000	-16.6245617129\\
520.0000000000	-16.7700432653\\
540.0000000000	-17.0130572820\\
560.0000000000	-17.1082753116\\
580.0000000000	-16.7326207510\\
600.0000000000	-15.8525488682\\
620.0000000000	-14.7992985921\\
640.0000000000	-13.8847238398\\
660.0000000000	-13.2171742102\\
680.0000000000	-12.7669717009\\
700.0000000000	-12.4522717622\\
720.0000000000	-12.1855923011\\
740.0000000000	-11.8932618939\\
760.0000000000	-11.5209594596\\
780.0000000000	-11.0316290600\\
800.0000000000	-10.3983529265\\
820.0000000000	-9.5930786373\\
840.0000000000	-8.5705458778\\
860.0000000000	-7.2433067693\\
880.0000000000	-5.4337551100\\
900.0000000000	-2.7462748013\\
920.0000000000	2.1812668818\\
940.0000000000	8.6311284964\\
960.0000000000	11.7752465964\\
980.0000000000	8.3220375788\\
1000.0000000000	12.6343816911\\
};
\addlegendentry{mic mid, $\lambda=0$}

\addplot [color=red, dashdotted, mark=x, mark options={solid, red}]
  table[row sep=crcr]{%
200.0000000000	-19.1953507257\\
220.0000000000	-19.2571074939\\
240.0000000000	-19.2941590625\\
260.0000000000	-19.3031909538\\
280.0000000000	-19.2770444594\\
300.0000000000	-19.2030856900\\
320.0000000000	-19.0649522381\\
340.0000000000	-18.8487415729\\
360.0000000000	-18.5507690247\\
380.0000000000	-18.1822423190\\
400.0000000000	-17.7686633468\\
420.0000000000	-17.3459560653\\
440.0000000000	-16.9568145660\\
460.0000000000	-16.6488870131\\
480.0000000000	-16.4731914384\\
500.0000000000	-16.4770101579\\
520.0000000000	-16.6781701357\\
540.0000000000	-16.9989074299\\
560.0000000000	-17.1717025621\\
580.0000000000	-16.8185447934\\
600.0000000000	-15.8838358557\\
620.0000000000	-14.7441302499\\
640.0000000000	-13.7534773496\\
660.0000000000	-13.0349057465\\
680.0000000000	-12.5614138554\\
700.0000000000	-12.2503727833\\
720.0000000000	-12.0111390455\\
740.0000000000	-11.7640326573\\
760.0000000000	-11.4464016342\\
780.0000000000	-11.0121853426\\
800.0000000000	-10.4266763777\\
820.0000000000	-9.6565617805\\
840.0000000000	-8.6543325208\\
860.0000000000	-7.3334228641\\
880.0000000000	-5.5215563127\\
900.0000000000	-2.8406616131\\
920.0000000000	1.9704381280\\
940.0000000000	10.5689698240\\
960.0000000000	14.7771418738\\
980.0000000000	7.7507608806\\
1000.0000000000	15.2032045779\\
};
\addlegendentry{mic mid, $\lambda=0.1$}

\addplot [color=blue, mark=o, mark options={solid, blue}]
  table[row sep=crcr]{%
200.0000000000	-19.9651037388\\
220.0000000000	-19.9433357191\\
240.0000000000	-19.8025334700\\
260.0000000000	-19.5426086247\\
280.0000000000	-19.1694069475\\
300.0000000000	-18.6935857974\\
320.0000000000	-18.1259893770\\
340.0000000000	-17.4727141886\\
360.0000000000	-16.7346308693\\
380.0000000000	-15.9111278339\\
400.0000000000	-15.0031765253\\
420.0000000000	-14.0129280320\\
440.0000000000	-12.9412543349\\
460.0000000000	-11.7850873406\\
480.0000000000	-10.5348210598\\
500.0000000000	-9.1717259120\\
520.0000000000	-7.6656646100\\
540.0000000000	-5.9740508078\\
560.0000000000	-4.0509769004\\
580.0000000000	-1.9165974412\\
600.0000000000	0.0645842325\\
620.0000000000	0.6387868167\\
640.0000000000	-0.8486706563\\
660.0000000000	-3.0152290030\\
680.0000000000	-4.9957306365\\
700.0000000000	-6.6158555799\\
720.0000000000	-7.8694396972\\
740.0000000000	-8.7702786483\\
760.0000000000	-9.3278040428\\
780.0000000000	-9.5472628045\\
800.0000000000	-9.4320383203\\
820.0000000000	-8.9810167936\\
840.0000000000	-8.1765046558\\
860.0000000000	-6.9585384619\\
880.0000000000	-5.1736067774\\
900.0000000000	-2.4378026988\\
920.0000000000	2.7472095504\\
940.0000000000	8.9731343831\\
960.0000000000	9.8705027637\\
980.0000000000	8.1475331801\\
1000.0000000000	19.4640964462\\
};
\addlegendentry{no mic mid, $\lambda=0$}

\addplot [color=blue, dashdotted, mark=x, mark options={solid, blue}]
  table[row sep=crcr]{%
200.0000000000	-18.5305485147\\
220.0000000000	-18.4521150989\\
240.0000000000	-18.3488102319\\
260.0000000000	-18.2171251642\\
280.0000000000	-18.0430451974\\
300.0000000000	-17.8019859539\\
320.0000000000	-17.4645486653\\
340.0000000000	-17.0060676570\\
360.0000000000	-16.4146962165\\
380.0000000000	-15.6932866500\\
400.0000000000	-14.8545805481\\
420.0000000000	-13.9134886497\\
440.0000000000	-12.8807541041\\
460.0000000000	-11.7596869593\\
480.0000000000	-10.5452034661\\
500.0000000000	-9.2238689551\\
520.0000000000	-7.7746173966\\
540.0000000000	-6.1719113072\\
560.0000000000	-4.3989462614\\
580.0000000000	-2.4900766342\\
600.0000000000	-0.6663413020\\
620.0000000000	0.0964379900\\
640.0000000000	-1.0507416122\\
660.0000000000	-3.0315612026\\
680.0000000000	-4.9385448338\\
700.0000000000	-6.5353816177\\
720.0000000000	-7.7892063109\\
740.0000000000	-8.7025458214\\
760.0000000000	-9.2802635112\\
780.0000000000	-9.5259083283\\
800.0000000000	-9.4409031018\\
820.0000000000	-9.0200587111\\
840.0000000000	-8.2404376562\\
860.0000000000	-7.0385957290\\
880.0000000000	-5.2629877729\\
900.0000000000	-2.5462466412\\
920.0000000000	2.4640618827\\
940.0000000000	9.3177149644\\
960.0000000000	12.0340832285\\
980.0000000000	10.0139952779\\
1000.0000000000	11.9296061371\\
};
\addlegendentry{no mic mid, $\lambda=0.1$}

\addplot [color=black, dashdotted]
  table[row sep=crcr]{%
200.0000000000	-25.9809522595\\
220.0000000000	-26.0397163748\\
240.0000000000	-25.8407092557\\
260.0000000000	-25.5111798765\\
280.0000000000	-25.1333802741\\
300.0000000000	-24.7406295517\\
320.0000000000	-24.3338188803\\
340.0000000000	-23.8995647710\\
360.0000000000	-23.4228613478\\
380.0000000000	-22.8964429915\\
400.0000000000	-22.3244947739\\
420.0000000000	-21.7189269054\\
440.0000000000	-21.0928779379\\
460.0000000000	-20.4561374283\\
480.0000000000	-19.8134226556\\
500.0000000000	-19.1645423991\\
520.0000000000	-18.5053982621\\
540.0000000000	-17.8291085569\\
560.0000000000	-17.1268097314\\
580.0000000000	-16.3879198058\\
600.0000000000	-15.5998226743\\
620.0000000000	-14.7469696653\\
640.0000000000	-13.8092729454\\
660.0000000000	-12.7593839426\\
680.0000000000	-11.5579019873\\
700.0000000000	-10.1443293959\\
720.0000000000	-8.4184264253\\
740.0000000000	-6.1974094954\\
760.0000000000	-3.1006928648\\
780.0000000000	1.9591917333\\
800.0000000000	7.2078944689\\
820.0000000000	7.6117651875\\
840.0000000000	2.9165956161\\
860.0000000000	-2.4917323607\\
880.0000000000	-5.7900921531\\
900.0000000000	-8.1352686389\\
920.0000000000	-9.8897504773\\
940.0000000000	-11.2158969431\\
960.0000000000	-12.1975628689\\
980.0000000000	-12.8826997233\\
1000.0000000000	-13.3027977799\\
};
\addlegendentry{square length $0.3\unit{m}$}

\end{axis}

\begin{axis}[%
width=1.226994\figurewidth,
height=1.226994\figureheight,
at={(-0.159509\figurewidth,-0.134969\figureheight)},
scale only axis,
xmin=0.0000000000,
xmax=1.0000000000,
ymin=0.0000000000,
ymax=1.0000000000,
axis line style={draw=none},
ticks=none,
axis x line*=bottom,
axis y line*=left,
scaled ticks=false, xticklabel style={/pgf/number format/fixed},yticklabel style={/pgf/number format/fixed}
]
\end{axis}
\end{tikzpicture}%
%}
	\caption{Comparison of results with a microphone in the middle and without a microphone in the middle. The wind speed is set to zero. For a better comparison the original setup's solution without microphone in the middle is also shown.}
	\label{fig:app_biggerSquareRE}
\end{figure}
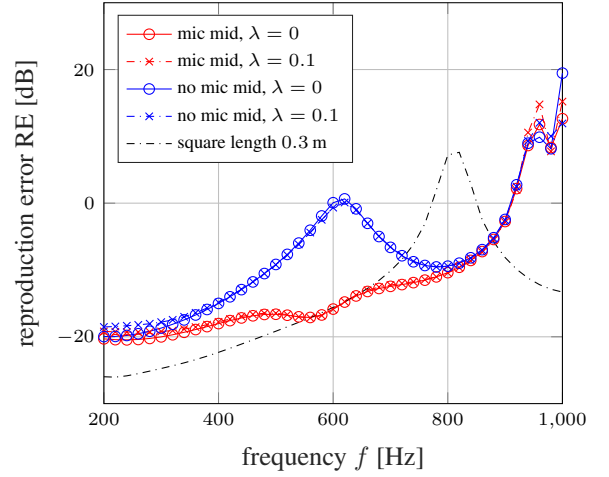
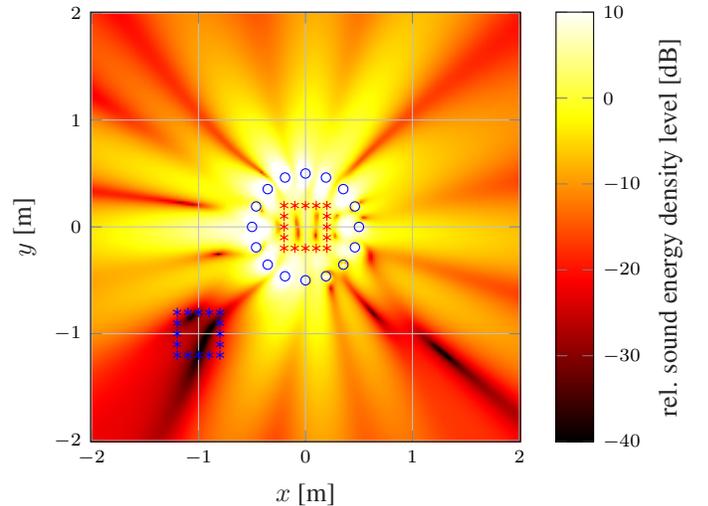
\begin{figure}
	\centering
	\setlength\figureheight{0.3\textwidth}
	\setlength\figurewidth{0.3\textwidth}
	%\resizebox{\linewidth}{!}{
    % This file was created by matlab2tikz.
%
\begin{tikzpicture}

\begin{axis}[%
width=\figurewidth,
height=\figureheight,
at={(0\figurewidth,0\figureheight)},
scale only axis,
point meta min=-40.0000000000,
point meta max=10.0000000000,
axis on top,
xmin=-2.0100502513,
xmax=2.0100502513,
xlabel style={font=\color{white!15!black}},
xlabel={$x$ [m]},
ymin=-2.0100502513,
xmajorgrids,
ymajorgrids,
ymax=2.0100502513,
ylabel style={font=\color{white!15!black}},
ylabel={$y$ [m]},
axis background/.style={fill=white},
colormap/hot2,
colorbar,
colorbar style={ylabel style={font=\color{white!15!black}}, ylabel={rel. sound energy density level [dB]}}
]
\addplot [forget plot] graphics [xmin=-2.0100502513, xmax=2.0100502513, ymin=-2.0100502513, ymax=2.0100502513] {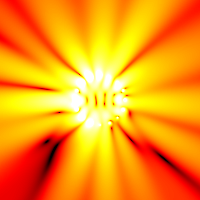};
\addplot[only marks, mark=asterisk, mark options={}, mark size=1.7678pt, draw=red, forget plot] table[row sep=crcr]{%
x	y\\
0.2000000000	-0.2000000000\\
0.2000000000	-0.1000000000\\
0.2000000000	0.0000000000\\
0.2000000000	0.1000000000\\
0.2000000000	0.2000000000\\
0.1000000000	0.2000000000\\
-0.0000000000	0.2000000000\\
-0.1000000000	0.2000000000\\
-0.2000000000	0.2000000000\\
-0.2000000000	0.1000000000\\
-0.2000000000	-0.0000000000\\
-0.2000000000	-0.1000000000\\
-0.2000000000	-0.2000000000\\
-0.1000000000	-0.2000000000\\
0.0000000000	-0.2000000000\\
0.1000000000	-0.2000000000\\
};
\addplot[only marks, mark=asterisk, mark options={}, mark size=1.7678pt, draw=blue, forget plot] table[row sep=crcr]{%
x	y\\
-0.8000000000	-1.2000000000\\
-0.8000000000	-1.1000000000\\
-0.8000000000	-1.0000000000\\
-0.8000000000	-0.9000000000\\
-0.8000000000	-0.8000000000\\
-0.9000000000	-0.8000000000\\
-1.0000000000	-0.8000000000\\
-1.1000000000	-0.8000000000\\
-1.2000000000	-0.8000000000\\
-1.2000000000	-0.9000000000\\
-1.2000000000	-1.0000000000\\
-1.2000000000	-1.1000000000\\
-1.2000000000	-1.2000000000\\
-1.1000000000	-1.2000000000\\
-1.0000000000	-1.2000000000\\
-0.9000000000	-1.2000000000\\
};
\addplot[only marks, mark=o, mark options={}, mark size=1.7678pt, draw=blue, forget plot] table[row sep=crcr]{%
x	y\\
0.5000000000	0.0000000000\\
0.4619397663	0.1913417162\\
0.3535533906	0.3535533906\\
0.1913417162	0.4619397663\\
0.0000000000	0.5000000000\\
-0.1913417162	0.4619397663\\
-0.3535533906	0.3535533906\\
-0.4619397663	0.1913417162\\
-0.5000000000	0.0000000000\\
-0.4619397663	-0.1913417162\\
-0.3535533906	-0.3535533906\\
-0.1913417162	-0.4619397663\\
-0.0000000000	-0.5000000000\\
0.1913417162	-0.4619397663\\
0.3535533906	-0.3535533906\\
0.4619397663	-0.1913417162\\
};
\end{axis}

\begin{axis}[%
width=1.226994\figurewidth,
height=1.226994\figureheight,
at={(-0.140681\figurewidth,-0.134969\figureheight)},
scale only axis,
point meta min=0.0000000000,
point meta max=1.0000000000,
xmin=0.0000000000,
xmax=1.0000000000,
ymin=0.0000000000,
ymax=1.0000000000,
axis line style={draw=none},
ticks=none,
axis x line*=bottom,
axis y line*=left
]
\end{axis}
\end{tikzpicture}%
%}
	\caption{The unregularised solution without wind at $\f=940\unit{Hz}$ is visualised to show the second resonance.}
	\label{fig:app_biggerSquareSolution}
\end{figure}
%%%%%%%%%%%%%%%%%%%%%%%%%%%%%%%%%%%%%%%%%%%%%%%%%%%%%%%%%%%%%%%%%%%%%%%%%%%%%%%%%%%%%%%%%%%%%%%%
\section{Effect of Noise on the Solution}
\label{sec:appNoise}
% \begin{itemize}
%     \item possible Gaussian noises examined
%     \item definition SNR
%     \item different mean values of SNR, different STD of SNR
% \end{itemize}
To get a feeling of the effects of noise on the solution an examination of the $\AC$ and the $\RE$ dependent on the signal to noise ratio SNR (in dB) and variance $\sigma^2$ of the noise is done. The SNR is distributed with a normal distribution with different mean values and different variances.

The noise gets added to the acoustic transfer functions in the following way:
\begin{linenomath}\begin{equation}
    g_\text{noise} = g + 10^{-\frac{\text{SNR}}{10\unit{dB}}}\cdot\left|g\right|\text{exp}\left(i\varphi\right)
\end{equation}\end{linenomath}
The phase angle $\varphi$ is randomly and uniformly distributed in the interval $[-\pi, \pi]$. For the uniformly distributed case the variance is set to 0 and the noise term gets scaled by an additional randomly and uniformly distributed scalar in the interval $[0, 1]$. The regularisation parameter is chosen according to the L-curve method using \cite{hansen_regtools_nodate}.

\begin{figure*}[htp!]
    \centering
	\begin{subfigure}[t]{0.48\textwidth}
    	\setlength\figureheight{0.38\textwidth}
    	\setlength\figurewidth{0.8\textwidth}
    	\centering
        % This file was created by matlab2tikz.
%
\definecolor{mycolor1}{rgb}{0.00000,0.44700,0.74100}%
\definecolor{mycolor2}{rgb}{0.85000,0.32500,0.09800}%
\definecolor{mycolor3}{rgb}{0.92900,0.69400,0.12500}%
\begin{tikzpicture}

\begin{axis}[%
width=\figurewidth,
height=\figureheight,
at={(0\figurewidth,0\figureheight)},
scale only axis,
xmin=200.0000000000,
xmax=1000.0000000000,
xlabel style={font=\color{white!15!black}},
xlabel={frequency $\f$ [Hz]},
ymin=-30.0000000000,
ymax=10.0000000000,
ylabel style={font=\color{white!15!black}},
ylabel={reproduction error $\RE$ [dB]},
axis background/.style={fill=white},
xmajorgrids,
ymajorgrids,
legend style={at={(0.03,0.97)}, anchor=north west, legend cell align=left, align=left, draw=white!15!black},
scaled ticks=false, xticklabel style={/pgf/number format/fixed},yticklabel style={/pgf/number format/fixed}
]
\addplot [color=mycolor1, mark=o, mark options={solid, mycolor1}]
  table[row sep=crcr]{%
200.0000000000	-24.2915639387\\
220.0000000000	-24.3701739488\\
240.0000000000	-25.0795724688\\
260.0000000000	-24.2170462370\\
280.0000000000	-24.6969827782\\
300.0000000000	-24.2855443449\\
320.0000000000	-24.0194042262\\
340.0000000000	-23.6396980570\\
360.0000000000	-23.2139165703\\
380.0000000000	-22.6320071546\\
400.0000000000	-22.1958484861\\
420.0000000000	-21.4684626908\\
440.0000000000	-20.8975231449\\
460.0000000000	-20.2063861533\\
480.0000000000	-19.6013847627\\
500.0000000000	-18.9013853062\\
520.0000000000	-18.2217660631\\
540.0000000000	-17.5563284254\\
560.0000000000	-17.0880142104\\
580.0000000000	-16.2298569293\\
600.0000000000	-15.5168947011\\
620.0000000000	-14.7083794080\\
640.0000000000	-13.8245073223\\
660.0000000000	-12.8305118399\\
680.0000000000	-11.5927930373\\
700.0000000000	-10.2810113022\\
720.0000000000	-8.6244447939\\
740.0000000000	-6.3720033754\\
760.0000000000	-3.2380807834\\
780.0000000000	1.8609625866\\
800.0000000000	8.0034866338\\
820.0000000000	-7.6904788130\\
840.0000000000	2.8587274116\\
860.0000000000	-2.7638771483\\
880.0000000000	-6.0383213776\\
900.0000000000	-8.4286901532\\
920.0000000000	-10.1846508278\\
940.0000000000	-11.4075980646\\
960.0000000000	-12.2532773009\\
980.0000000000	-12.8496639598\\
1000.0000000000	-13.1671812656\\
};
\addlegendentry{SNR $30\unit{dB}$}

\addplot [color=mycolor2, mark=o, mark options={solid, mycolor2}]
  table[row sep=crcr]{%
200.0000000000	-23.6423154509\\
220.0000000000	-24.5614253044\\
240.0000000000	-24.1818876458\\
260.0000000000	-24.1293288605\\
280.0000000000	-22.9765712895\\
300.0000000000	-23.8120381656\\
320.0000000000	-22.7289934347\\
340.0000000000	-21.5942832828\\
360.0000000000	-21.2896437904\\
380.0000000000	-20.5737808891\\
400.0000000000	-19.9905819856\\
420.0000000000	-21.5260841275\\
440.0000000000	-21.2934423380\\
460.0000000000	-19.7313597716\\
480.0000000000	-19.7436370513\\
500.0000000000	-18.9300956383\\
520.0000000000	-18.0294803924\\
540.0000000000	-17.4395291373\\
560.0000000000	-16.9602072798\\
580.0000000000	-16.5565845280\\
600.0000000000	-15.7267156360\\
620.0000000000	-14.5012071087\\
640.0000000000	-14.1198728937\\
660.0000000000	-13.2614911369\\
680.0000000000	-12.0451281542\\
700.0000000000	-10.1778120906\\
720.0000000000	-8.7423424537\\
740.0000000000	-6.6858139384\\
760.0000000000	-2.6728157040\\
780.0000000000	-9.4842311316\\
800.0000000000	-8.1307222050\\
820.0000000000	9.1252327134\\
840.0000000000	3.5002186556\\
860.0000000000	-2.4264706463\\
880.0000000000	-6.2478192330\\
900.0000000000	-8.2511329648\\
920.0000000000	-10.3474976957\\
940.0000000000	-11.4784211288\\
960.0000000000	-12.3828277905\\
980.0000000000	-12.5767190561\\
1000.0000000000	-13.2812732832\\
};
\addlegendentry{SNR $20\unit{dB}$}

\addplot [color=mycolor3, mark=o, mark options={solid, mycolor3}]
  table[row sep=crcr]{%
200.0000000000	-18.8496128110\\
220.0000000000	-19.8431158069\\
240.0000000000	-19.5284863922\\
260.0000000000	-19.8696699565\\
280.0000000000	-20.4222317767\\
300.0000000000	-22.6797939781\\
320.0000000000	-21.2756032443\\
340.0000000000	-15.4768686010\\
360.0000000000	-20.3677472851\\
380.0000000000	-18.2224619756\\
400.0000000000	-18.1355171849\\
420.0000000000	-18.7020129277\\
440.0000000000	-18.1738812196\\
460.0000000000	-17.7275609330\\
480.0000000000	-15.0689371936\\
500.0000000000	-15.9385375210\\
520.0000000000	-16.1784100512\\
540.0000000000	-17.8015729696\\
560.0000000000	-16.0815975601\\
580.0000000000	-13.9797833573\\
600.0000000000	-17.6906406217\\
620.0000000000	-13.1328635501\\
640.0000000000	-16.2583530098\\
660.0000000000	-13.9714269253\\
680.0000000000	-15.5742259185\\
700.0000000000	-10.6101497904\\
720.0000000000	-14.1429850309\\
740.0000000000	-9.6613005140\\
760.0000000000	-7.3618581536\\
780.0000000000	-7.6257441592\\
800.0000000000	-7.8770316434\\
820.0000000000	2.8116742697\\
840.0000000000	-12.2502148056\\
860.0000000000	-6.7970132769\\
880.0000000000	-8.7137277236\\
900.0000000000	-5.5875844732\\
920.0000000000	-11.2615793662\\
940.0000000000	-13.4185653344\\
960.0000000000	-12.1353248936\\
980.0000000000	-12.0950902672\\
1000.0000000000	-14.3507160019\\
};
\addlegendentry{SNR $10\unit{dB}$}

\addplot [color=black, dashdotted]
  table[row sep=crcr]{%
200.0000000000	-16.3622000000\\
220.0000000000	-16.0273000000\\
240.0000000000	-16.4708000000\\
260.0000000000	-17.0856000000\\
280.0000000000	-17.5057000000\\
300.0000000000	-17.2642000000\\
320.0000000000	-17.7077000000\\
340.0000000000	-17.6686000000\\
360.0000000000	-17.3337000000\\
380.0000000000	-17.5592000000\\
400.0000000000	-16.6716000000\\
420.0000000000	-16.3756000000\\
440.0000000000	-16.3754000000\\
460.0000000000	-16.2196000000\\
480.0000000000	-15.4176000000\\
500.0000000000	-15.3240000000\\
520.0000000000	-14.8646000000\\
540.0000000000	-14.0003000000\\
560.0000000000	-14.3193000000\\
580.0000000000	-12.6144000000\\
600.0000000000	-15.1207000000\\
620.0000000000	-10.7613000000\\
640.0000000000	-15.4394000000\\
660.0000000000	-10.2394000000\\
680.0000000000	-9.3284900000\\
700.0000000000	-9.3438600000\\
720.0000000000	-10.1766000000\\
740.0000000000	-10.4566000000\\
760.0000000000	-10.8767000000\\
780.0000000000	-11.6628000000\\
800.0000000000	-12.3086000000\\
820.0000000000	-12.0827000000\\
840.0000000000	-12.8998000000\\
860.0000000000	-12.5727000000\\
880.0000000000	-12.0043000000\\
900.0000000000	-11.7939000000\\
920.0000000000	-11.2488000000\\
940.0000000000	-10.8906000000\\
960.0000000000	-10.5245000000\\
980.0000000000	-9.5746600000\\
1000.0000000000	-8.9517400000\\
};
\addlegendentry{reference}

\end{axis}
\end{tikzpicture}%

		\caption{standard deviation $\sigma=0\unit{dB}$}
		\label{fig:app_noiseRESigma0}
	\end{subfigure}
~
	\begin{subfigure}[t]{0.48\textwidth}
    	\setlength\figureheight{0.38\textwidth}
    	\setlength\figurewidth{0.8\textwidth}
    	\centering
        % This file was created by matlab2tikz.
%
\definecolor{mycolor1}{rgb}{0.00000,0.44700,0.74100}%
\definecolor{mycolor2}{rgb}{0.85000,0.32500,0.09800}%
\definecolor{mycolor3}{rgb}{0.92900,0.69400,0.12500}%
\begin{tikzpicture}

\begin{axis}[%
width=\figurewidth,
height=\figureheight,
at={(0\figurewidth,0\figureheight)},
scale only axis,
xmin=200.0000000000,
xmax=1000.0000000000,
xlabel style={font=\color{white!15!black}},
xlabel={frequency $\f$ [Hz]},
ymin=-30.0000000000,
ymax=10.0000000000,
ylabel style={font=\color{white!15!black}},
ylabel={reproduction error $\RE$ [dB]},
axis background/.style={fill=white},
xmajorgrids,
ymajorgrids,
legend style={at={(0.03,0.97)}, anchor=north west, legend cell align=left, align=left, draw=white!15!black},
scaled ticks=false, xticklabel style={/pgf/number format/fixed},yticklabel style={/pgf/number format/fixed}
]
\addplot [color=mycolor1, mark=o, mark options={solid, mycolor1}]
  table[row sep=crcr]{%
200.0000000000	-24.2894706456\\
220.0000000000	-24.3490604687\\
240.0000000000	-25.3697557424\\
260.0000000000	-24.7875712763\\
280.0000000000	-24.7322751163\\
300.0000000000	-24.4089026723\\
320.0000000000	-23.8952302655\\
340.0000000000	-23.7437165261\\
360.0000000000	-23.1637222754\\
380.0000000000	-22.6854228101\\
400.0000000000	-22.1463140141\\
420.0000000000	-21.5765358062\\
440.0000000000	-20.9154719373\\
460.0000000000	-20.2364637346\\
480.0000000000	-19.5656832310\\
500.0000000000	-18.9206035146\\
520.0000000000	-18.2100668977\\
540.0000000000	-17.5736921273\\
560.0000000000	-16.8527018757\\
580.0000000000	-16.2377232112\\
600.0000000000	-15.6585768459\\
620.0000000000	-14.7014155778\\
640.0000000000	-13.7551856668\\
660.0000000000	-12.7441390800\\
680.0000000000	-11.5616689795\\
700.0000000000	-10.3441256696\\
720.0000000000	-8.5411098580\\
740.0000000000	-6.4257565026\\
760.0000000000	-3.1831463526\\
780.0000000000	1.9211432400\\
800.0000000000	7.7697630228\\
820.0000000000	8.7732861920\\
840.0000000000	2.8967420622\\
860.0000000000	-2.6405606120\\
880.0000000000	-6.0675729995\\
900.0000000000	-8.4060172622\\
920.0000000000	-10.1645295674\\
940.0000000000	-11.4039698360\\
960.0000000000	-12.2775537385\\
980.0000000000	-12.8387315694\\
1000.0000000000	-13.1929302637\\
};
\addlegendentry{SNR $30\unit{dB}$}

\addplot [color=mycolor2, mark=o, mark options={solid, mycolor2}]
  table[row sep=crcr]{%
200.0000000000	-23.8015684144\\
220.0000000000	-23.6198746403\\
240.0000000000	-24.4193732096\\
260.0000000000	-24.1714071588\\
280.0000000000	-24.3363872275\\
300.0000000000	-22.3702914323\\
320.0000000000	-21.8917842806\\
340.0000000000	-21.6885894165\\
360.0000000000	-22.5904694588\\
380.0000000000	-20.9083620746\\
400.0000000000	-21.9602942943\\
420.0000000000	-19.3053556726\\
440.0000000000	-20.6624315702\\
460.0000000000	-19.9940286148\\
480.0000000000	-19.4797412192\\
500.0000000000	-18.3207199838\\
520.0000000000	-17.8389859007\\
540.0000000000	-17.3978132463\\
560.0000000000	-16.4169282520\\
580.0000000000	-16.6675010726\\
600.0000000000	-15.7282780584\\
620.0000000000	-14.5740902094\\
640.0000000000	-14.1029398936\\
660.0000000000	-13.1320562530\\
680.0000000000	-12.4875081458\\
700.0000000000	-10.4013581535\\
720.0000000000	-8.1898221562\\
740.0000000000	-6.5569993490\\
760.0000000000	-3.0884816381\\
780.0000000000	1.0643587392\\
800.0000000000	-7.8628680274\\
820.0000000000	8.9516079885\\
840.0000000000	2.1028594974\\
860.0000000000	-2.3953175243\\
880.0000000000	-6.1834016861\\
900.0000000000	-8.3896382277\\
920.0000000000	-10.4373520813\\
940.0000000000	-11.3209050606\\
960.0000000000	-11.9559347270\\
980.0000000000	-12.8407289861\\
1000.0000000000	-11.2574232404\\
};
\addlegendentry{SNR $20\unit{dB}$}

\addplot [color=mycolor3, mark=o, mark options={solid, mycolor3}]
  table[row sep=crcr]{%
200.0000000000	-19.1940256943\\
220.0000000000	-21.2875806301\\
240.0000000000	-19.4186276750\\
260.0000000000	-21.4595879358\\
280.0000000000	-18.1512126621\\
300.0000000000	-20.9960498922\\
320.0000000000	-17.8076794559\\
340.0000000000	-15.1198546027\\
360.0000000000	-20.4486973275\\
380.0000000000	-21.8334169850\\
400.0000000000	-18.3380271649\\
420.0000000000	-17.7842473515\\
440.0000000000	-19.0148115958\\
460.0000000000	-14.8122547336\\
480.0000000000	-18.9078653101\\
500.0000000000	-15.0704763696\\
520.0000000000	-14.7333752384\\
540.0000000000	-12.0017884191\\
560.0000000000	-16.8969718425\\
580.0000000000	-13.6904423345\\
600.0000000000	-15.4160000569\\
620.0000000000	-14.1285652855\\
640.0000000000	-16.6397504870\\
660.0000000000	-11.6530394261\\
680.0000000000	-10.8377724282\\
700.0000000000	-9.8352947986\\
720.0000000000	-9.7141868157\\
740.0000000000	-11.0097814422\\
760.0000000000	-13.6525754845\\
780.0000000000	-7.0306760912\\
800.0000000000	-6.6786267909\\
820.0000000000	-8.1784441479\\
840.0000000000	-6.4772352511\\
860.0000000000	-11.6244220233\\
880.0000000000	-10.6456428892\\
900.0000000000	-12.6242905637\\
920.0000000000	-11.7919903325\\
940.0000000000	-11.7814012612\\
960.0000000000	-10.4821562717\\
980.0000000000	-11.2179025338\\
1000.0000000000	-13.0864137631\\
};
\addlegendentry{SNR $10\unit{dB}$}

\addplot [color=black, dashdotted]
  table[row sep=crcr]{%
200.0000000000	-16.3622000000\\
220.0000000000	-16.0273000000\\
240.0000000000	-16.4708000000\\
260.0000000000	-17.0856000000\\
280.0000000000	-17.5057000000\\
300.0000000000	-17.2642000000\\
320.0000000000	-17.7077000000\\
340.0000000000	-17.6686000000\\
360.0000000000	-17.3337000000\\
380.0000000000	-17.5592000000\\
400.0000000000	-16.6716000000\\
420.0000000000	-16.3756000000\\
440.0000000000	-16.3754000000\\
460.0000000000	-16.2196000000\\
480.0000000000	-15.4176000000\\
500.0000000000	-15.3240000000\\
520.0000000000	-14.8646000000\\
540.0000000000	-14.0003000000\\
560.0000000000	-14.3193000000\\
580.0000000000	-12.6144000000\\
600.0000000000	-15.1207000000\\
620.0000000000	-10.7613000000\\
640.0000000000	-15.4394000000\\
660.0000000000	-10.2394000000\\
680.0000000000	-9.3284900000\\
700.0000000000	-9.3438600000\\
720.0000000000	-10.1766000000\\
740.0000000000	-10.4566000000\\
760.0000000000	-10.8767000000\\
780.0000000000	-11.6628000000\\
800.0000000000	-12.3086000000\\
820.0000000000	-12.0827000000\\
840.0000000000	-12.8998000000\\
860.0000000000	-12.5727000000\\
880.0000000000	-12.0043000000\\
900.0000000000	-11.7939000000\\
920.0000000000	-11.2488000000\\
940.0000000000	-10.8906000000\\
960.0000000000	-10.5245000000\\
980.0000000000	-9.5746600000\\
1000.0000000000	-8.9517400000\\
};
\addlegendentry{reference}

\end{axis}
\end{tikzpicture}%
        \caption{standard deviation $\sigma=2\unit{dB}$}
        \label{fig:app_noiseRESigma2}
    \end{subfigure}
	\begin{subfigure}[t]{0.48\textwidth}
    	\setlength\figureheight{0.38\textwidth}
    	\setlength\figurewidth{0.8\textwidth}
    	\centering
        % This file was created by matlab2tikz.
%
\definecolor{mycolor1}{rgb}{0.00000,0.44700,0.74100}%
\definecolor{mycolor2}{rgb}{0.85000,0.32500,0.09800}%
\definecolor{mycolor3}{rgb}{0.92900,0.69400,0.12500}%
\begin{tikzpicture}

\begin{axis}[%
width=\figurewidth,
height=\figureheight,
at={(0\figurewidth,0\figureheight)},
scale only axis,
xmin=200.0000000000,
xmax=1000.0000000000,
xlabel style={font=\color{white!15!black}},
xlabel={frequency $\f$ [Hz]},
ymin=-25.0000000000,
ymax=15.0000000000,
ylabel style={font=\color{white!15!black}},
ylabel={reproduction error $\RE$ [dB]},
axis background/.style={fill=white},
xmajorgrids,
ymajorgrids,
legend style={at={(0.03,0.97)}, anchor=north west, legend cell align=left, align=left, draw=white!15!black},
scaled ticks=false, xticklabel style={/pgf/number format/fixed},yticklabel style={/pgf/number format/fixed}
]
\addplot [color=mycolor1, mark=o, mark options={solid, mycolor1}]
  table[row sep=crcr]{%
200.0000000000	-24.1427440909\\
220.0000000000	-24.2249575544\\
240.0000000000	-24.3253156883\\
260.0000000000	-24.8480935154\\
280.0000000000	-24.3897097284\\
300.0000000000	-24.2840845352\\
320.0000000000	-22.2341479552\\
340.0000000000	-23.5807451087\\
360.0000000000	-23.0682983378\\
380.0000000000	-22.6877295859\\
400.0000000000	-22.3026104273\\
420.0000000000	-21.7241573291\\
440.0000000000	-20.9170203260\\
460.0000000000	-20.1961093871\\
480.0000000000	-19.4349000040\\
500.0000000000	-18.8509853823\\
520.0000000000	-18.2208245240\\
540.0000000000	-17.7394914322\\
560.0000000000	-17.0240224882\\
580.0000000000	-16.2732082163\\
600.0000000000	-15.5984063968\\
620.0000000000	-14.8953357973\\
640.0000000000	-14.2497311571\\
660.0000000000	-12.8550801803\\
680.0000000000	-11.6288680360\\
700.0000000000	-10.1762397198\\
720.0000000000	-8.9258195243\\
740.0000000000	-6.2003483404\\
760.0000000000	-2.9971026121\\
780.0000000000	1.7248834537\\
800.0000000000	7.1832562634\\
820.0000000000	-7.7430991693\\
840.0000000000	3.2373928821\\
860.0000000000	-2.6532137271\\
880.0000000000	-5.9475368640\\
900.0000000000	-8.4761119967\\
920.0000000000	-10.0457111105\\
940.0000000000	-11.4816527335\\
960.0000000000	-12.3332753144\\
980.0000000000	-12.9982481281\\
1000.0000000000	-13.1224146081\\
};
\addlegendentry{SNR $30\unit{dB}$}

\addplot [color=mycolor2, mark=o, mark options={solid, mycolor2}]
  table[row sep=crcr]{%
200.0000000000	-23.3089620413\\
220.0000000000	-20.8913036947\\
240.0000000000	-19.4832614337\\
260.0000000000	-23.3342774296\\
280.0000000000	-22.4946302302\\
300.0000000000	-22.4254697502\\
320.0000000000	-21.9808452719\\
340.0000000000	-21.2847936022\\
360.0000000000	-20.9997140394\\
380.0000000000	-20.5680749957\\
400.0000000000	-21.2178427668\\
420.0000000000	-19.9466026118\\
440.0000000000	-18.4744217617\\
460.0000000000	-17.5569950250\\
480.0000000000	-17.4716857431\\
500.0000000000	-18.2402906798\\
520.0000000000	-17.2606672255\\
540.0000000000	-16.4656902122\\
560.0000000000	-17.0748853139\\
580.0000000000	-15.3996124387\\
600.0000000000	-15.0514526140\\
620.0000000000	-15.2916394314\\
640.0000000000	-12.7319097723\\
660.0000000000	-14.2016565413\\
680.0000000000	-13.4747770229\\
700.0000000000	-13.7129700190\\
720.0000000000	-10.4166463452\\
740.0000000000	-8.9951426262\\
760.0000000000	-9.8718078828\\
780.0000000000	-9.2996483902\\
800.0000000000	-9.2342169272\\
820.0000000000	-7.6423118326\\
840.0000000000	-7.8008326024\\
860.0000000000	-1.1397981503\\
880.0000000000	-6.5854921103\\
900.0000000000	-10.0498390004\\
920.0000000000	-10.6165086054\\
940.0000000000	-12.5677341267\\
960.0000000000	-11.6061431348\\
980.0000000000	-12.4749468794\\
1000.0000000000	-13.0687422434\\
};
\addlegendentry{SNR $20\unit{dB}$}

\addplot [color=mycolor3, mark=o, mark options={solid, mycolor3}]
  table[row sep=crcr]{%
200.0000000000	-17.1416181871\\
220.0000000000	-16.4133484362\\
240.0000000000	-15.7352286743\\
260.0000000000	-14.4120983656\\
280.0000000000	-17.0208538447\\
300.0000000000	-18.0731505655\\
320.0000000000	-15.2662603385\\
340.0000000000	-18.2785892222\\
360.0000000000	-16.6539776556\\
380.0000000000	-12.1442852512\\
400.0000000000	-20.1455051546\\
420.0000000000	-17.3476840684\\
440.0000000000	-16.4800381856\\
460.0000000000	-4.3345403851\\
480.0000000000	-13.6432019939\\
500.0000000000	-6.1673603031\\
520.0000000000	-12.3559068094\\
540.0000000000	-11.9080612431\\
560.0000000000	-9.1641171145\\
580.0000000000	-14.4271617916\\
600.0000000000	-16.1209262779\\
620.0000000000	-8.4838327139\\
640.0000000000	-12.1833962965\\
660.0000000000	-11.9225562360\\
680.0000000000	-8.8835096175\\
700.0000000000	-7.4026590311\\
720.0000000000	-9.1730719903\\
740.0000000000	-10.0806109709\\
760.0000000000	-10.4118415473\\
780.0000000000	-6.6469822411\\
800.0000000000	-10.8004930324\\
820.0000000000	12.6314857411\\
840.0000000000	-8.4617474345\\
860.0000000000	-7.6417031166\\
880.0000000000	-7.8047469754\\
900.0000000000	-5.9888101641\\
920.0000000000	-5.6144146259\\
940.0000000000	-7.7181907443\\
960.0000000000	-9.9418622653\\
980.0000000000	-9.8511938324\\
1000.0000000000	-8.0126853246\\
};
\addlegendentry{SNR $10\unit{dB}$}

\addplot [color=black, dashdotted]
  table[row sep=crcr]{%
200.0000000000	-16.3622000000\\
220.0000000000	-16.0273000000\\
240.0000000000	-16.4708000000\\
260.0000000000	-17.0856000000\\
280.0000000000	-17.5057000000\\
300.0000000000	-17.2642000000\\
320.0000000000	-17.7077000000\\
340.0000000000	-17.6686000000\\
360.0000000000	-17.3337000000\\
380.0000000000	-17.5592000000\\
400.0000000000	-16.6716000000\\
420.0000000000	-16.3756000000\\
440.0000000000	-16.3754000000\\
460.0000000000	-16.2196000000\\
480.0000000000	-15.4176000000\\
500.0000000000	-15.3240000000\\
520.0000000000	-14.8646000000\\
540.0000000000	-14.0003000000\\
560.0000000000	-14.3193000000\\
580.0000000000	-12.6144000000\\
600.0000000000	-15.1207000000\\
620.0000000000	-10.7613000000\\
640.0000000000	-15.4394000000\\
660.0000000000	-10.2394000000\\
680.0000000000	-9.3284900000\\
700.0000000000	-9.3438600000\\
720.0000000000	-10.1766000000\\
740.0000000000	-10.4566000000\\
760.0000000000	-10.8767000000\\
780.0000000000	-11.6628000000\\
800.0000000000	-12.3086000000\\
820.0000000000	-12.0827000000\\
840.0000000000	-12.8998000000\\
860.0000000000	-12.5727000000\\
880.0000000000	-12.0043000000\\
900.0000000000	-11.7939000000\\
920.0000000000	-11.2488000000\\
940.0000000000	-10.8906000000\\
960.0000000000	-10.5245000000\\
980.0000000000	-9.5746600000\\
1000.0000000000	-8.9517400000\\
};
\addlegendentry{reference}

\end{axis}
\end{tikzpicture}%
        \caption{standard deviation $\sigma=5\unit{dB}$}
        \label{fig:app_noiseRESigma5}
    \end{subfigure}
~
	\begin{subfigure}[t]{0.48\textwidth}
    	\setlength\figureheight{0.38\textwidth}
    	\setlength\figurewidth{0.8\textwidth}
    	\centering
        % This file was created by matlab2tikz.
%
\definecolor{mycolor1}{rgb}{0.00000,0.44700,0.74100}%
\definecolor{mycolor2}{rgb}{0.85000,0.32500,0.09800}%
\definecolor{mycolor3}{rgb}{0.92900,0.69400,0.12500}%
\begin{tikzpicture}

\begin{axis}[%
width=\figurewidth,
height=\figureheight,
at={(0\figurewidth,0\figureheight)},
scale only axis,
xmin=200.0000000000,
xmax=1000.0000000000,
xlabel style={font=\color{white!15!black}},
xlabel={frequency $\f$ [Hz]},
ymin=-25.0000000000,
ymax=10.0000000000,
ylabel style={font=\color{white!15!black}},
ylabel={reproduction error $\RE$ [dB]},
axis background/.style={fill=white},
xmajorgrids,
ymajorgrids,
legend style={at={(0.03,0.97)}, anchor=north west, legend cell align=left, align=left, draw=white!15!black},
scaled ticks=false, xticklabel style={/pgf/number format/fixed},yticklabel style={/pgf/number format/fixed}
]
\addplot [color=mycolor1, mark=o, mark options={solid, mycolor1}]
  table[row sep=crcr]{%
200.0000000000	-24.2848875740\\
220.0000000000	-24.4021874843\\
240.0000000000	-24.3496384781\\
260.0000000000	-24.9925926445\\
280.0000000000	-24.6389488941\\
300.0000000000	-24.4229460263\\
320.0000000000	-24.0278716593\\
340.0000000000	-23.7133591896\\
360.0000000000	-23.1862359000\\
380.0000000000	-22.6523379923\\
400.0000000000	-22.1160043847\\
420.0000000000	-21.5325899857\\
440.0000000000	-20.8906301310\\
460.0000000000	-20.2435963228\\
480.0000000000	-19.5740427902\\
500.0000000000	-18.9059028507\\
520.0000000000	-18.3582365921\\
540.0000000000	-17.5562724058\\
560.0000000000	-16.9043004902\\
580.0000000000	-16.2480377750\\
600.0000000000	-15.5083768034\\
620.0000000000	-14.6927982405\\
640.0000000000	-13.7523460294\\
660.0000000000	-12.7844634522\\
680.0000000000	-11.6416330185\\
700.0000000000	-10.2408865441\\
720.0000000000	-8.5242048907\\
740.0000000000	-6.3143113020\\
760.0000000000	-3.2546448232\\
780.0000000000	1.7871863141\\
800.0000000000	7.5910703066\\
820.0000000000	8.5101914560\\
840.0000000000	2.7448891825\\
860.0000000000	-2.7165795383\\
880.0000000000	-6.0466196660\\
900.0000000000	-8.4249565822\\
920.0000000000	-10.1663380354\\
940.0000000000	-11.4040142334\\
960.0000000000	-12.2852734035\\
980.0000000000	-12.8799691836\\
1000.0000000000	-13.1807078635\\
};
\addlegendentry{SNR $30\unit{dB}$}

\addplot [color=mycolor2, mark=o, mark options={solid, mycolor2}]
  table[row sep=crcr]{%
200.0000000000	-24.2539195050\\
220.0000000000	-24.3300481573\\
240.0000000000	-24.3953696313\\
260.0000000000	-24.5520536243\\
280.0000000000	-24.1962387263\\
300.0000000000	-22.7466737417\\
320.0000000000	-23.7126635172\\
340.0000000000	-23.8628950043\\
360.0000000000	-21.2186752902\\
380.0000000000	-22.4619672901\\
400.0000000000	-22.0567977638\\
420.0000000000	-21.4698366188\\
440.0000000000	-20.9814819599\\
460.0000000000	-20.2621591514\\
480.0000000000	-19.4897229811\\
500.0000000000	-19.0965405366\\
520.0000000000	-18.6041390543\\
540.0000000000	-17.6284389924\\
560.0000000000	-16.9082655365\\
580.0000000000	-16.1572444889\\
600.0000000000	-15.4529581099\\
620.0000000000	-15.0099393056\\
640.0000000000	-14.0314377790\\
660.0000000000	-13.2872561353\\
680.0000000000	-11.5795649315\\
700.0000000000	-10.2842871141\\
720.0000000000	-8.5884316822\\
740.0000000000	-6.6326235556\\
760.0000000000	-3.5674883239\\
780.0000000000	1.8144832653\\
800.0000000000	-7.9569238178\\
820.0000000000	7.1350872441\\
840.0000000000	1.9258840246\\
860.0000000000	-2.7649846264\\
880.0000000000	-6.1879908366\\
900.0000000000	-8.4891894895\\
920.0000000000	-10.1230074975\\
940.0000000000	-11.4114783983\\
960.0000000000	-12.2937804067\\
980.0000000000	-12.7989339236\\
1000.0000000000	-13.2512733949\\
};
\addlegendentry{SNR $20\unit{dB}$}

\addplot [color=mycolor3, mark=o, mark options={solid, mycolor3}]
  table[row sep=crcr]{%
200.0000000000	-20.4870303269\\
220.0000000000	-20.9927610370\\
240.0000000000	-20.5023222100\\
260.0000000000	-21.1042147744\\
280.0000000000	-21.9000862485\\
300.0000000000	-23.1805693051\\
320.0000000000	-21.6548040864\\
340.0000000000	-21.8517920443\\
360.0000000000	-20.4329529707\\
380.0000000000	-20.4593539373\\
400.0000000000	-20.1858960212\\
420.0000000000	-18.9085981406\\
440.0000000000	-18.2298746231\\
460.0000000000	-18.1520219779\\
480.0000000000	-16.9010318136\\
500.0000000000	-18.7925921998\\
520.0000000000	-16.9575713828\\
540.0000000000	-16.5888821569\\
560.0000000000	-15.6073174063\\
580.0000000000	-16.4108038655\\
600.0000000000	-15.2889507767\\
620.0000000000	-14.4299171451\\
640.0000000000	-13.7447613415\\
660.0000000000	-11.8786249379\\
680.0000000000	-13.2122725976\\
700.0000000000	-12.8025284392\\
720.0000000000	-12.0793016397\\
740.0000000000	-10.9040050604\\
760.0000000000	-2.6709192059\\
780.0000000000	-10.3529260614\\
800.0000000000	-9.9237647246\\
820.0000000000	-7.1590270261\\
840.0000000000	-7.2234522502\\
860.0000000000	-0.2547883317\\
880.0000000000	-4.8701841345\\
900.0000000000	-10.4068825236\\
920.0000000000	-10.8903634553\\
940.0000000000	-10.5140546686\\
960.0000000000	-13.9964618446\\
980.0000000000	-12.2782086953\\
1000.0000000000	-11.5102582730\\
};
\addlegendentry{SNR $10\unit{dB}$}

\addplot [color=black, dashdotted]
  table[row sep=crcr]{%
200.0000000000	-16.3622000000\\
220.0000000000	-16.0273000000\\
240.0000000000	-16.4708000000\\
260.0000000000	-17.0856000000\\
280.0000000000	-17.5057000000\\
300.0000000000	-17.2642000000\\
320.0000000000	-17.7077000000\\
340.0000000000	-17.6686000000\\
360.0000000000	-17.3337000000\\
380.0000000000	-17.5592000000\\
400.0000000000	-16.6716000000\\
420.0000000000	-16.3756000000\\
440.0000000000	-16.3754000000\\
460.0000000000	-16.2196000000\\
480.0000000000	-15.4176000000\\
500.0000000000	-15.3240000000\\
520.0000000000	-14.8646000000\\
540.0000000000	-14.0003000000\\
560.0000000000	-14.3193000000\\
580.0000000000	-12.6144000000\\
600.0000000000	-15.1207000000\\
620.0000000000	-10.7613000000\\
640.0000000000	-15.4394000000\\
660.0000000000	-10.2394000000\\
680.0000000000	-9.3284900000\\
700.0000000000	-9.3438600000\\
720.0000000000	-10.1766000000\\
740.0000000000	-10.4566000000\\
760.0000000000	-10.8767000000\\
780.0000000000	-11.6628000000\\
800.0000000000	-12.3086000000\\
820.0000000000	-12.0827000000\\
840.0000000000	-12.8998000000\\
860.0000000000	-12.5727000000\\
880.0000000000	-12.0043000000\\
900.0000000000	-11.7939000000\\
920.0000000000	-11.2488000000\\
940.0000000000	-10.8906000000\\
960.0000000000	-10.5245000000\\
980.0000000000	-9.5746600000\\
1000.0000000000	-8.9517400000\\
};
\addlegendentry{reference}

\end{axis}
\end{tikzpicture}%
        \caption{uniformly distributed}
        \label{fig:app_noiseREUniform}
    \end{subfigure}
    \caption{$\RE$ dependent on the signal to noise ratio SNR, regularisation according to L-curve method}
    \label{fig:app_noiseRE}
\end{figure*}
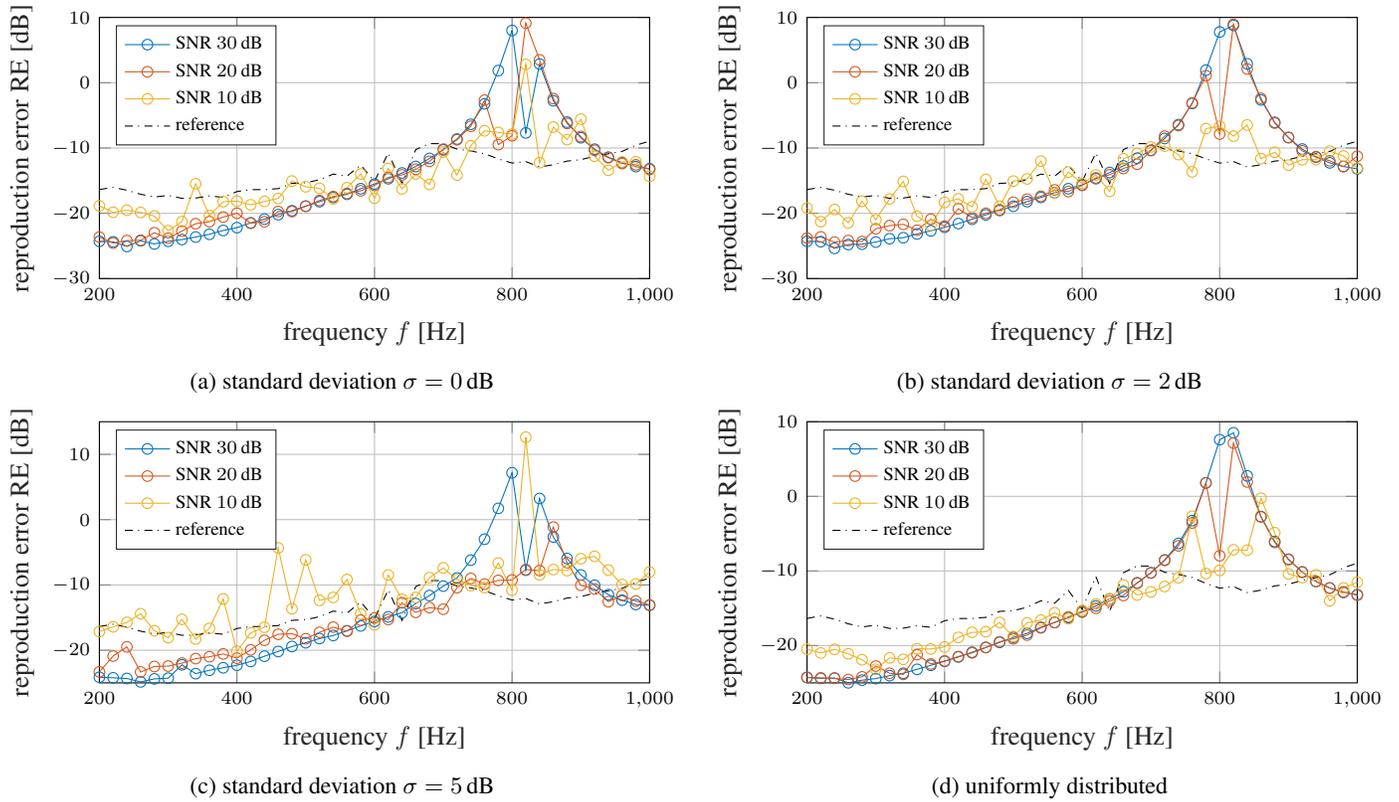
\begin{figure*}
    \centering
	\begin{subfigure}[t]{0.48\textwidth}
    	\setlength\figureheight{0.38\textwidth}
    	\setlength\figurewidth{0.8\textwidth}
    	\centering
        % This file was created by matlab2tikz.
%
\definecolor{mycolor1}{rgb}{0.00000,0.44700,0.74100}%
\definecolor{mycolor2}{rgb}{0.85000,0.32500,0.09800}%
\definecolor{mycolor3}{rgb}{0.92900,0.69400,0.12500}%
\begin{tikzpicture}

\begin{axis}[%
width=\figurewidth,
height=\figureheight,
at={(0\figurewidth,0\figureheight)},
scale only axis,
xmin=200.0000000000,
xmax=1000.0000000000,
xlabel style={font=\color{white!15!black}},
xlabel={frequency $\f$ [Hz]},
ymin=15.0000000000,
ymax=80.0000000000,
ylabel style={font=\color{white!15!black}},
ylabel={acoustic contrast $\AC$ [dB]},
axis background/.style={fill=white},
xmajorgrids,
ymajorgrids,
legend style={at={(0.03,0.97)}, anchor=north west, legend cell align=left, align=left, draw=white!15!black},
scaled ticks=false, xticklabel style={/pgf/number format/fixed},yticklabel style={/pgf/number format/fixed}
]
\addplot [color=mycolor1, mark=o, mark options={solid, mycolor1}]
  table[row sep=crcr]{%
200.0000000000	31.7123054342\\
220.0000000000	32.3223503701\\
240.0000000000	31.7288129058\\
260.0000000000	33.0426450781\\
280.0000000000	32.7582857891\\
300.0000000000	33.2208477778\\
320.0000000000	33.7008336809\\
340.0000000000	34.9532904903\\
360.0000000000	35.1542136447\\
380.0000000000	36.2424877061\\
400.0000000000	37.3451887257\\
420.0000000000	38.4394806333\\
440.0000000000	39.1492042587\\
460.0000000000	39.4636190458\\
480.0000000000	39.0929066022\\
500.0000000000	38.3812103619\\
520.0000000000	37.4243685083\\
540.0000000000	36.3118493966\\
560.0000000000	48.3938199882\\
580.0000000000	34.4906501775\\
600.0000000000	46.8304236831\\
620.0000000000	46.6368584082\\
640.0000000000	45.0375673429\\
660.0000000000	44.1129427994\\
680.0000000000	43.1206894518\\
700.0000000000	42.0629457563\\
720.0000000000	41.3315222016\\
740.0000000000	40.4520064125\\
760.0000000000	39.9395786439\\
780.0000000000	39.6155651655\\
800.0000000000	39.7429492084\\
820.0000000000	36.0900713124\\
840.0000000000	38.5338002799\\
860.0000000000	37.6204702071\\
880.0000000000	36.9353849074\\
900.0000000000	36.6983632110\\
920.0000000000	36.6512328767\\
940.0000000000	36.8240387988\\
960.0000000000	37.4791648924\\
980.0000000000	38.3987212491\\
1000.0000000000	39.5071658063\\
};
\addlegendentry{SNR $30\unit{dB}$}

\addplot [color=mycolor2, mark=o, mark options={solid, mycolor2}]
  table[row sep=crcr]{%
200.0000000000	31.6448477883\\
220.0000000000	32.6339462692\\
240.0000000000	32.9249787196\\
260.0000000000	33.4131400944\\
280.0000000000	32.7458845135\\
300.0000000000	33.1506183273\\
320.0000000000	33.7070433688\\
340.0000000000	33.7984029149\\
360.0000000000	33.7268006309\\
380.0000000000	33.6687757432\\
400.0000000000	33.3575087286\\
420.0000000000	40.8559160094\\
440.0000000000	39.8626808521\\
460.0000000000	36.9456600857\\
480.0000000000	38.5231055813\\
500.0000000000	36.8000354094\\
520.0000000000	37.4440125863\\
540.0000000000	35.9440855406\\
560.0000000000	35.5613341980\\
580.0000000000	34.1775923980\\
600.0000000000	33.5639494336\\
620.0000000000	33.0130953667\\
640.0000000000	32.4782180130\\
660.0000000000	32.2847460949\\
680.0000000000	31.8655797445\\
700.0000000000	40.6731668791\\
720.0000000000	38.7638749090\\
740.0000000000	39.3381675739\\
760.0000000000	38.4339998459\\
780.0000000000	34.8347675216\\
800.0000000000	36.1681267452\\
820.0000000000	39.3503921385\\
840.0000000000	39.0150808512\\
860.0000000000	37.5488712568\\
880.0000000000	37.6018089764\\
900.0000000000	36.1794690307\\
920.0000000000	36.9981915662\\
940.0000000000	36.7016311924\\
960.0000000000	37.2381156665\\
980.0000000000	38.8871421619\\
1000.0000000000	39.4696097571\\
};
\addlegendentry{SNR $20\unit{dB}$}

\addplot [color=mycolor3, mark=o, mark options={solid, mycolor3}]
  table[row sep=crcr]{%
200.0000000000	24.6964790022\\
220.0000000000	23.4968553879\\
240.0000000000	23.4783515502\\
260.0000000000	23.5537996767\\
280.0000000000	29.5296512194\\
300.0000000000	31.0550610489\\
320.0000000000	30.7684990209\\
340.0000000000	21.9192263426\\
360.0000000000	30.3606635783\\
380.0000000000	31.5933398620\\
400.0000000000	33.0828571293\\
420.0000000000	32.0920179599\\
440.0000000000	32.3486341357\\
460.0000000000	33.6759562259\\
480.0000000000	28.9446594182\\
500.0000000000	37.0338650988\\
520.0000000000	36.7069188050\\
540.0000000000	35.6539927559\\
560.0000000000	34.2214279091\\
580.0000000000	34.4894366438\\
600.0000000000	35.5981532611\\
620.0000000000	29.9814732735\\
640.0000000000	31.4280176403\\
660.0000000000	29.1529869137\\
680.0000000000	30.6183663988\\
700.0000000000	29.8108978014\\
720.0000000000	32.6891779313\\
740.0000000000	33.1797916451\\
760.0000000000	31.0860868800\\
780.0000000000	33.3368424763\\
800.0000000000	34.7220157684\\
820.0000000000	34.7830557175\\
840.0000000000	34.4891899125\\
860.0000000000	33.2424048989\\
880.0000000000	31.2696665952\\
900.0000000000	28.3131168264\\
920.0000000000	32.0182187515\\
940.0000000000	29.2778667265\\
960.0000000000	28.6002534337\\
980.0000000000	29.3616555475\\
1000.0000000000	34.1447872346\\
};
\addlegendentry{SNR $10\unit{dB}$}

\addplot [color=black, dashdotted]
  table[row sep=crcr]{%
200.0000000000	19.8396000000\\
220.0000000000	24.3606000000\\
240.0000000000	26.3177000000\\
260.0000000000	26.7274000000\\
280.0000000000	29.1851000000\\
300.0000000000	29.6402000000\\
320.0000000000	27.1370000000\\
340.0000000000	31.6428000000\\
360.0000000000	30.6567000000\\
380.0000000000	34.1006000000\\
400.0000000000	34.1916000000\\
420.0000000000	34.1309000000\\
440.0000000000	36.0577000000\\
460.0000000000	34.6316000000\\
480.0000000000	37.8023000000\\
500.0000000000	32.8414000000\\
520.0000000000	33.1751000000\\
540.0000000000	39.5622000000\\
560.0000000000	32.6593000000\\
580.0000000000	35.8149000000\\
600.0000000000	31.9007000000\\
620.0000000000	31.3849000000\\
640.0000000000	33.6454000000\\
660.0000000000	30.5657000000\\
680.0000000000	29.9285000000\\
700.0000000000	31.7339000000\\
720.0000000000	32.0221000000\\
740.0000000000	34.4950000000\\
760.0000000000	35.9666000000\\
780.0000000000	36.5583000000\\
800.0000000000	37.8327000000\\
820.0000000000	38.9402000000\\
840.0000000000	35.9211000000\\
860.0000000000	38.6519000000\\
880.0000000000	39.1071000000\\
900.0000000000	38.4699000000\\
920.0000000000	38.8340000000\\
940.0000000000	37.8934000000\\
960.0000000000	35.7846000000\\
980.0000000000	36.8314000000\\
1000.0000000000	33.7213000000\\
};
\addlegendentry{reference}

\end{axis}
\end{tikzpicture}%

		\caption{standard deviation $\sigma=0\unit{dB}$}
		\label{fig:app_noiseACSigma0}
	\end{subfigure}
~
	\begin{subfigure}[t]{0.48\textwidth}
    	\setlength\figureheight{0.38\textwidth}
    	\setlength\figurewidth{0.8\textwidth}
    	\centering
        % This file was created by matlab2tikz.
%
\definecolor{mycolor1}{rgb}{0.00000,0.44700,0.74100}%
\definecolor{mycolor2}{rgb}{0.85000,0.32500,0.09800}%
\definecolor{mycolor3}{rgb}{0.92900,0.69400,0.12500}%
\begin{tikzpicture}

\begin{axis}[%
width=\figurewidth,
height=\figureheight,
at={(0\figurewidth,0\figureheight)},
scale only axis,
xmin=200.0000000000,
xmax=1000.0000000000,
xlabel style={font=\color{white!15!black}},
xlabel={frequency $\f$ [Hz]},
ymin=15.0000000000,
ymax=80.0000000000,
ylabel style={font=\color{white!15!black}},
ylabel={acoustic contrast $\AC$ [dB]},
axis background/.style={fill=white},
xmajorgrids,
ymajorgrids,
legend style={at={(0.03,0.97)}, anchor=north west, legend cell align=left, align=left, draw=white!15!black},
scaled ticks=false, xticklabel style={/pgf/number format/fixed},yticklabel style={/pgf/number format/fixed}
]
\addplot [color=mycolor1, mark=o, mark options={solid, mycolor1}]
  table[row sep=crcr]{%
200.0000000000	31.6714081168\\
220.0000000000	32.3725326826\\
240.0000000000	32.1101017980\\
260.0000000000	32.1513887336\\
280.0000000000	32.4351578926\\
300.0000000000	33.3554641832\\
320.0000000000	33.6898602262\\
340.0000000000	34.5443975935\\
360.0000000000	36.2450171495\\
380.0000000000	36.2694032040\\
400.0000000000	37.3784743761\\
420.0000000000	40.8996597856\\
440.0000000000	39.2238681045\\
460.0000000000	39.5095752695\\
480.0000000000	39.1952596380\\
500.0000000000	38.3169404701\\
520.0000000000	37.4280760377\\
540.0000000000	36.3740798610\\
560.0000000000	35.3325271775\\
580.0000000000	34.4449156831\\
600.0000000000	46.6355190057\\
620.0000000000	45.6133680686\\
640.0000000000	45.1836696272\\
660.0000000000	43.9857727355\\
680.0000000000	43.2210973670\\
700.0000000000	42.2537307632\\
720.0000000000	40.8588814099\\
740.0000000000	40.2205840619\\
760.0000000000	40.1301044658\\
780.0000000000	39.6516306975\\
800.0000000000	39.6479696180\\
820.0000000000	39.6713639927\\
840.0000000000	38.4950427535\\
860.0000000000	37.6716142951\\
880.0000000000	37.0448464768\\
900.0000000000	36.5747015990\\
920.0000000000	36.5465824126\\
940.0000000000	36.8080942087\\
960.0000000000	37.3955373238\\
980.0000000000	38.3633148225\\
1000.0000000000	39.5140677227\\
};
\addlegendentry{SNR $30\unit{dB}$}

\addplot [color=mycolor2, mark=o, mark options={solid, mycolor2}]
  table[row sep=crcr]{%
200.0000000000	31.7838940952\\
220.0000000000	31.4345034742\\
240.0000000000	33.2703380658\\
260.0000000000	33.4134554997\\
280.0000000000	33.7533388776\\
300.0000000000	32.8931472433\\
320.0000000000	33.5231111015\\
340.0000000000	33.3206755591\\
360.0000000000	34.0639767369\\
380.0000000000	33.5395008262\\
400.0000000000	35.8292734321\\
420.0000000000	33.6161801432\\
440.0000000000	38.3702575078\\
460.0000000000	38.7473923677\\
480.0000000000	38.8210621619\\
500.0000000000	37.3889826370\\
520.0000000000	37.2274501021\\
540.0000000000	36.2504779429\\
560.0000000000	36.6993070477\\
580.0000000000	34.3871051120\\
600.0000000000	33.9208122091\\
620.0000000000	33.3278959234\\
640.0000000000	32.2489225634\\
660.0000000000	31.8997749889\\
680.0000000000	31.6235130631\\
700.0000000000	38.3957518592\\
720.0000000000	41.3686141791\\
740.0000000000	40.2800872888\\
760.0000000000	38.4318049190\\
780.0000000000	38.9304175968\\
800.0000000000	36.0627952526\\
820.0000000000	38.5454575964\\
840.0000000000	39.4491565178\\
860.0000000000	37.0488840812\\
880.0000000000	37.5170651714\\
900.0000000000	36.1623658251\\
920.0000000000	36.0934677688\\
940.0000000000	36.2411424868\\
960.0000000000	37.3589203650\\
980.0000000000	39.4004000228\\
1000.0000000000	29.9626508977\\
};
\addlegendentry{SNR $20\unit{dB}$}

\addplot [color=mycolor3, mark=o, mark options={solid, mycolor3}]
  table[row sep=crcr]{%
200.0000000000	26.1189935077\\
220.0000000000	24.9743160478\\
240.0000000000	23.7165018701\\
260.0000000000	27.3747090008\\
280.0000000000	22.8105856663\\
300.0000000000	32.8622958635\\
320.0000000000	22.6586517638\\
340.0000000000	20.7209358359\\
360.0000000000	32.7115009466\\
380.0000000000	28.5665324773\\
400.0000000000	28.9097749017\\
420.0000000000	33.1803114010\\
440.0000000000	35.5686385271\\
460.0000000000	31.5707518451\\
480.0000000000	31.0017042893\\
500.0000000000	31.2888025806\\
520.0000000000	38.0528286928\\
540.0000000000	28.1583298609\\
560.0000000000	35.8795574145\\
580.0000000000	36.0326177545\\
600.0000000000	30.4267274648\\
620.0000000000	32.1501304638\\
640.0000000000	33.9594022727\\
660.0000000000	29.0021243687\\
680.0000000000	30.5256164934\\
700.0000000000	31.2874886695\\
720.0000000000	28.0495508060\\
740.0000000000	33.2158992251\\
760.0000000000	30.2374037783\\
780.0000000000	31.7197261912\\
800.0000000000	32.6949570917\\
820.0000000000	35.5844815404\\
840.0000000000	32.8054080694\\
860.0000000000	30.9163037852\\
880.0000000000	31.2783564603\\
900.0000000000	31.6950987012\\
920.0000000000	30.6990390779\\
940.0000000000	31.2474761516\\
960.0000000000	30.1967948322\\
980.0000000000	33.0836007850\\
1000.0000000000	28.3384687791\\
};
\addlegendentry{SNR $10\unit{dB}$}

\addplot [color=black, dashdotted]
  table[row sep=crcr]{%
200.0000000000	19.8396000000\\
220.0000000000	24.3606000000\\
240.0000000000	26.3177000000\\
260.0000000000	26.7274000000\\
280.0000000000	29.1851000000\\
300.0000000000	29.6402000000\\
320.0000000000	27.1370000000\\
340.0000000000	31.6428000000\\
360.0000000000	30.6567000000\\
380.0000000000	34.1006000000\\
400.0000000000	34.1916000000\\
420.0000000000	34.1309000000\\
440.0000000000	36.0577000000\\
460.0000000000	34.6316000000\\
480.0000000000	37.8023000000\\
500.0000000000	32.8414000000\\
520.0000000000	33.1751000000\\
540.0000000000	39.5622000000\\
560.0000000000	32.6593000000\\
580.0000000000	35.8149000000\\
600.0000000000	31.9007000000\\
620.0000000000	31.3849000000\\
640.0000000000	33.6454000000\\
660.0000000000	30.5657000000\\
680.0000000000	29.9285000000\\
700.0000000000	31.7339000000\\
720.0000000000	32.0221000000\\
740.0000000000	34.4950000000\\
760.0000000000	35.9666000000\\
780.0000000000	36.5583000000\\
800.0000000000	37.8327000000\\
820.0000000000	38.9402000000\\
840.0000000000	35.9211000000\\
860.0000000000	38.6519000000\\
880.0000000000	39.1071000000\\
900.0000000000	38.4699000000\\
920.0000000000	38.8340000000\\
940.0000000000	37.8934000000\\
960.0000000000	35.7846000000\\
980.0000000000	36.8314000000\\
1000.0000000000	33.7213000000\\
};
\addlegendentry{reference}

\end{axis}
\end{tikzpicture}%
        \caption{standard deviation $\sigma=2\unit{dB}$}
        \label{fig:app_noiseACSigma2}
    \end{subfigure}
	\begin{subfigure}[t]{0.48\textwidth}
    	\setlength\figureheight{0.38\textwidth}
    	\setlength\figurewidth{0.8\textwidth}
    	\centering
        % This file was created by matlab2tikz.
%
\definecolor{mycolor1}{rgb}{0.00000,0.44700,0.74100}%
\definecolor{mycolor2}{rgb}{0.85000,0.32500,0.09800}%
\definecolor{mycolor3}{rgb}{0.92900,0.69400,0.12500}%
\begin{tikzpicture}

\begin{axis}[%
width=\figurewidth,
height=\figureheight,
at={(0\figurewidth,0\figureheight)},
scale only axis,
xmin=200.0000000000,
xmax=1000.0000000000,
xlabel style={font=\color{white!15!black}},
xlabel={frequency $\f$ [Hz]},
ymin=15.0000000000,
ymax=80.0000000000,
ylabel style={font=\color{white!15!black}},
ylabel={acoustic contrast $\AC$ [dB]},
axis background/.style={fill=white},
xmajorgrids,
ymajorgrids,
legend style={at={(0.03,0.97)}, anchor=north west, legend cell align=left, align=left, draw=white!15!black},
scaled ticks=false, xticklabel style={/pgf/number format/fixed},yticklabel style={/pgf/number format/fixed}
]
\addplot [color=mycolor1, mark=o, mark options={solid, mycolor1}]
  table[row sep=crcr]{%
200.0000000000	31.8064720316\\
220.0000000000	32.3607576766\\
240.0000000000	32.7621271254\\
260.0000000000	33.4018477567\\
280.0000000000	33.5135020398\\
300.0000000000	32.9837368079\\
320.0000000000	33.3307298970\\
340.0000000000	33.8221993959\\
360.0000000000	35.8607349450\\
380.0000000000	36.5773123750\\
400.0000000000	37.2771539642\\
420.0000000000	39.0946371126\\
440.0000000000	42.6461562176\\
460.0000000000	39.7563665289\\
480.0000000000	38.9992918072\\
500.0000000000	38.2727322640\\
520.0000000000	37.2689695260\\
540.0000000000	36.2804820624\\
560.0000000000	35.6014794473\\
580.0000000000	34.5159608782\\
600.0000000000	33.6728368341\\
620.0000000000	32.9099955279\\
640.0000000000	32.4533251112\\
660.0000000000	45.1824873661\\
680.0000000000	41.7624361011\\
700.0000000000	42.2528430810\\
720.0000000000	40.9611731873\\
740.0000000000	39.7120700169\\
760.0000000000	38.9576285611\\
780.0000000000	39.1576282913\\
800.0000000000	39.3643000448\\
820.0000000000	36.0482027290\\
840.0000000000	38.3418810853\\
860.0000000000	37.4761174785\\
880.0000000000	36.8421338500\\
900.0000000000	36.4528658321\\
920.0000000000	36.9001743097\\
940.0000000000	36.9026924395\\
960.0000000000	37.5062726885\\
980.0000000000	38.5200543332\\
1000.0000000000	39.3589621453\\
};
\addlegendentry{SNR $30\unit{dB}$}

\addplot [color=mycolor2, mark=o, mark options={solid, mycolor2}]
  table[row sep=crcr]{%
200.0000000000	30.4216093325\\
220.0000000000	24.0610111976\\
240.0000000000	23.0231621382\\
260.0000000000	32.2124881557\\
280.0000000000	32.1704422748\\
300.0000000000	35.0621179273\\
320.0000000000	32.3146855992\\
340.0000000000	33.0899384741\\
360.0000000000	34.6497346736\\
380.0000000000	33.4740950847\\
400.0000000000	35.2899402297\\
420.0000000000	31.5537043216\\
440.0000000000	33.2285292075\\
460.0000000000	33.1223015098\\
480.0000000000	32.3281373351\\
500.0000000000	36.9239124508\\
520.0000000000	37.9218395706\\
540.0000000000	37.2189355404\\
560.0000000000	35.4288417125\\
580.0000000000	34.5777477677\\
600.0000000000	35.0251640326\\
620.0000000000	32.9930522468\\
640.0000000000	33.8728363136\\
660.0000000000	30.6724295363\\
680.0000000000	30.9643214596\\
700.0000000000	30.2655035708\\
720.0000000000	32.3286814832\\
740.0000000000	33.6823866488\\
760.0000000000	33.7847216018\\
780.0000000000	34.7797754493\\
800.0000000000	36.4008543884\\
820.0000000000	35.8143869298\\
840.0000000000	35.0919886109\\
860.0000000000	38.9383303530\\
880.0000000000	35.5913812209\\
900.0000000000	32.8836274193\\
920.0000000000	36.3490507060\\
940.0000000000	32.9093169519\\
960.0000000000	31.6106201836\\
980.0000000000	37.0910411879\\
1000.0000000000	37.6122607850\\
};
\addlegendentry{SNR $20\unit{dB}$}

\addplot [color=mycolor3, mark=o, mark options={solid, mycolor3}]
  table[row sep=crcr]{%
200.0000000000	23.8726270324\\
220.0000000000	22.0569134155\\
240.0000000000	19.6681220278\\
260.0000000000	22.6108378411\\
280.0000000000	22.3023262681\\
300.0000000000	29.6468141409\\
320.0000000000	22.6972383111\\
340.0000000000	31.1056465789\\
360.0000000000	26.1633191320\\
380.0000000000	17.9189097774\\
400.0000000000	31.3001217355\\
420.0000000000	27.8369926068\\
440.0000000000	24.8287880886\\
460.0000000000	19.6728323399\\
480.0000000000	31.5648269495\\
500.0000000000	22.9558502231\\
520.0000000000	25.8192942853\\
540.0000000000	28.8219175329\\
560.0000000000	27.4731837512\\
580.0000000000	30.6727845527\\
600.0000000000	28.9920320361\\
620.0000000000	26.2829057923\\
640.0000000000	26.6338714328\\
660.0000000000	31.6410677676\\
680.0000000000	24.5297102012\\
700.0000000000	21.1322005123\\
720.0000000000	30.5123300821\\
740.0000000000	26.1268820858\\
760.0000000000	31.2824806092\\
780.0000000000	27.9658069299\\
800.0000000000	31.8740602407\\
820.0000000000	24.7359923289\\
840.0000000000	30.7025666818\\
860.0000000000	30.5631479194\\
880.0000000000	27.3559884500\\
900.0000000000	28.0046744120\\
920.0000000000	33.0538223215\\
940.0000000000	26.4126364685\\
960.0000000000	25.2652898610\\
980.0000000000	25.8634784898\\
1000.0000000000	26.2215601771\\
};
\addlegendentry{SNR $10\unit{dB}$}

\addplot [color=black, dashdotted]
  table[row sep=crcr]{%
200.0000000000	19.8396000000\\
220.0000000000	24.3606000000\\
240.0000000000	26.3177000000\\
260.0000000000	26.7274000000\\
280.0000000000	29.1851000000\\
300.0000000000	29.6402000000\\
320.0000000000	27.1370000000\\
340.0000000000	31.6428000000\\
360.0000000000	30.6567000000\\
380.0000000000	34.1006000000\\
400.0000000000	34.1916000000\\
420.0000000000	34.1309000000\\
440.0000000000	36.0577000000\\
460.0000000000	34.6316000000\\
480.0000000000	37.8023000000\\
500.0000000000	32.8414000000\\
520.0000000000	33.1751000000\\
540.0000000000	39.5622000000\\
560.0000000000	32.6593000000\\
580.0000000000	35.8149000000\\
600.0000000000	31.9007000000\\
620.0000000000	31.3849000000\\
640.0000000000	33.6454000000\\
660.0000000000	30.5657000000\\
680.0000000000	29.9285000000\\
700.0000000000	31.7339000000\\
720.0000000000	32.0221000000\\
740.0000000000	34.4950000000\\
760.0000000000	35.9666000000\\
780.0000000000	36.5583000000\\
800.0000000000	37.8327000000\\
820.0000000000	38.9402000000\\
840.0000000000	35.9211000000\\
860.0000000000	38.6519000000\\
880.0000000000	39.1071000000\\
900.0000000000	38.4699000000\\
920.0000000000	38.8340000000\\
940.0000000000	37.8934000000\\
960.0000000000	35.7846000000\\
980.0000000000	36.8314000000\\
1000.0000000000	33.7213000000\\
};
\addlegendentry{reference}

\end{axis}
\end{tikzpicture}%
        \caption{standard deviation $\sigma=5\unit{dB}$}
        \label{fig:app_noiseACSigma5}
    \end{subfigure}
~
	\begin{subfigure}[t]{0.48\textwidth}
    	\setlength\figureheight{0.38\textwidth}
    	\setlength\figurewidth{0.8\textwidth}
    	\centering
        % This file was created by matlab2tikz.
%
\definecolor{mycolor1}{rgb}{0.00000,0.44700,0.74100}%
\definecolor{mycolor2}{rgb}{0.85000,0.32500,0.09800}%
\definecolor{mycolor3}{rgb}{0.92900,0.69400,0.12500}%
\begin{tikzpicture}

\begin{axis}[%
width=\figurewidth,
height=\figureheight,
at={(0\figurewidth,0\figureheight)},
scale only axis,
xmin=200.0000000000,
xmax=1000.0000000000,
xlabel style={font=\color{white!15!black}},
xlabel={frequency $\f$ [Hz]},
ymin=15.0000000000,
ymax=80.0000000000,
ylabel style={font=\color{white!15!black}},
ylabel={acoustic contrast $\AC$ [dB]},
axis background/.style={fill=white},
xmajorgrids,
ymajorgrids,
legend style={at={(0.03,0.97)}, anchor=north west, legend cell align=left, align=left, draw=white!15!black},
scaled ticks=false, xticklabel style={/pgf/number format/fixed},yticklabel style={/pgf/number format/fixed}
]
\addplot [color=mycolor1, mark=o, mark options={solid, mycolor1}]
  table[row sep=crcr]{%
200.0000000000	31.6741384753\\
220.0000000000	32.3918400604\\
240.0000000000	32.7948793695\\
260.0000000000	31.9995559286\\
280.0000000000	32.5757108500\\
300.0000000000	33.1400788562\\
320.0000000000	33.8148897115\\
340.0000000000	34.9815913060\\
360.0000000000	35.2645490268\\
380.0000000000	36.2177655652\\
400.0000000000	37.3486543127\\
420.0000000000	38.4570705952\\
440.0000000000	39.2570671974\\
460.0000000000	39.4846500698\\
480.0000000000	39.1433046108\\
500.0000000000	38.3934328498\\
520.0000000000	48.6979769645\\
540.0000000000	36.3458449223\\
560.0000000000	35.3579224195\\
580.0000000000	34.4463812279\\
600.0000000000	46.8765988444\\
620.0000000000	45.8200902841\\
640.0000000000	45.1213704689\\
660.0000000000	44.2486991959\\
680.0000000000	43.2757185288\\
700.0000000000	42.1281898034\\
720.0000000000	41.3776125936\\
740.0000000000	40.4241556600\\
760.0000000000	39.8930711574\\
780.0000000000	39.6455905608\\
800.0000000000	39.6653906880\\
820.0000000000	39.4845870120\\
840.0000000000	38.6077513598\\
860.0000000000	37.6242418008\\
880.0000000000	36.9685052017\\
900.0000000000	36.6075475030\\
920.0000000000	36.5869242783\\
940.0000000000	36.8513272327\\
960.0000000000	37.4869796772\\
980.0000000000	38.4227728307\\
1000.0000000000	39.5210267417\\
};
\addlegendentry{SNR $30\unit{dB}$}

\addplot [color=mycolor2, mark=o, mark options={solid, mycolor2}]
  table[row sep=crcr]{%
200.0000000000	32.2916854703\\
220.0000000000	32.3039463725\\
240.0000000000	32.8502919419\\
260.0000000000	33.2431620398\\
280.0000000000	33.3963830503\\
300.0000000000	33.3341563389\\
320.0000000000	33.4820771613\\
340.0000000000	34.0327119416\\
360.0000000000	33.5809545166\\
380.0000000000	37.0741830487\\
400.0000000000	37.4000065050\\
420.0000000000	38.1018286346\\
440.0000000000	39.1132699714\\
460.0000000000	39.1182848041\\
480.0000000000	38.6494632402\\
500.0000000000	38.3747277570\\
520.0000000000	43.5985796421\\
540.0000000000	36.1114766910\\
560.0000000000	35.2958958674\\
580.0000000000	34.5222052030\\
600.0000000000	33.4148246652\\
620.0000000000	32.9549770016\\
640.0000000000	32.6245196492\\
660.0000000000	32.1656705686\\
680.0000000000	43.6452297530\\
700.0000000000	40.1555574186\\
720.0000000000	39.8835671508\\
740.0000000000	39.6467131822\\
760.0000000000	39.5265611193\\
780.0000000000	38.9757982288\\
800.0000000000	36.1035924318\\
820.0000000000	38.9541014412\\
840.0000000000	38.7994077608\\
860.0000000000	37.8126847444\\
880.0000000000	37.0314931706\\
900.0000000000	36.5934957411\\
920.0000000000	36.6276329987\\
940.0000000000	36.7431210784\\
960.0000000000	37.4618375016\\
980.0000000000	38.0755007697\\
1000.0000000000	39.5647066049\\
};
\addlegendentry{SNR $20\unit{dB}$}

\addplot [color=mycolor3, mark=o, mark options={solid, mycolor3}]
  table[row sep=crcr]{%
200.0000000000	24.0346324814\\
220.0000000000	24.5847750023\\
240.0000000000	23.9181495629\\
260.0000000000	33.3324799050\\
280.0000000000	33.6662769997\\
300.0000000000	31.7498270512\\
320.0000000000	32.1357164098\\
340.0000000000	32.4056142277\\
360.0000000000	32.5630820695\\
380.0000000000	33.2538400496\\
400.0000000000	32.8905628638\\
420.0000000000	32.5852375504\\
440.0000000000	33.7135317978\\
460.0000000000	35.3413244116\\
480.0000000000	31.8400181856\\
500.0000000000	36.9828553774\\
520.0000000000	36.8767911564\\
540.0000000000	37.8923051946\\
560.0000000000	35.3223246447\\
580.0000000000	36.1151971458\\
600.0000000000	35.0786474806\\
620.0000000000	34.1888892478\\
640.0000000000	32.0511656706\\
660.0000000000	31.1466609534\\
680.0000000000	29.8201357250\\
700.0000000000	30.3843906329\\
720.0000000000	29.6231775215\\
740.0000000000	32.9784122523\\
760.0000000000	38.1803887999\\
780.0000000000	34.6021930586\\
800.0000000000	35.9901745892\\
820.0000000000	34.6613749956\\
840.0000000000	36.2349424907\\
860.0000000000	33.5617600524\\
880.0000000000	36.3671115050\\
900.0000000000	33.8808987888\\
920.0000000000	31.9651755981\\
940.0000000000	35.0172872821\\
960.0000000000	35.2370414014\\
980.0000000000	29.0106738511\\
1000.0000000000	29.7641177244\\
};
\addlegendentry{SNR $10\unit{dB}$}

\addplot [color=black, dashdotted]
  table[row sep=crcr]{%
200.0000000000	19.8396000000\\
220.0000000000	24.3606000000\\
240.0000000000	26.3177000000\\
260.0000000000	26.7274000000\\
280.0000000000	29.1851000000\\
300.0000000000	29.6402000000\\
320.0000000000	27.1370000000\\
340.0000000000	31.6428000000\\
360.0000000000	30.6567000000\\
380.0000000000	34.1006000000\\
400.0000000000	34.1916000000\\
420.0000000000	34.1309000000\\
440.0000000000	36.0577000000\\
460.0000000000	34.6316000000\\
480.0000000000	37.8023000000\\
500.0000000000	32.8414000000\\
520.0000000000	33.1751000000\\
540.0000000000	39.5622000000\\
560.0000000000	32.6593000000\\
580.0000000000	35.8149000000\\
600.0000000000	31.9007000000\\
620.0000000000	31.3849000000\\
640.0000000000	33.6454000000\\
660.0000000000	30.5657000000\\
680.0000000000	29.9285000000\\
700.0000000000	31.7339000000\\
720.0000000000	32.0221000000\\
740.0000000000	34.4950000000\\
760.0000000000	35.9666000000\\
780.0000000000	36.5583000000\\
800.0000000000	37.8327000000\\
820.0000000000	38.9402000000\\
840.0000000000	35.9211000000\\
860.0000000000	38.6519000000\\
880.0000000000	39.1071000000\\
900.0000000000	38.4699000000\\
920.0000000000	38.8340000000\\
940.0000000000	37.8934000000\\
960.0000000000	35.7846000000\\
980.0000000000	36.8314000000\\
1000.0000000000	33.7213000000\\
};
\addlegendentry{reference}

\end{axis}
\end{tikzpicture}%
        \caption{uniformly distributed}
        \label{fig:app_noiseACUniform}
    \end{subfigure}
    \caption{Dependance of $\AC$ on the signal to noise ratio SNR with regularisation according to L-curve method as in \cite{du_multizone_2021}.}
    \label{fig:app_noiseAC}
\end{figure*}
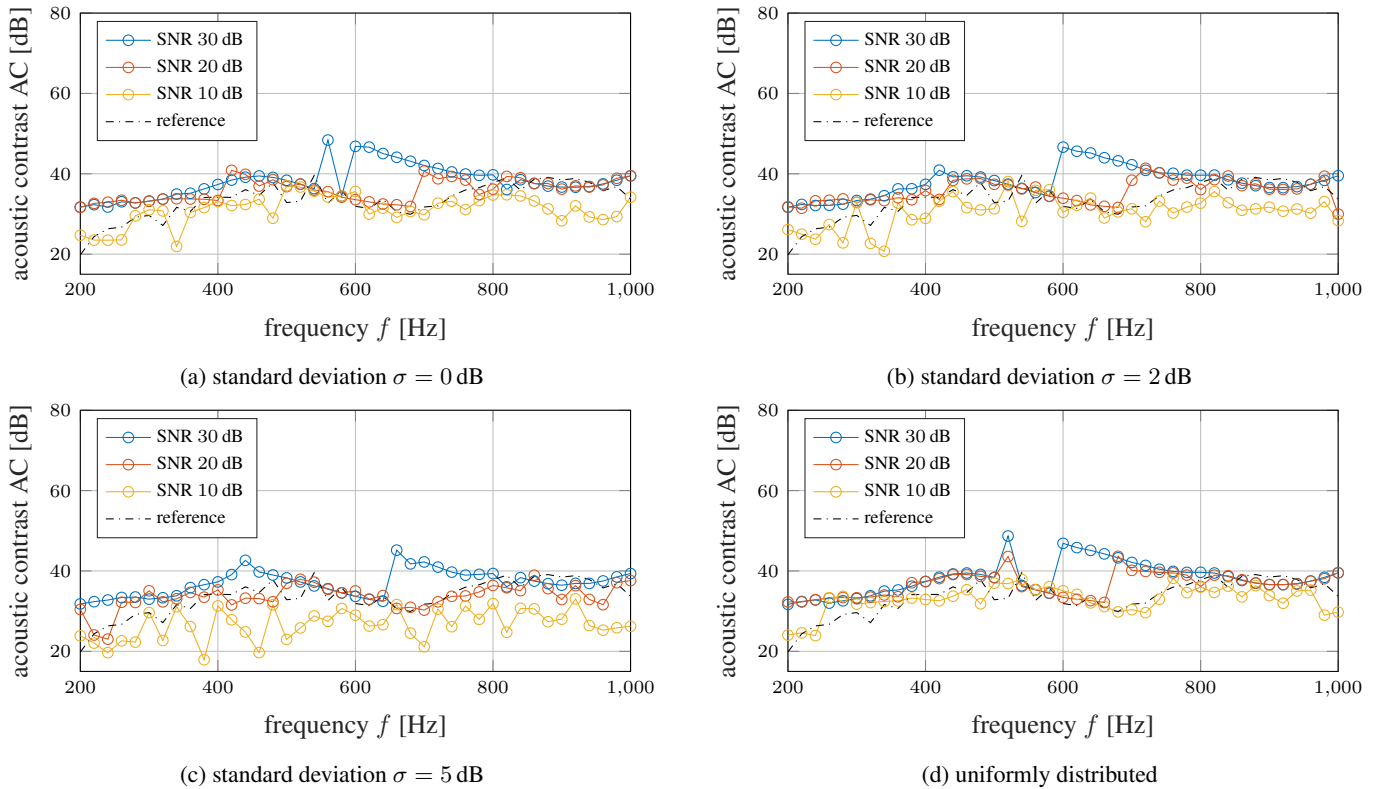

The results are shown in \Cref{fig:app_noiseAC} for the $\AC$ and in \Cref{fig:app_noiseRE} for the $\RE$. Since the $\RE$ is quite resistant to changes in the regularisation parameter, the effect of the noise on the $\RE$ can be assessed. It is visible that a SNR of $30\unit{dB}$ only leads to minor changes in the $\RE$ where a smaller SNR of $10\unit{dB}$ leads to bigger changes and an approximation of the $\RE$ to the reference values.

The $\AC$ however shows that even with a big SNR of $30\unit{dB}$ big changes in the $\AC$ are notable. The regularisation parameter is chosen according to the L-curve method. In \Cref{fig:benchNoWind_acousticContrast} it is visible, that the regularisation parameter has a big influence on the $\AC$. This leads to the possibility that changes in the $\AC$ can also be induced by the noise's impact on the L-curve method or fluctuations of the L-curve method itself.

\section{Parameters for the 2D Finite Difference Euler Equation Simulation}
\label{sec:appParamsEuler}
The \philipp{spatial} and temporal discretisation schemes and boundary conditions are described in \cite{lemke_adjoint_2015}.

The simulation is done in a 8m $\times$ 8m free field domain (non-reflecting boundary conditions) discretised with 500 uniformly spaced points with the grid size $h$ in $x$ and $y$ direction. The domain size is chosen in a way to prevent errors in the non-reflecting boundary conditions at the boundary of the domain to influence the results.

The time step $\Delta t=1/(48\unit{kHz})$ is chosen to match the common sampling frequency of $48\unit{kHz}$. We use 800 time steps in total, where the last 100 time steps are treated as the steady state solution.

A speed of sound $c=343\unit{m/s}$ is realised at $y=0\unit{m}$. The adiabatic index is set to $\gamma=1.4$ and the ambient values are as following: The ambient pressure $\pScalar_\infty=101325\unit{Pa}$, the specific gas constant of air $R_s=287.058\unit{J/(kgK)}$ and the temperature $T=19.597\unit{°C} + \vert\nabla T\vert y$. The horizontal component of the wind profile is introduced as $u_x = 0.4673\Ma\cdot\speedSound\cdot\text{ln}\left((y + 2.3\unit{m})/0.3\unit{m} \right)$. Scaled to the wind speed of $5\unit{Bft}$($9.35\unit{m/s}$ at $y=0\unit{m}$ this leads to a maximum wind speed of $11.63\unit{m/s}$ at $y=2\unit{m}$. Below of $y=-2\unit{m}$ the component is set to $u_x=0\unit{m/s}$. In the vertical direction the wind is set to $u_y=0$. The horizontal wind component is smoothed around the kink at $y=-2.3\unit{m}$. This is done using a circle with radius 0.4.

% in den veröffentlichten Papern des Journals ist die Bibliography hinter dem Appendix

\end{document}